\documentclass[twocolumn]{aastex63}
\usepackage[utf8]{inputenc}
\usepackage{hyperref}
\usepackage{float}
\usepackage{verbatim}
\usepackage{natbib}
\usepackage[caption=false]{subfig}
\usepackage{multirow}
\usepackage{dcolumn}
\usepackage[nomarkers,nolists]{endfloat}
\newcolumntype{d}[1]{D{.}{.}{#1}}

\usepackage{xcolor}

\begin{document}

\title{97 Eclipsing Quadruple Star Candidates Discovered in {\em TESS} Full Frame Images}

\correspondingauthor{Veselin B. Kostov}
\email{veselin.b.kostov@nasa.gov}
\author[0000-0001-9786-1031]{Veselin~B.~Kostov}
\affiliation{NASA Goddard Space Flight Center, 8800 Greenbelt Road, Greenbelt, MD 20771, USA}
\affiliation{SETI Institute, 189 Bernardo Ave, Suite 200, Mountain View, CA 94043, USA}
\affiliation{NASA Goddard Space Flight Center Sellers Exoplanet Environments Collaboration}
\author[0000-0003-0501-2636]{Brian P. Powell}
\affiliation{NASA Goddard Space Flight Center, 8800 Greenbelt Road, Greenbelt, MD 20771, USA}
\author[0000-0003-3182-5569]{Saul A. Rappaport}
\affiliation{Department of Physics, Kavli Institute for Astrophysics and Space Research, M.I.T., Cambridge, MA 02139, USA}
\author[0000-0002-8806-496X]{Tam\'as Borkovits}
\affiliation{Baja Astronomical Observatory of University of Szeged, H-6500 Baja, Szegedi út, Kt. 766, Hungary}
\affiliation{Konkoly Observatory, Research Centre for Astronomy and Earth Sciences, H-1121 Budapest, Konkoly Thege Miklós út 15-17, Hungary}
\affiliation{ELTE Gothard Astrophysical Observatory, H-9700 Szombathely, Szent Imre h. u. 112, Hungary}
\author[0000-0002-5665-1879]{Robert Gagliano}
\affiliation{Amateur Astronomer, Glendale, AZ 85308}
\author[0000-0003-3988-3245]{Thomas L. Jacobs}
\affiliation{Amateur Astronomer, 12812 SE 69th Place, Bellevue, WA 98006}
\author[0000-0002-2607-138X]{Martti~H.~Kristiansen}
\affil{Brorfelde Observatory, Observator Gyldenkernes Vej 7, DK-4340 T\o{}ll\o{}se, Denmark}
%
\author[0000-0002-8527-2114]{Daryll M. LaCourse}
\affiliation{Amateur Astronomer, 7507 52nd Place NE Marysville, WA 98270}
\author{Mark Omohundro}
\affiliation{Citizen Scientist, c/o Zooniverse, Department of Physics, University of Oxford, Denys Wilkinson Building, Keble Road, Oxford, OX13RH, UK}
\author{Jerome Orosz}
\affiliation{Department of Astronomy, San Diego State University, 5500 Campanile Drive, San Diego, CA 92182, USA}
\author[0000-0002-5034-0949]{Allan R. Schmitt}
\affiliation{Citizen Scientist, 616 W. 53rd. St., Apt. 101, Minneapolis, MN 55419, USA}
\author{Hans M. Schwengeler}
\affiliation{Citizen Scientist, Planet Hunter, Bottmingen, Switzerland}
\author{Ivan A. Terentev}
\affiliation{Citizen Scientist, Planet Hunter, Petrozavodsk, Russia}
\author[0000-0002-5286-0251]{Guillermo Torres}
\affiliation{Center for Astrophysics $\vert$ Harvard \& Smithsonian, 60 Garden St, Cambridge, MA, 02138, USA}
\author{Thomas Barclay}
\affiliation{NASA Goddard Space Flight Center, 8800 Greenbelt Road, Greenbelt, MD 20771, USA}
\affiliation{University of Maryland, Baltimore County, 1000 Hilltop Circle, Baltimore, MD 21250, USA}
\author{Adam H. Friedman}
\affiliation{University of Michigan, 500 S State St, Ann Arbor, MI 48109}
\affiliation{NASA Goddard Space Flight Center, 8800 Greenbelt Road, Greenbelt, MD 20771, USA}
\author{Ethan Kruse}
\affiliation{NASA Goddard Space Flight Center, 8800 Greenbelt Road, Greenbelt, MD 20771, USA}
\author[0000-0001-8472-2219]{Greg Olmschenk}
\affiliation{NASA Goddard Space Flight Center, 8800 Greenbelt Road, Greenbelt, MD 20771, USA}
\affiliation{Universities Space Research Association, 7178 Columbia Gateway Drive, Columbia, MD 21046}
\author[0000-0001-7246-5438]{Andrew Vanderburg}
\affiliation{Department of Astronomy, University of Wisconsin-Madison, Madison, WI 53706, USA}
\author{William Welsh}
\affiliation{Department of Astronomy, San Diego State University, 5500 Campanile Drive, San Diego, CA 92182, USA}
\begin{abstract}
We present a catalog of 97 uniformly-vetted candidates for quadruple star systems. The candidates were identified in {\em TESS} Full Frame Image data from Sectors 1 through 42 through a combination of machine learning techniques and visual examination, with major contributions from a dedicated group of citizen scientists. All targets exhibit two sets of eclipses with two different periods, both of which pass photocenter tests confirming that the eclipses are on-target. This catalog outlines the statistical properties of the sample, nearly doubles the number of known multiply-eclipsing quadruple systems, and provides the basis for detailed future studies of individual systems. Several important discoveries have already resulted from this effort, including the first sextuply-eclipsing sextuple stellar system and the first transiting circumbinary planet detected from one sector of {\em TESS} data. 
\end{abstract}

\accepted{ApJS}


\section{Introduction}\label{sec:intro}

Half of Sun-like stars are members of binary systems \citep{2010ApJS..190....1R} and about one in ten of these are triples and quadruples \citep{2018ApJS..235....6T}. The higher-order multiplicity fraction increases with stellar mass, to the point where massive single stars are an exception rather than the norm \citep[e.g.][]{2017ApJS..230...15M}. Multiple stellar systems are important tracers of stellar formation, can experience rich interactions such as Lidov-Kozai oscillations \citep{1962P&SS....9..719L,1962AJ.....67..591K} or dynamical instability, and provide pathways for important stages of stellar evolution such as short-period binaries, common-envelope events, Type Ia Supernovae, black hole mergers \citep[e.g.][]{2013MNRAS.435..943P,2018MNRAS.476.4234F,2021MNRAS.502.4479H,2019MNRAS.486.4781F,2019MNRAS.483.4060L}. For example, the mass ratios between the individual components of a quadruple system, the period ratios between the constituent binary systems and the mutual inclination provide important insight into whether the system formed through a `top-down' scenario via core or disk fragmentation, or `bottom-up' aggregation via gravitational capture \citep[e.g.][]{1994ARA&A..32..465M,2015Natur.518..213P,2016Natur.538..483T,Tokovinin2021,Whitworth2001}. In addition, the evolution of a compact quadruple stellar system can include a combination of single-star evolution, interactions between the two stars in the constituent binary systems, as well as dynamical interactions between the two binary systems. Detection and characterization of eclipsing binary stars in quadruple stellar systems provide an excellent opportunity to explore these processes \citep[e.g.][]{2016MNRAS.455.4136B,2016MNRAS.462.1812R,2017MNRAS.467.2160R,Tokovinin2021}.

To study multiple stellar systems, we have been performing a search for eclipsing binary stars (EBs) utilizing the long-cadence {\em TESS} lightcurves (Kruse et al. in prep). The lightcurves were created using the {\tt eleanor} pipeline (Feinstein et al. 2019), which uses the Full-Frame Image {\em TESS} data to extract photometry on a target-by-target basis. A natural by-product of this search is the identification of 97 candidates for quadruple stellar systems based on the presence of (at least) two sets of eclipses following two distinct periods and/or measured eclipse-timing variations (ETVs). These eclipses indicate quadruple candidates with a 2+2 hierarchical configuration. As part of this effort, we have already discovered the first sextuply-eclipsing sextuple system \citep[TIC 168789840,][]{2021AJ....161..162P}, a compact and coplanar quadruply-eclipsing quadruple system \citep[TIC 454140642,][]{2021ApJ...917...93K}, and a transiting circumbinary planet \citep[TIC 172900988,][]{2021arXiv210508614K}. 

Multiply-eclipsing stellar quadruples are highly valuable as they provide precise measurements of orbital periods, relative stellar sizes, temperatures, and orbital inclinations -- yet are quite rare as they require fortuitous alignment with the observer. At the time of writing there are about 150 ${\it candidates}$ for such systems, many of which could be false positives caused by two unrelated binaries, and only a handful of ${\it confirmed}$ systems \citep[e.g.][]{2019A&A...630A.128Z}. Thus the catalog of uniformly-vetted quadruple candidates presented here nearly doubles their numbers.

We note that the targets listed in this catalog are quadruple candidates that each originate from a single {\em TESS} source, i.e. the two component EBs are unresolved in {\em TESS} data. The reason is that for the purposes of this work our interests are in close quadruple systems that can exhibit dynamically-interesting interactions on a human timescale (months to years). Thus we deliberately exclude quadruple candidates that originate from two resolved sources in {\em TESS} data (either within the same pixel or separated by multiple pixels) yet have parallaxes and proper motions that may be consistent within their mutual uncertainties. For example, two EBs that are located on two different point sources separated on the sky by half a TESS pixel ($\sim10$ arcsec) and are both at a distance of 200 pc have a sky-projected separation of 2,000 AU; if the two EBs are separated by 2 {\em TESS} pixels and the distance is 500 pc, the sky-projected separation is 20,000 AU. If such EBs have the same proper motion they may indeed represent genuine, wide quadruple systems. However, even if there are significant dynamical interactions in such systems, the very long timescales are beyond the interest and scope of this work.

The organization of this paper is as follows:  Section 2 provides an overview of detection methods, Section 3 outlines the vetting process, and Section 4 describes the ephemerides determination. We present the catalog and discuss the results in Section 5, and draw our conclusions in Section 6. 

\section{Detection Methods}

A collaboration between NASA Goddard Space Flight Center (GSFC) Astrophysics Science Division, the MIT Kavli Institute, and seven experienced citizen scientists has arisen in order to fully exploit the {\em TESS} Full-Frame Images (FFIs) in search of interesting lightcurves. Though many different types of stellar systems as well as planets have been discovered in this pursuit, the aforementioned collaboration has specialized in the identification of triple and quadruple star systems. To rule out false positives due to nearby field stars or systematic effects, we evaluate the pixel-by-pixel lightcurve of the target as well as the motion of center-of-light during each set of detected eclipses, also taking into account the stellar size and contamination ratio according to the TIC, as well as the astrometric noise measured by Gaia. 

Following initial visual identification and inspection of the lightcurve for known artefacts, each candidate quadruple system must undergo further analysis to ensure that the eclipse signals are originating from the indicated source. Of the candidate quadruple systems identified through visual analysis, only about 5-10 percent pass photocenter vetting\footnote{A catalog of quadruple false positives is beyond the scope of this work but we plan to present such a catalog in the future}. The remainder are dismissed as being caused by contamination of the lightcurve by two eclipsing binaries (EBs) that are either physically unrelated or too widely separated on the sky (as discussed above). We note that some of these contaminated sources may actually be wide physical quadruples as per the measured parallaxes and proper motions from Gaia. Each system presented in this catalog has undergone and passed photocenter vetting, indicating, at the very least, the source of each of the independent periods of eclipses is visually inseparable. We assess that a substantial majority of the systems presented here will be further confirmed as being gravitationally-bound hierarchical quadruple systems.  

Prior to this collaboration, each organization pursued these candidates through different means. The NASA GSFC group pursued machine learning methods to find eclipses, then visually examined the lightcurves containing high-confidence eclipses. The machine learning method is described in \citet{2021AJ....161..162P}.  Briefly, we use a convolutional neural network (CNN) adapted from the ResNet \citep{2015arXiv151203385H} structure to accept one-dimensional lightcurves as input.  The CNN was trained to find the feature of the eclipse in the lightcurves, using over 40,000 training examples and a binary cross entropy loss function. After constructing all the FFI lightcurves from Sectors 1-40 for targets brighter than 15th magnitude ($\sim115$ million), we used the neural network to perform inference and positively identify those lightcurves containing eclipses. Altogether, about 450,000 EB candidates were identified (Kruse et al. in prep). Finally, we conducted a visual inspection of those lightcurves identified by the neural network.  Given that the final step of our process was visual inspection, a collaboration with the VSG was natural.  Since the start of our collaboration, we have continued the process of constructing the FFI lightcurves for every newly released sector of {\em TESS} data, inference through the neural network, then a final visual survey by the VSG. 

\begin{figure*}
    \centering
    \includegraphics[width=0.85\linewidth]{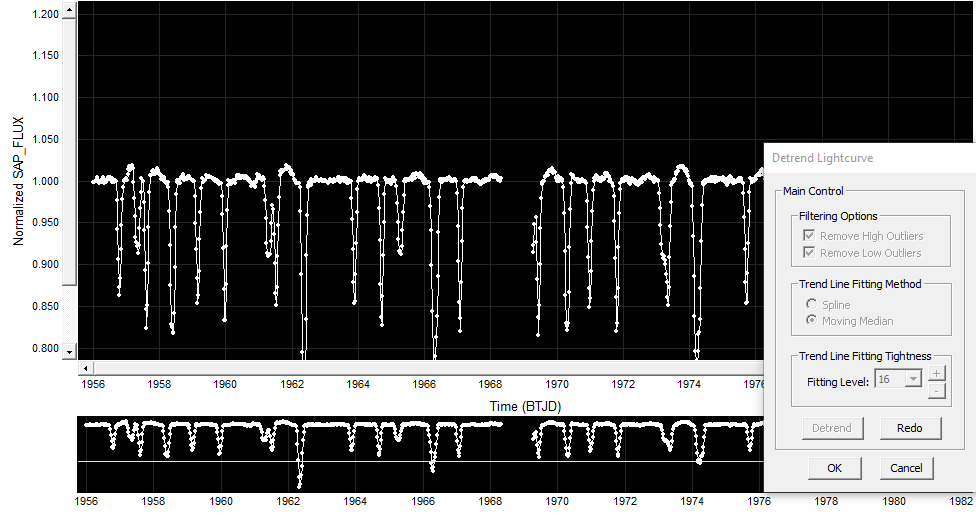}
    \includegraphics[width=0.85\linewidth]{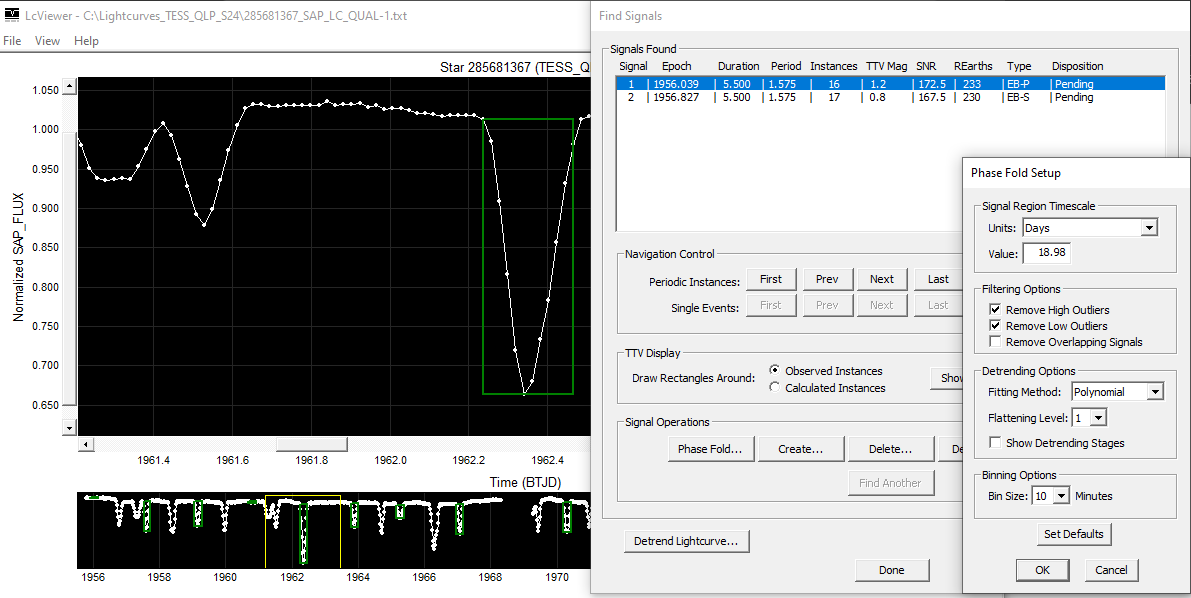}
    \caption{Illustrative example of a screenshot of LcTools used to detect and analyze the quadruple candidate TIC 285681367. The upper panels show the detrended {\em TESS} QLP lightcurve of the target, along with the detrending options. The lower panels show the phase-folding setup, options and results, as well as a zoomed-in section of the lightcurve centered on a particular eclipse used to fix the ephemeris. }
    \label{fig:lctools_example}
\end{figure*}

The MIT Visual Survey Group (VSG, Kristiansen et al. submitted) visually examined outputs of the MIT Quick Look Pipeline \citep[QLP,][]{2020RNAAS...4..204H}. The VSG discoveries were all made using standard personal computers with Linux, Macintosh, or Windows operating systems. The visual surveyors made use of the LcTools software system \citep{2019arXiv191008034S,2021arXiv210310285S} -- an interactive set of tools designed for lightcurve analysis -- and custom software written in Python, C, or JavaScript. The most common method of detection required scanning through millions of public domain lightcurves using LcTools. Where necessary, the data were detrended and filtered for additional qualification to remove systematic noise and glitches when possible. LcTools was often used to also check eclipse depth and periodicity using a built-in BLS (Box-fitting Least Squares) or the built-in QuickFind method to further qualify the candidate lightcurves. The candidates were then visually inspected for dips from possible multiple eclipsing binaries. An example of this process is shown in Figure \ref{fig:lctools_example}, highlighting the preliminary analysis of quadruple candidate TIC 285681367. 

We note that trained visual inspection for {\it specific features} (like additional eclipses superimposed on an otherwise regular pattern) using tools specifically-designed for the task can be quite fast and efficient. The three main reasons are that (i) LcViewer is extremely fast with close to zero lag time between light curve presentations, and presents the lightcurves in a consistent, uniform, and homogeneous format; (ii) the seasoned visual surveyor is experienced in knowing what an object of interest looks like as well as whether it is a known pattern or not; and (iii) human perception is exceptionally good at recognizing a change in a known pattern or the emergence of a new pattern. For example, showing a trained surveyor 99 consecutive images of trees planted at regular intervals (like eclipses) is unlikely to trigger a reaction if asked to identify a new pattern. While the types of trees and the intervals change between images, the size, shape and color pattern of the images remains the same so that, in essence, the trees (eclipses) practically {\it become} the background. If, however, the hundredth image contains a tiger hiding behind a new set of trees planted at a regular interval, the surveyor will raise a red flag in the matter of seconds. 

From our experience with Kepler, K2 and TESS data, members of our team can inspect a particular lightcurve in about five to ten seconds, and potentially much faster. Thus assuming a typical `cruising speed' of 10 sec per lightcurve, ten dedicated visual surveyors can inspect 1 million light curves in two years spending less than 25 minutes a day. For context, over the last 10 years VSG members have visually surveyed more than 15 million light curves from Kepler, K2 and TESS (Kristiansen et al. submitted). In comparison, the subsequent vetting process (described below) is orders of magnitude slower.

\section{Vetting Methods}

Due to the large pixel size of the {\em TESS} photometer ($\approx20$ arcsec), false positives due to nearby field stars are a common occurrence. To account for this, we evaluate the motion of the measured center-of-light during each set of eclipses detected in the lightcurve of each target. We also take into account the presence of nearby field stars and their respective magnitude differences with the target star, contamination ratio according to the {\em TESS} Input Catalog (TIC) where available, as well as information from the Gaia EDR3 catalog. In addition, we pursue follow-up photometry observations for a subset of targets as part of the {\em TESS} Follow-up Observing Program (TFOP), as well as dedicated spectroscopy on the 1.5m telescope at the F.\ L.\ Whipple Observatory in Arizona with the Tillinghast Reflector Echelle Spectrograph (TRES; Szentgyorgyi \& Fuŕesz 2007; Furesz 2008). 

The vast majority of our quadruple candidates were unknown as EBs prior to their detection with {\em TESS}. As a result, they were not on the list of {\em TESS} targets observed at 2-min cadence and no data validation reports were available. Thus we used a center-of-light analysis based on the photocenter module of the {\tt DAVE} vetting pipeline \citep{2019yCat..51570124K} to evaluate the source of the detected EBs. Briefly, we investigate the center-of-light motion for each eclipse of each EB for each sector of available data by fitting to the difference image (out-of-eclipse image minus in-eclipse image) a Point-Spread Function (PSF) and a Pixel-Response Function (PRF), and measuring the corresponding photocenter. When the eclipses of two EBs are too close to each other in time, so that there is little to no out-of-eclipse section of the lightcurve available for photocenter measurement, we exclude said eclipses from the analysis. To evaluate whether there is a significant motion during the detected eclipses, we compare the measured average difference image photocenter to the pixel position of the target as provided in the corresponding FITS header. We note that comparing the photocenter measured from the average difference image to the photocenter measured from the average out-of-eclipse image, as used for the analysis of Kepler and K2 data, is not optimal for {\em TESS}. The reason for this difference is that the {\em TESS} aperture is much larger and often contains multiple field stars that are as bright as the target itself (or even brighter). As a result, these field stars ``pull'' the measured out-of-eclipse photocenters away from the position of the target star, effectively preventing a meaningful comparison between the difference image and the out-of-eclipse image. While this was a known issue for K2 data (e.g. Kostov et al. 2019), it is much more prevalent for {\em TESS} data. 

Based on our experience, the resolution limit of the photocenter analysis depends on multiple factors, which typically vary not only on a target-by-target but also on a sector-by-sector basis. Specifically, the limiting factors are the (i) magnitude difference between the target and nearby field stars; (ii) overall contamination ratio; (iii) out-of-eclipse lightcurve variability; (iv) number and depth of clean, un-blended eclipses; (v) quality of the difference images used to measure the photocenters; and (vi) the peculiarities of the systematic effects. For a typical pair of sources, measuring a photocenter separation of $\sim5-10$ arcsec ($\sim0.25-0.5$ pixels) is relatively easy, whereas a separation of $\sim1$ arcsec ($\sim0.05$ pixels) is highly challenging.

We note that for some targets the TIC and/or Gaia EDR3 catalogs show that there is indeed a field star within $\sim1$ arcsec of the target, and the photocenter measurements may not be sufficiently precise to pinpoint the true source of the eclipses. In these cases, we evaluate whether the eclipses can be produced by a field star using the eclipse depth (${\rm d_e}$) and the magnitude difference in the {\em TESS} bandpass (${\rm \Delta T}$) between the field star and the target star, ${\rm \Delta T = -2.5\log_{10}(2d_e)}$ mag. For example, for a field star to produce 10\%-deep eclipses as contamination in the target's {\em TESS} lightcurve, ${\rm \Delta T}$ has to be smaller than about 1.75 mag.

\begin{figure*}
    \centering
    \includegraphics[width=0.95\linewidth]{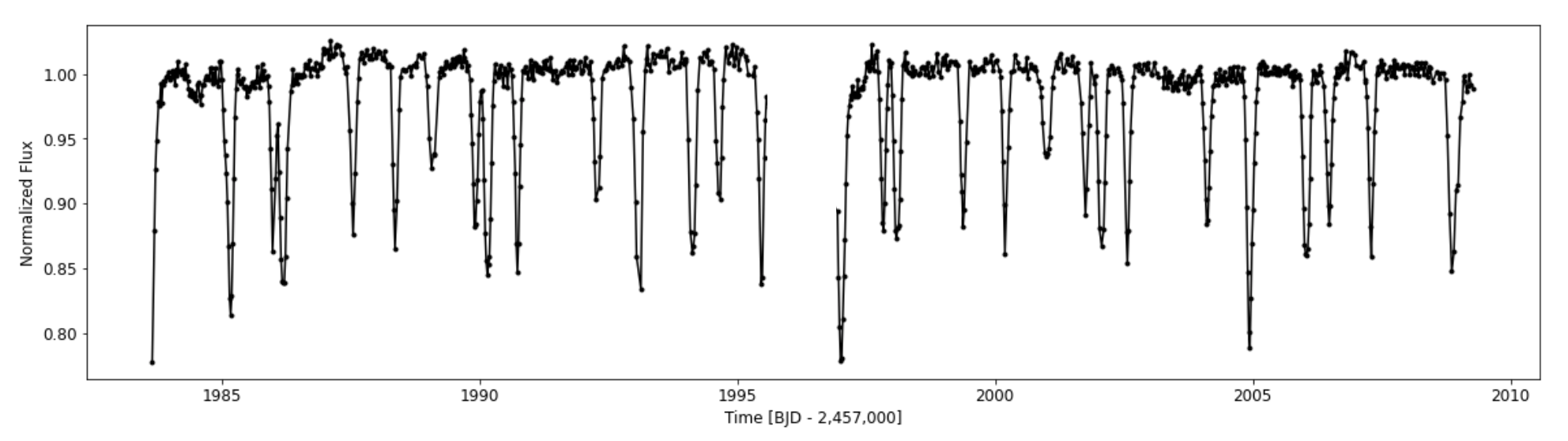}
    \includegraphics[width=0.95\linewidth]{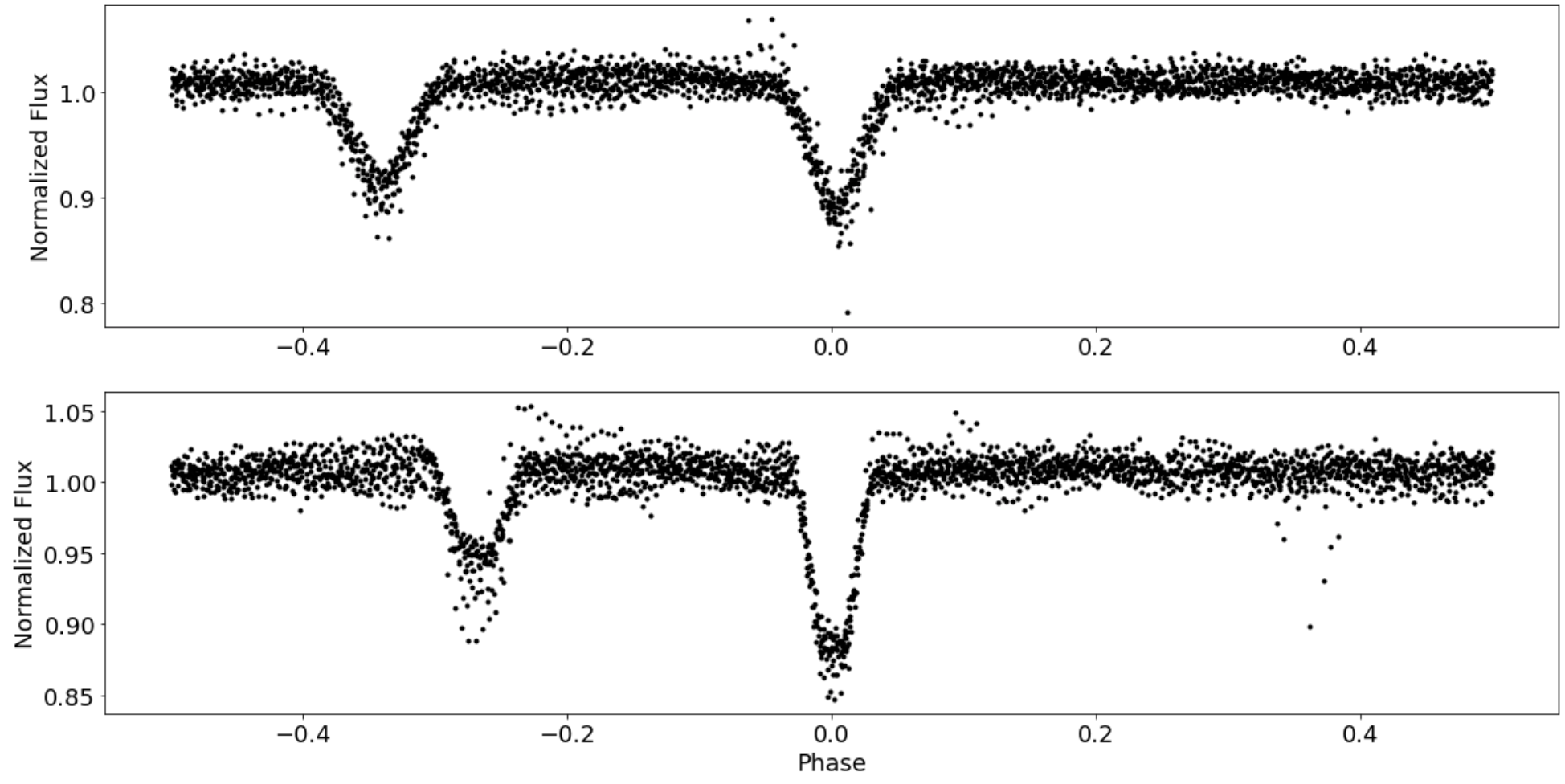}
    
    \subfloat{
        \includegraphics[width=0.31\linewidth]{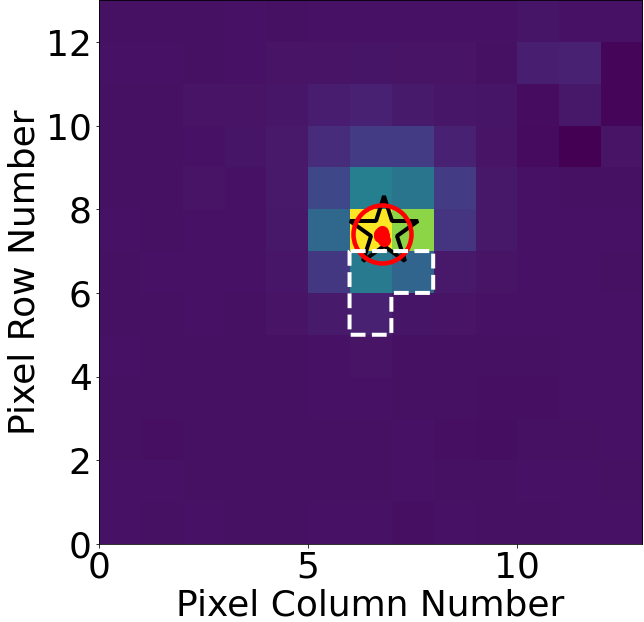}
    }
    \hfill
    \subfloat{
    \centering
        \includegraphics[width=0.31\linewidth]{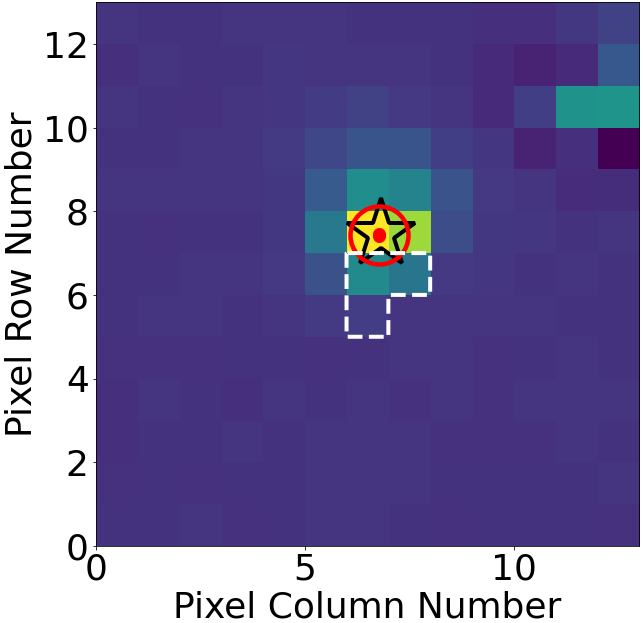}
    }
    \hfill
    \subfloat{
        \includegraphics[width=0.34\linewidth]{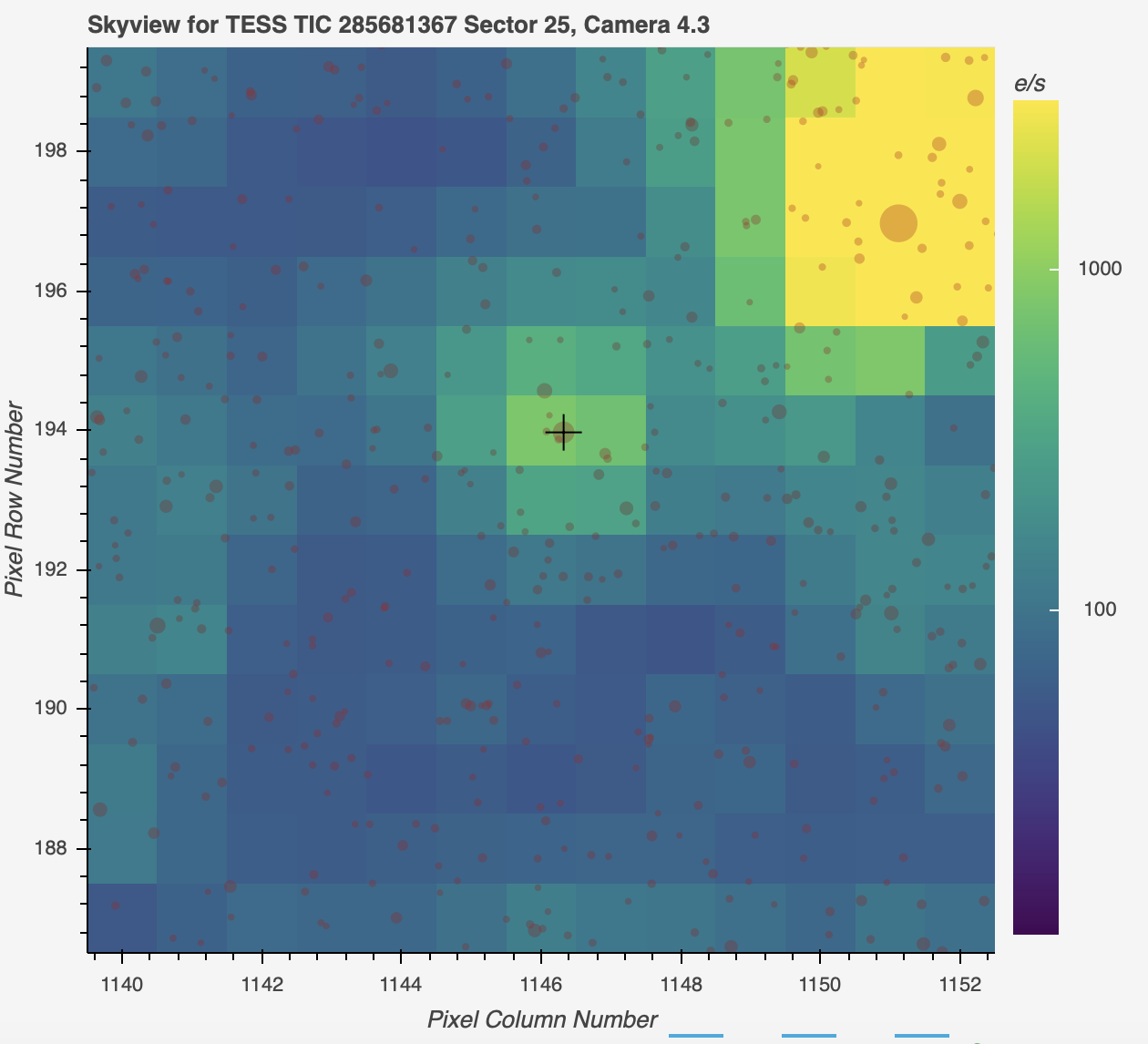}
    }
    \caption{Upper panel: long-cadence {\em TESS} {\tt eleanor} lightcurve of TIC 285681367, showing two sets of eclipses with period ${\rm P_A}$ = 2.37 days and ${\rm P_B}$ = 3.97 days. Second and third panels from top: Phase-folded lightcurve for binary A and binary B, respectively. For clarity, the lightcurve was disentangled such that the eclipses of the other binary were removed (Powell et al. 2021). Lower left and middle panels: {\em TESS} difference images ($13 \times 13$ pixels) for ${\rm P_A}$ and PB, respectively, showing the pixels changing during the corresponding primary eclipses. The axes represent the corresponding relative pixel number along the x-axis (pixel column number) and y-axis (pixel row number). The small red symbols represent the measured photocenter for the individual primary eclipses of each respective binary and the large red circle shows the corresponding average photocenters. The black star is the catalog position of the target and the dashed contour represent eleanor's aperture used to extract the target's FFI lightcurve. Lower right panel: Skyview image (also $13 \times 13$ pixels) of the target's {\em TESS} aperture showing known nearby sources down to G = 21 mag. The photocenter analysis shows that both sets of eclipses are on-target. }
    \label{fig:on_on_target}
\end{figure*}

An example of a quadruple candidate (TIC 285681367) passing these vetting tests is shown in Figure \ref{fig:on_on_target}. The target was observed in Sectors 18, 24 and 25, and produced two sets of eclipses with periods of ${\rm P_A}$ = 2.3660 days, and ${\rm P_B}$ = 3.9703 days, each showing primary and secondary eclipses. The FFI lightcurve of the target for Sector 25 is shown in the upper panel of the figure. The lower left and middle panels of the figure show the average difference images for each set of primary eclipses; the red symbols represent the measured photocenters and the black star represents the catalog position of the target. Our photocenter analysis shows that the target is the source of both sets of eclipses. We note that the aperture {\tt eleanor} used for TIC 285681367 in Sector 25 (dashed contour in lower left and middle panels) does not include the target itself. This is a relatively rare (and sector-dependent) occurrence, and indicates that the eclipses are actually deeper than what we see in the {\tt eleanor} lightcurve. 

Throughout this work, we used {\tt eleanor}'s default $aperture\_mode = `normal'$ option which, by design, already tests various aperture sizes depending on the target's magnitude and contamination ratio. Further adjusting the photometric aperture in order to extract a custom lightcurve would require modifying the source code itself. This is beyond the scope of our work and would likely require working closely with the creators of {\tt eleanor}. However, the aperture only affects the measured eclipse depths. It does not affect our photocenter analysis at all because we uses our own codes and all available pixels as directly provided by TESS (typically a 13x13 pixels cutout, see lower left panel in Figure \ref{fig:on_on_target}).

The lower right panel of Figure \ref{fig:on_on_target} represents a Skyview image of the target's {\em TESS} aperture (with the same size as the difference images, $13 \times 13$ pixels) showing all stars down to G = 21 mag. This is a crowded field -- the target is blended with TIC 627730721 (separation ${\approx 2.7}$ arcsec, ${\rm \Delta T \approx 7.3}$ mag) and there are two more field stars inside the central pixel for Sector 25 -- TIC 627730803 (separation ${\approx 5.1}$ arcsec, ${\rm \Delta T \approx 7.3}$ mag) and TIC 627730804 (separation ${\approx 6.5}$ arcsec, ${\rm \Delta T \approx 8.6}$ mag). However, none of these field stars is bright enough to produce the detected eclipses. 

\subsection{Pixel-by-pixel analysis}
\label{sec:pixel-by-pixel}

Once a multi-stellar candidate is identified, we utilize the interactive feature in Lightkurve (Lightkurve Collaboration et al. 2018) to inspect the target pixel file as an initial test prior to the photocenter analysis discussed above. As a standard, we compute a $15 \times 15$ pixel cutout which normally encompasses both EBs in question. The size of the cutout is based on a compromise between reliability and efficiency. In terms of reliability, our experience shows that this pixel mask allows consistent identification of the location of both EBs, regardless of their brightness. In terms of efficiency, the computational time is negligible compared to the time needed to adjust the aperture in an attempt to find the perfect match for each target. We also note that a contaminating star does not need to be exceptionally bright -- only bright enough compared to the target star itself (see Section 3 for details).

\begin{figure}
    \centering
    \includegraphics[width=0.65\linewidth]{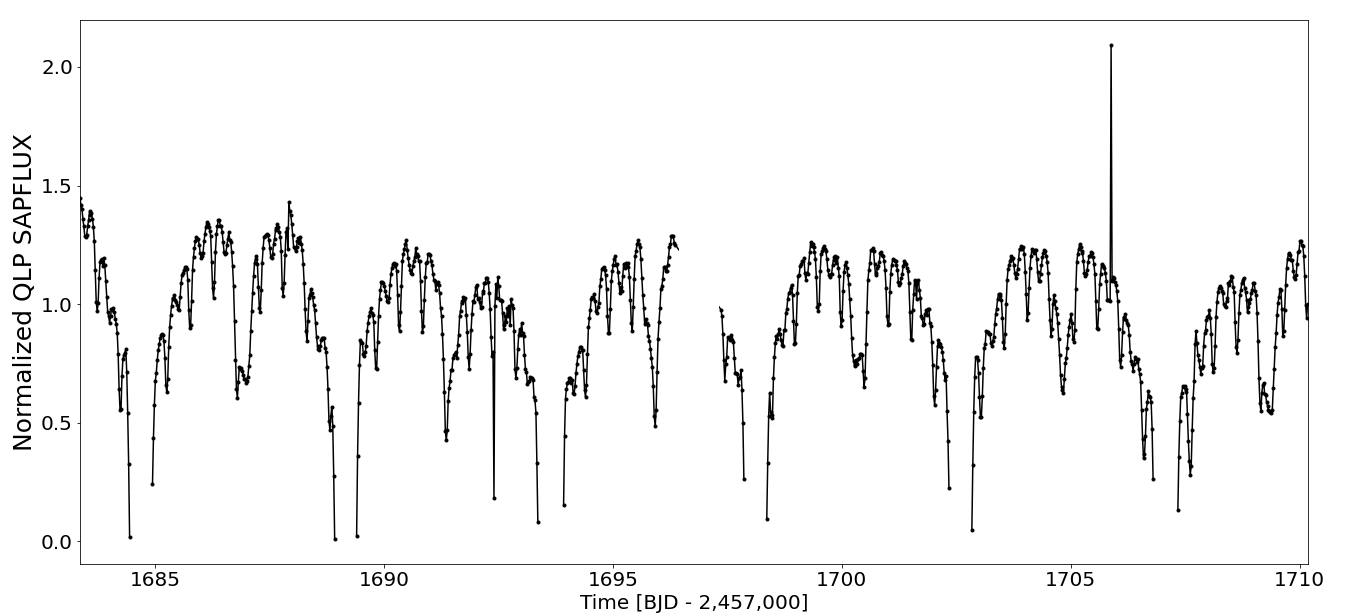}
    \includegraphics[width=0.33\linewidth]{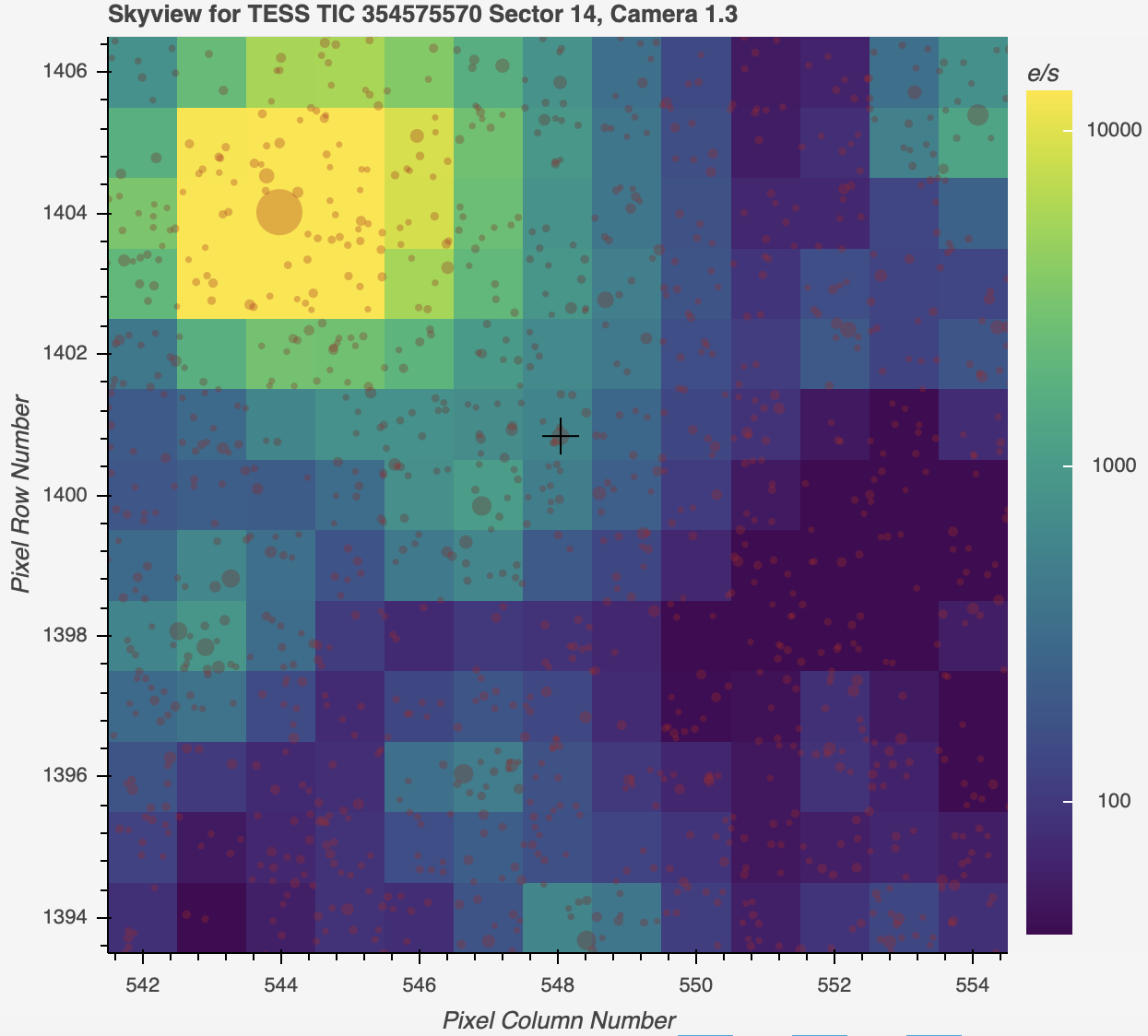}
    \includegraphics[width=0.99\linewidth]{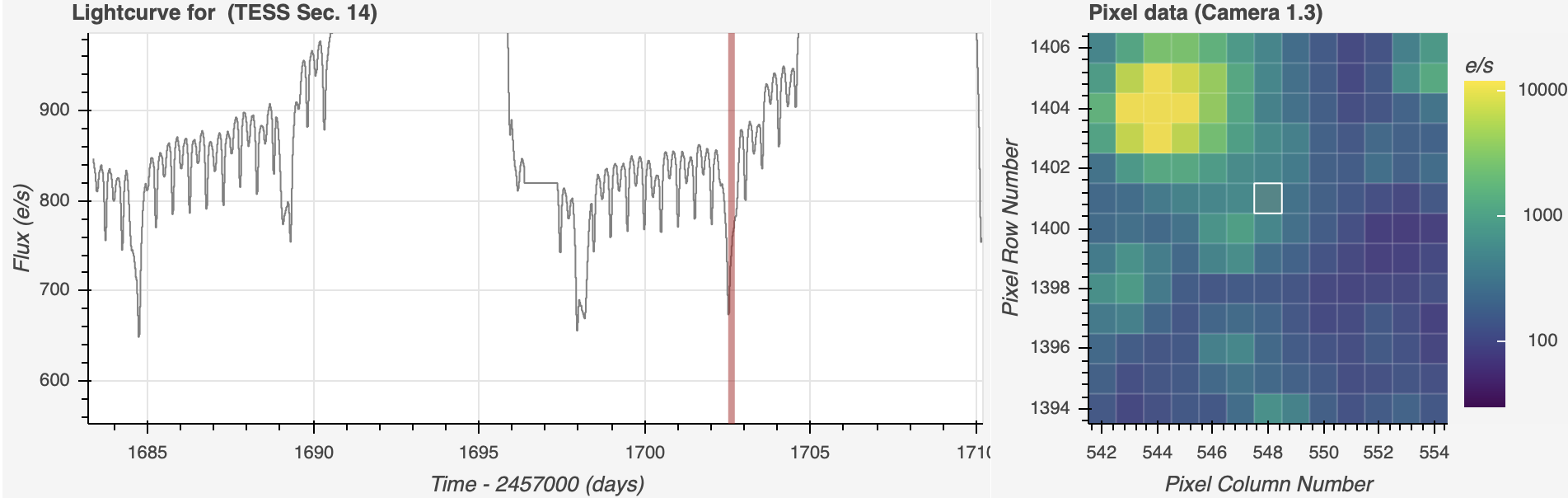}
    \includegraphics[width=0.99\linewidth]{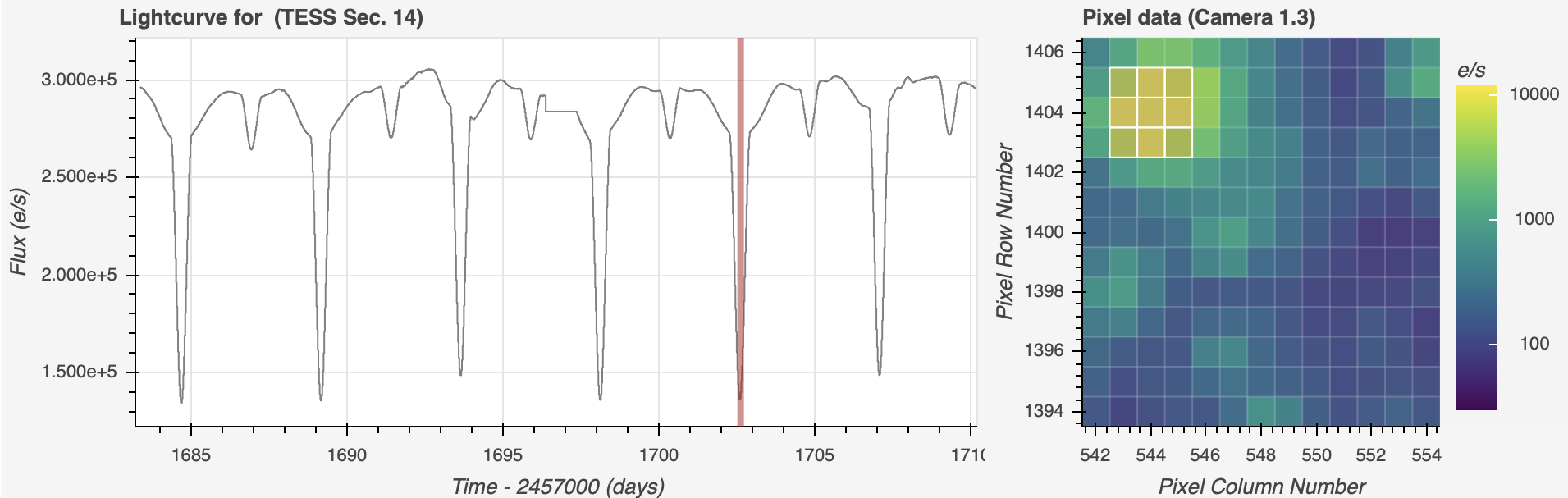}
    \caption{Example quadruple false positive for scenario (i) described in \ref{sec:pixel-by-pixel}. First row, left panel: Sector 14 QLP lightcurve of TIC 354575570 as used by the visual surveyors for the detection of two apparent sets of eclipses. First row, right panel: $15 \times 15$ pixels Skyview image of the target’s {\em TESS} aperture showing known nearby sources. Middle and lower rows: Lighkurve visualization of the selected pixels (white contours in right panels) and the corresponding lightcurve (left panels). The target is a quadruple false positive as one of the EBs is clearly off-target.}
    \label{fig:pixel_by_pixel_TIC_354575570}
\end{figure}

\begin{figure}
    \centering
    \includegraphics[width=0.65\linewidth]{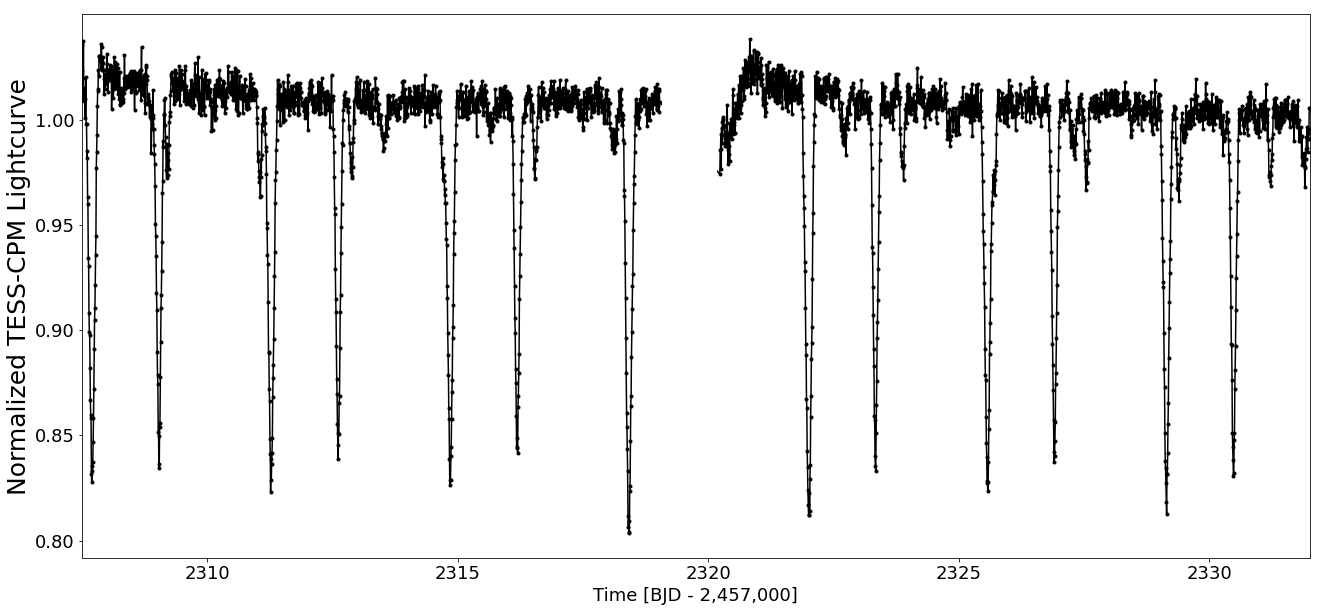}
    \includegraphics[width=0.33\linewidth]{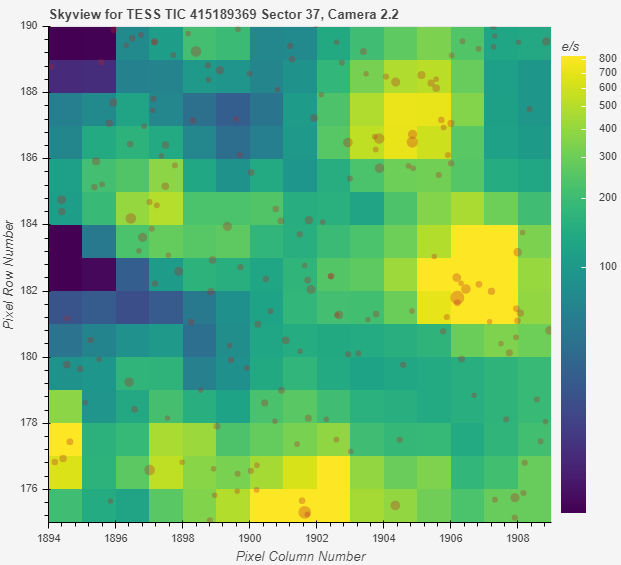}
    \includegraphics[width=0.99\linewidth]{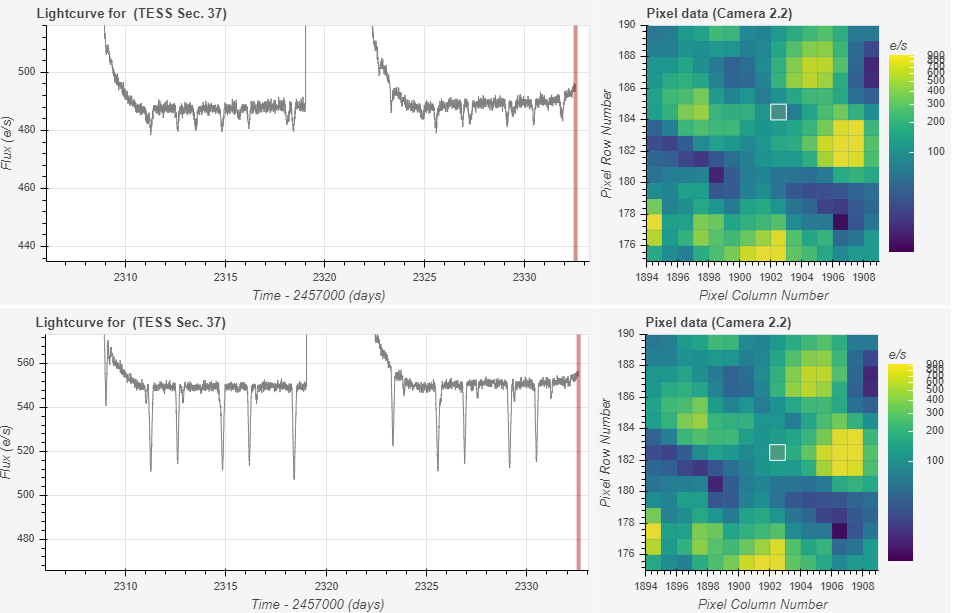}
    \caption{Same as Fig. \ref{fig:pixel_by_pixel_TIC_354575570} but for false positive scenario (ii). First row, left panel: TESS-CPM lightcurve for TIC 415189369 (Hattori et al. 2020, AAS, 23515705H). The target is a quadruple false positive as the relative depths of the two sets of eclipses scale differently depending on the selected pixel. Middle and lower rows: Lighkurve visualization of the selected pixels (white contours in right panels) and the corresponding lightcurve (left panels).}
    \label{fig:pixel_by_pixel_TIC_415189369}
\end{figure}

In most cases, we experience one of the following scenarios: (i) The two EBs are sufficiently separated on the sky with at least one EB being off-target; or (ii) the positions of the two EBs are in adjacent pixels (as inferred with {\tt Lightkurve}) yet different apertures show that the corresponding eclipse depths scale differently as a function of the aperture size. This indicates that the EBs originate from two resolved targets, which may or may not be a wide quadruple system. These two scenarios can be further evaluated by comparing parallax and proper motion values which we find via the Swarthmore Finding Chart through ExoFOP-TESS. Examples of these scenarios are shown in Fig. \ref{fig:pixel_by_pixel_TIC_354575570} and \ref{fig:pixel_by_pixel_TIC_415189369}.  

We note that there can be more than two unrelated stars in a target's {\em TESS} aperture (i.e., with different Gaia distances and/or proper motions), coming from clearly distinct pixels of the aperture. For example, there are four unrelated EBs in the {\em TESS} aperture of quadruple false positive TIC 28553336. Finally, sometimes two or more stars are located very close to each other within the same pixel. In cases like this, pinpointing the source of each EB needs further analysis as discussed above.  

\subsection{Types of False Positives}

All candidate quadruples listed in this catalog have passed the photocenter vetting process described above. Those that did not pass the process fall into several categories, described below for completeness. 

\begin{figure}[h]
    \centering
    \includegraphics[width=1.05\textwidth]{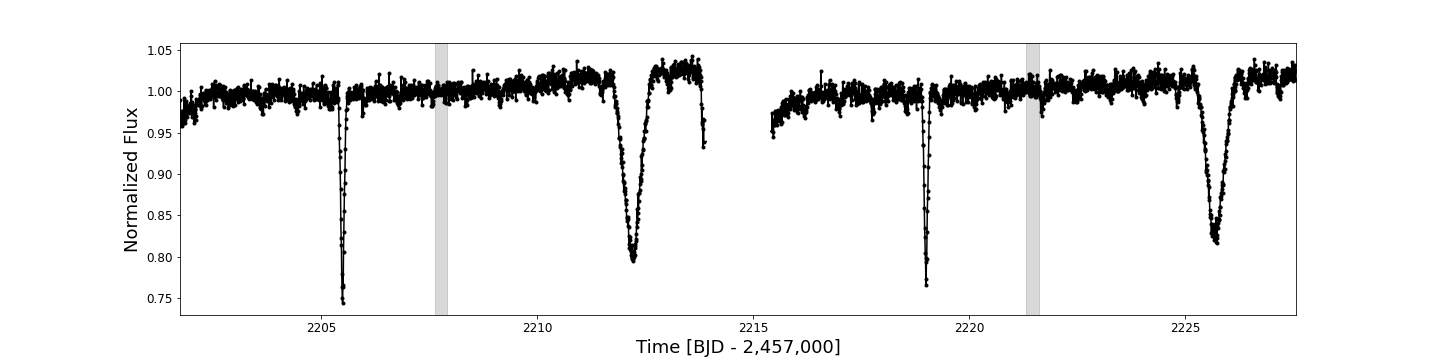}
    \subfloat{
        \includegraphics[width=0.31\linewidth]{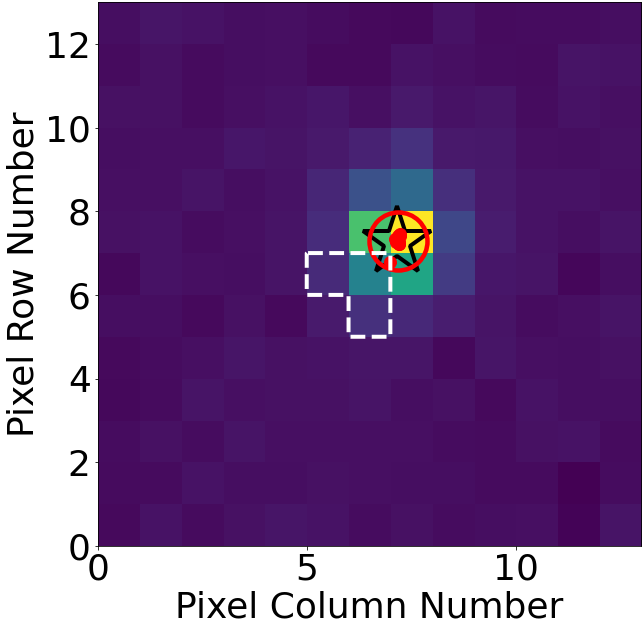}
    }
    \hfill
    \subfloat{
    \centering
        \includegraphics[width=0.31\linewidth]{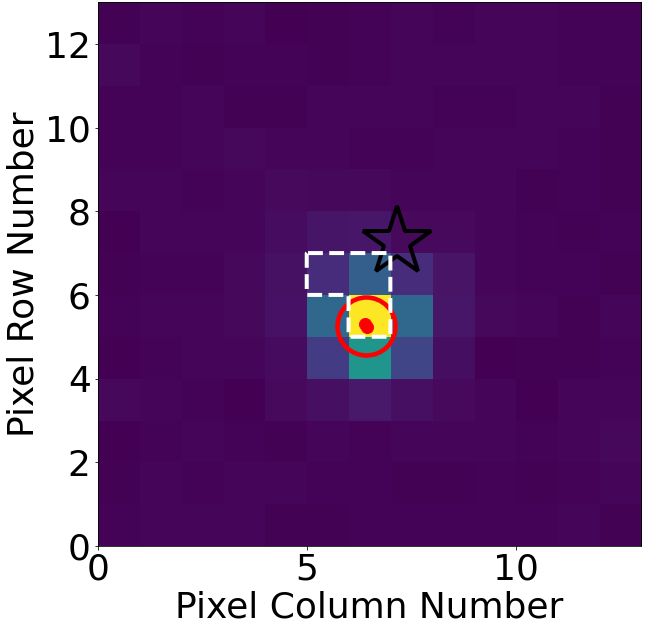}
    }
    \hfill
    \subfloat{
        \includegraphics[width=0.34\linewidth]{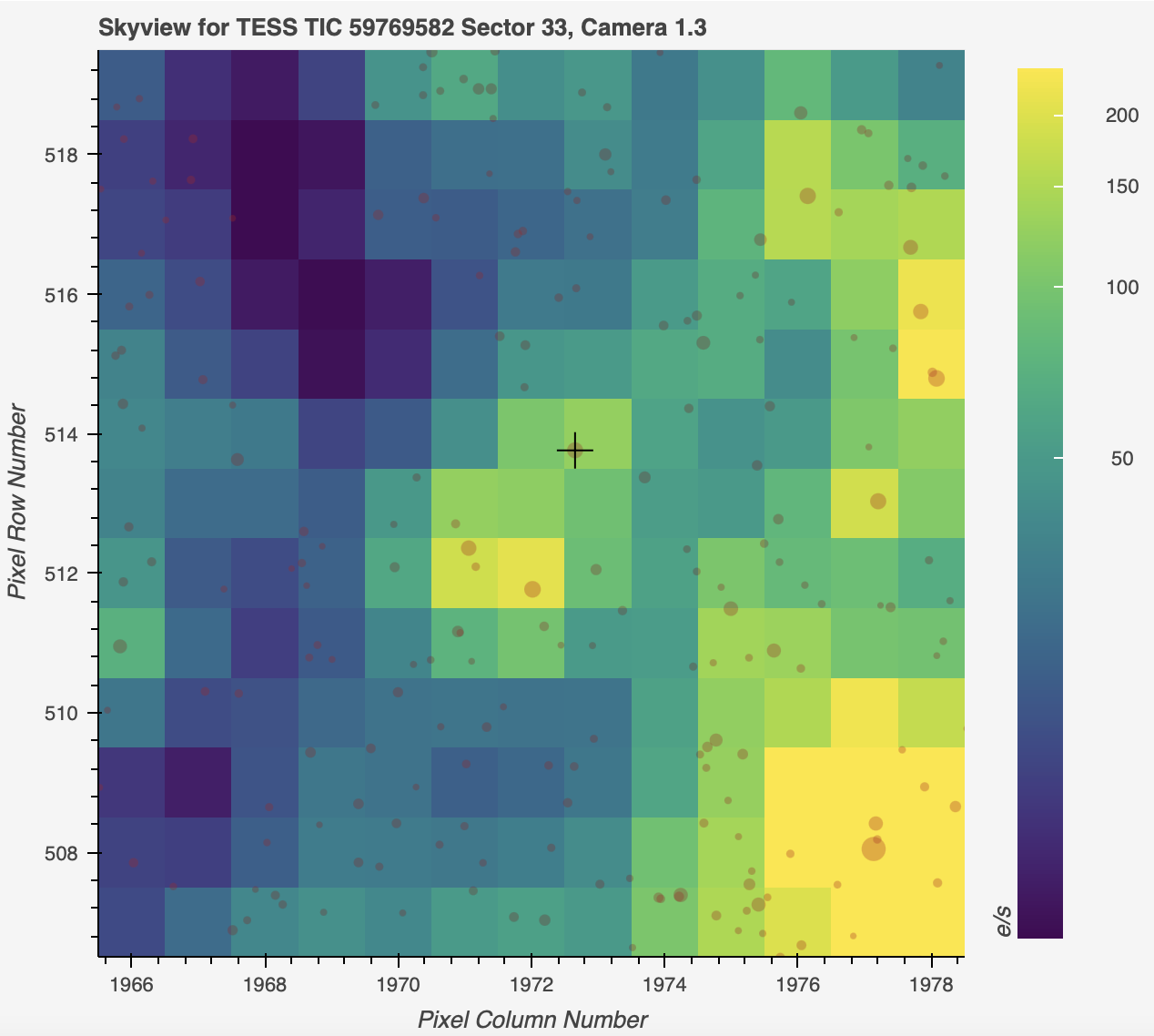}
    }
    \caption{Long-cadence {\em TESS} {\tt eleanor} lightcurve (upper panel) and photocenter analysis (lower left and middle panels) of TIC 59769582. The analysis shows that binary A with ${\rm P_A}$ = 1.57 days is on-target (lower left panel), while binary B with ${\rm P_B}$ = 13.5 days is off-target (lower middle panel), about 2 pixels SW of TIC 59769582. The lower right panel shows Skyview image of the {\em TESS} aperture for TIC 59769582 (at the center of the image, marked with a plus symbol), highlighting the source of the ${\rm P_B}$ eclipses (red arrow). This is an example of on-target EB + Field Star EB false positive.}
    \label{fig:on_off_target}
\end{figure}

\begin{figure}[h]
    \centering
    \includegraphics[width=0.95\textwidth]{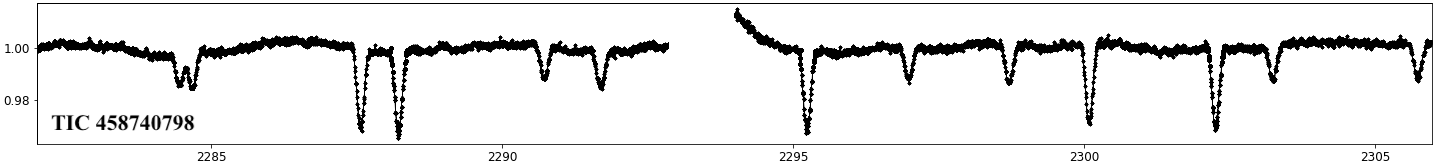}
    \includegraphics[width=0.95\textwidth]{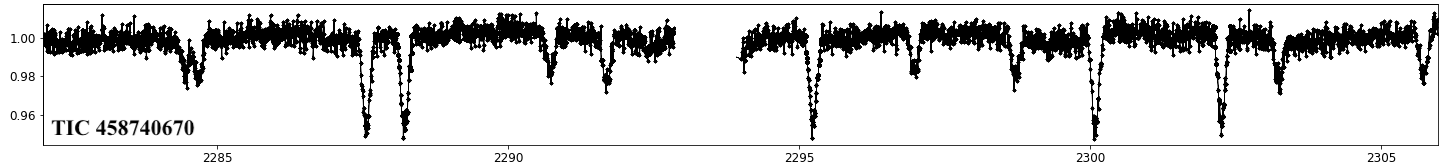}
    
    \subfloat{
        \includegraphics[width=0.255\textwidth]{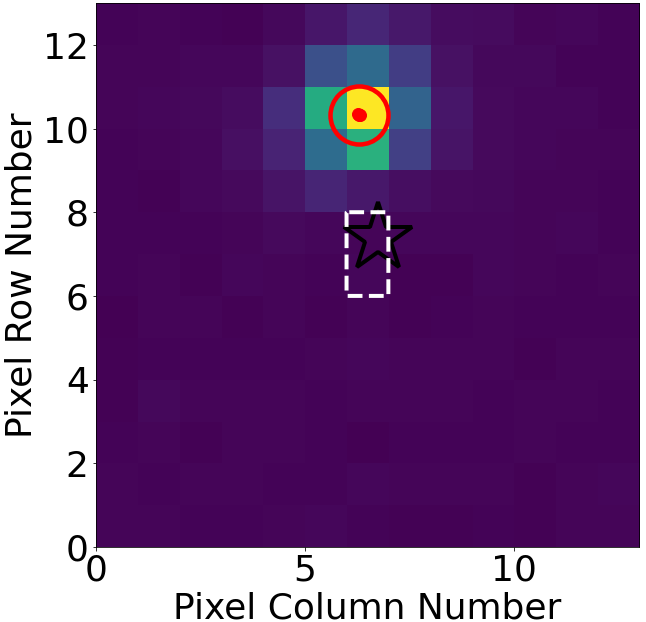}
    }
    \hfill
    \subfloat{
        \includegraphics[width=0.255\textwidth]{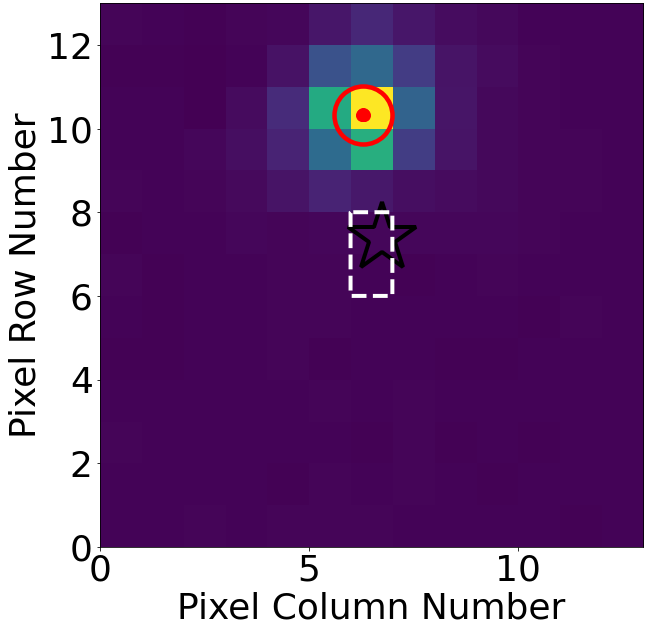}
    }
    \hfill
    \subfloat{
        \includegraphics[width=0.285\textwidth]{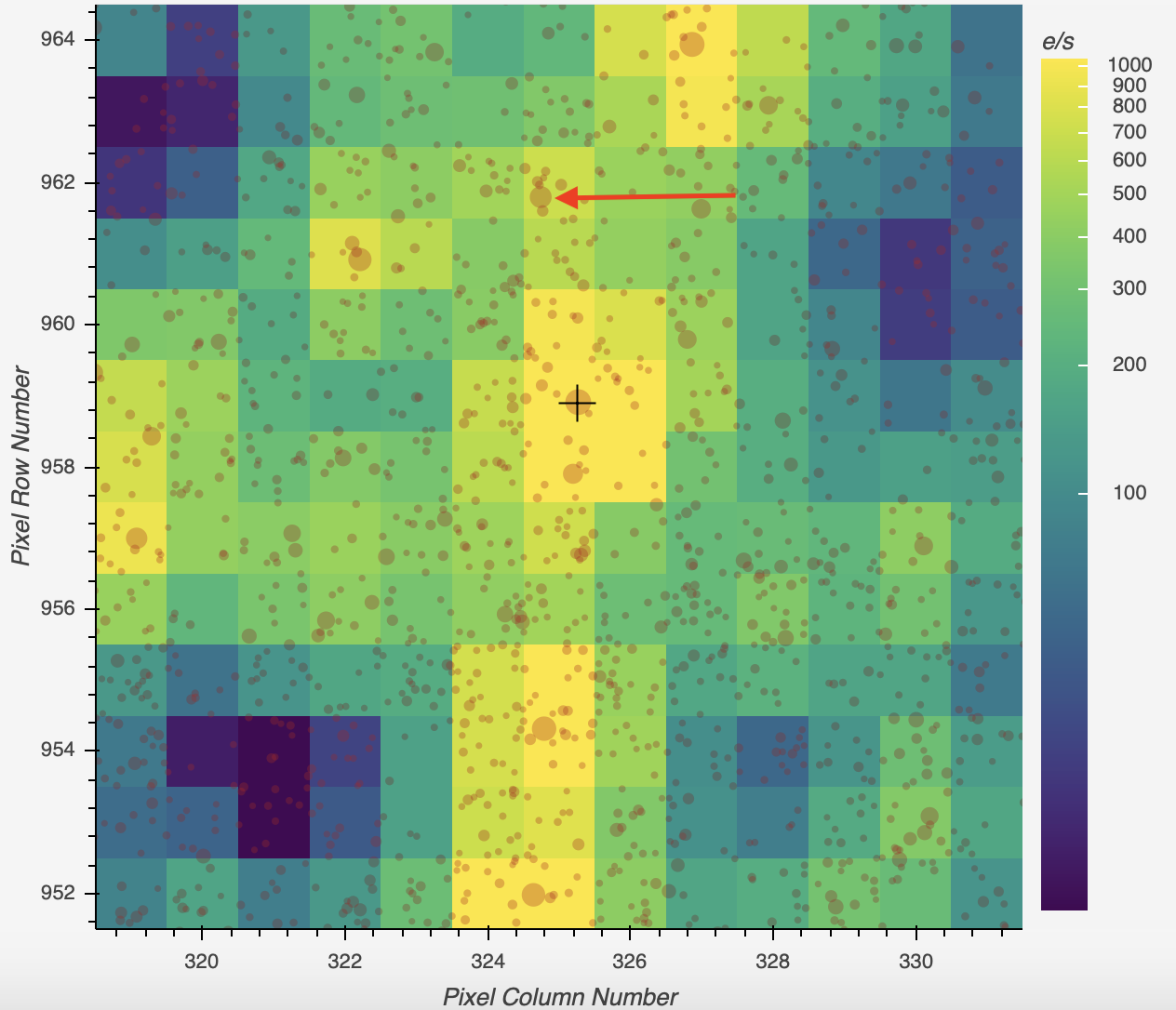}
    }
    \\
    \subfloat{
        \includegraphics[width=0.255\textwidth]{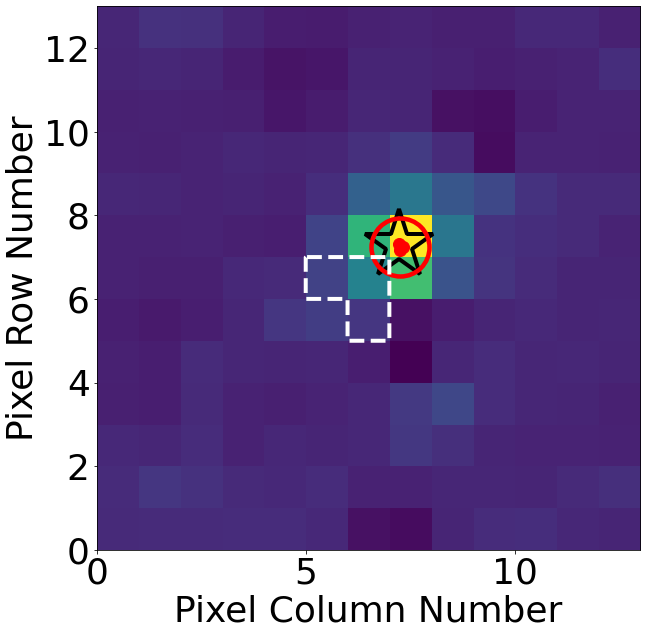}
    }
    \hfill
    \subfloat{
        \includegraphics[width=0.255\textwidth]{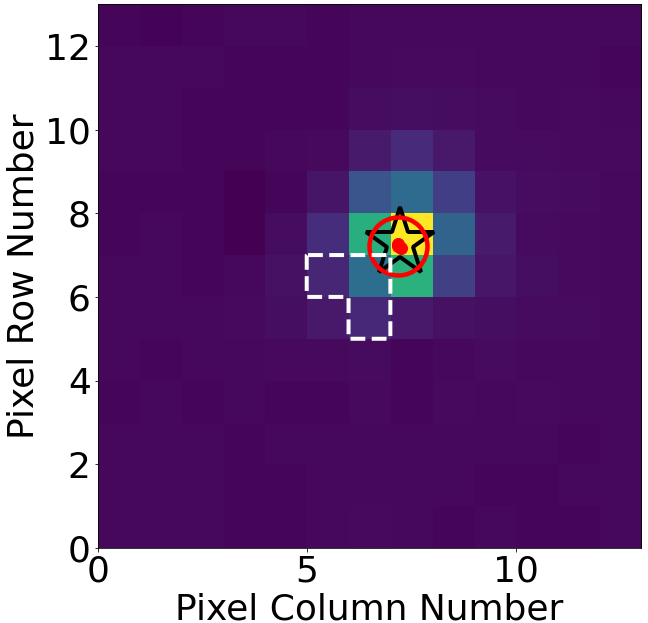}
    }
    \hfill
    \subfloat{
        \includegraphics[width=0.285\textwidth]{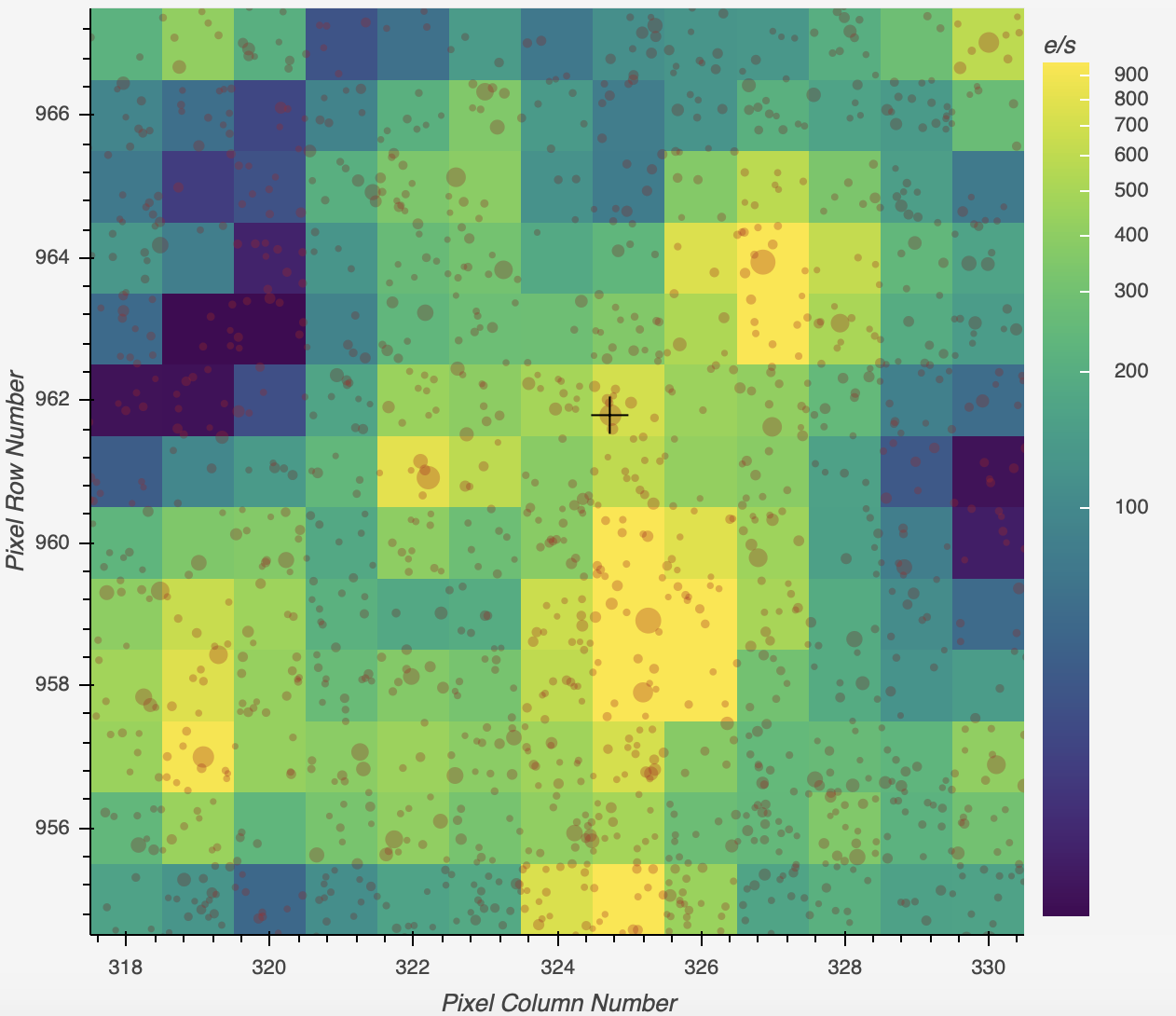}
    }
    \caption{An example of a nominal false positive (TIC 458740798) caused by a genuine quadruple candidate (TIC 458740670) in the field. First row: Long-cadence {\em TESS} {\tt eleanor} lightcurve of TIC 458740798 for Sector 36 (the false positive). Second row: same as first row but for TIC 458740670 (the quadruple). Note how all eclipses are deeper on TIC 458740670, already indicating that it is the source of the eclipses. Third row, left and middle panels: photocenter analysis for TIC 458740798, showing the difference images (left for binary A with P = 6.26-days, middle for binary B with P = 7.02-days), along with the PRF-based measurements of the individual photocenters (small red dots) and average photocenter (large open circle). The black star symbol indicates the catalog position of TIC 458740798. The measured photocenters are centered on a location about 3 pixels above TIC 458740798 -- near the position of TIC 458740670. Third row, right panel: Skyview image of the {\em TESS} aperture for TIC 458740798 (at the center of the image, marked with a black plus symbol), highlighting the true source of the eclipses TIC 458740670 (red arrow). Fourth row: same as third row but for TIC 458740670 -- a genuine quadruple candidate -- showing that the measured photocenters coincide with the location of TIC 458740670. Note that Eleanor's aperture for TIC 458740670 is adjacent to the target and as a result the true eclipse depths are larger than those seen in the lightcurve (second row).}
    \label{fig:off_target_quad}
\end{figure}

\begin{figure}[h]
    \centering
    \includegraphics[width=0.95\textwidth]{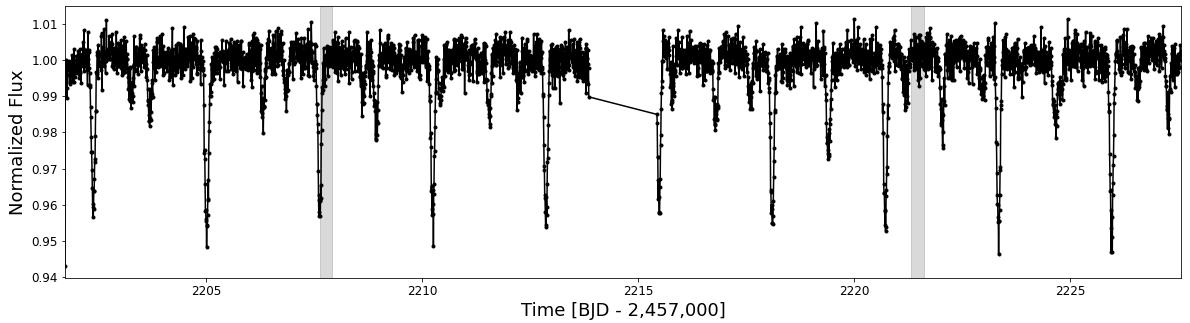}
    \\
    \hfill
    \subfloat{
        \includegraphics[width=0.255\textwidth]{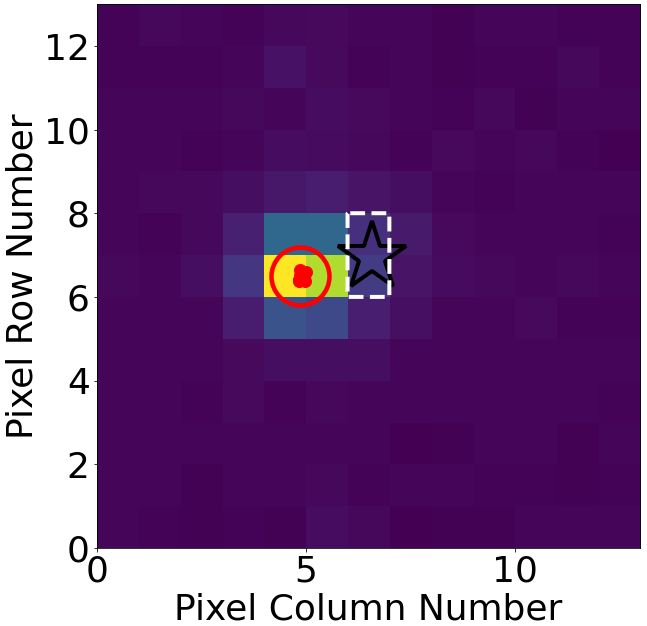}
    }
    \hfill
    \subfloat{
        \includegraphics[width=0.255\textwidth]{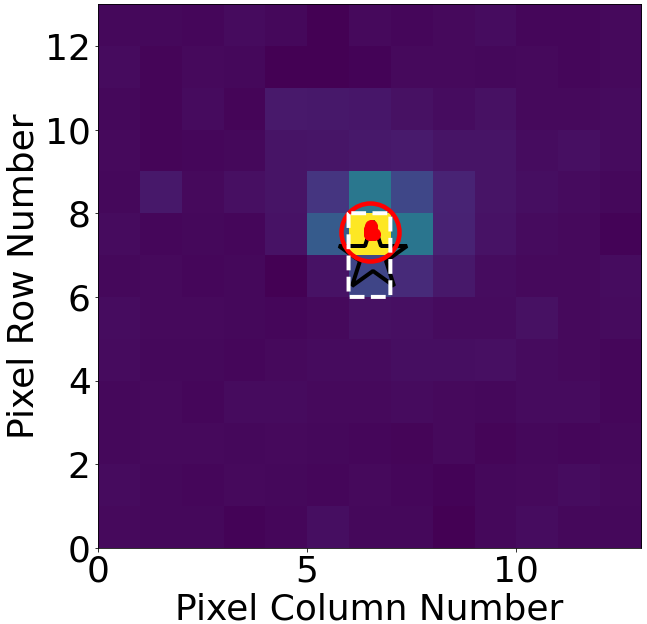}
    }
    \hfill
    \subfloat{
        \includegraphics[width=0.285\textwidth]{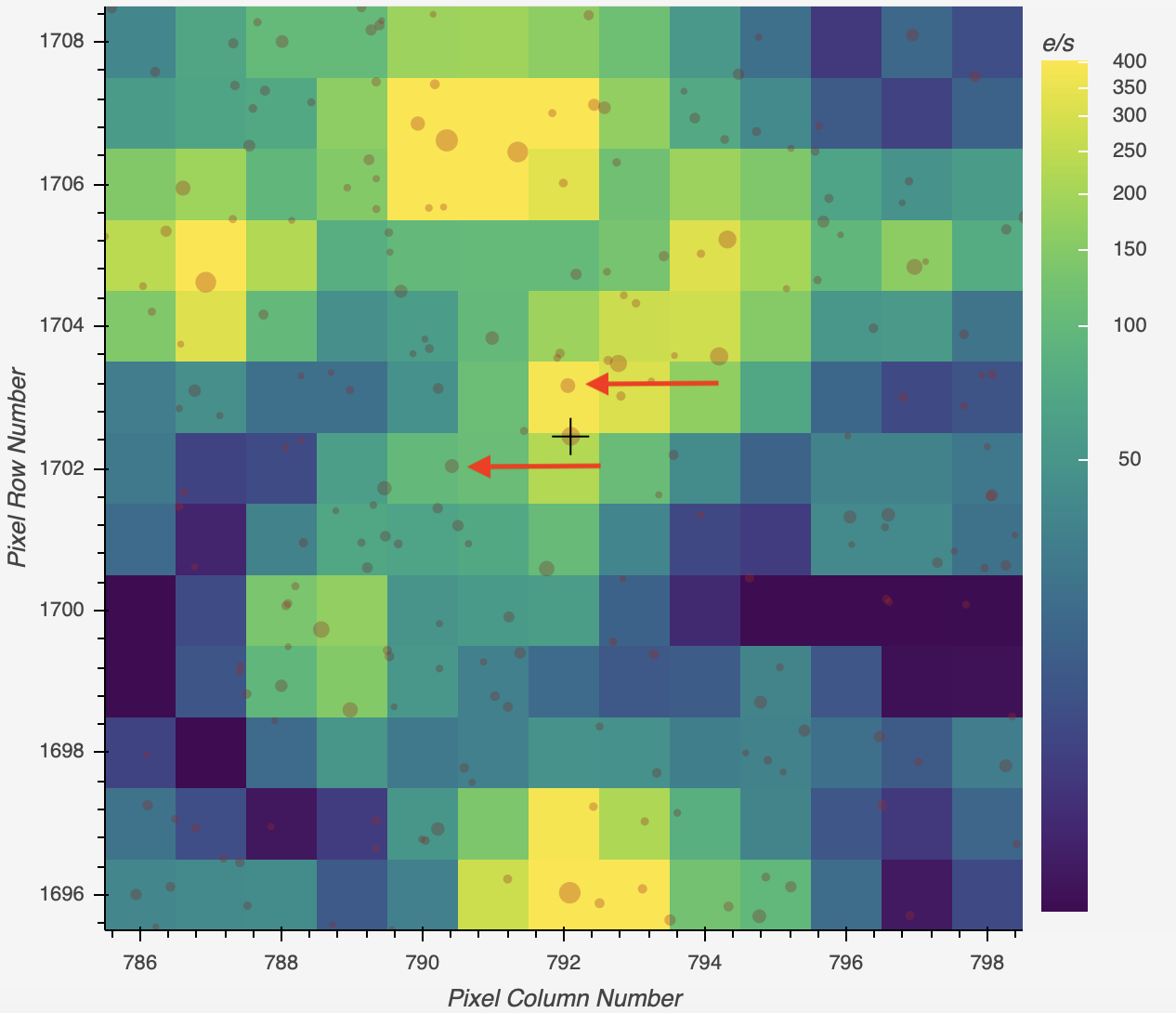}
    }
    \hfill
    \null
    \caption{Same as Figure \ref{fig:on_off_target} but for TIC 262535168. The two EBs seen in the lightcurve originate from two different targets, neither of which is TIC 262535168, as highlighted on the Skyview image. This is an example of a false positive due to field EB (star 1) + field EB (star 2).}
    \label{fig:off_off_target}
\end{figure}

\begin{figure}[h]
    \centering
    \includegraphics[width=0.89\textwidth]{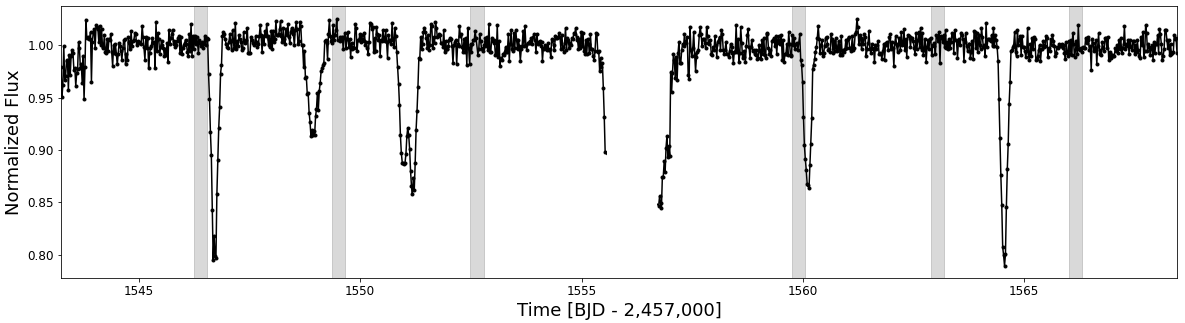}
    \includegraphics[width=0.89\textwidth]{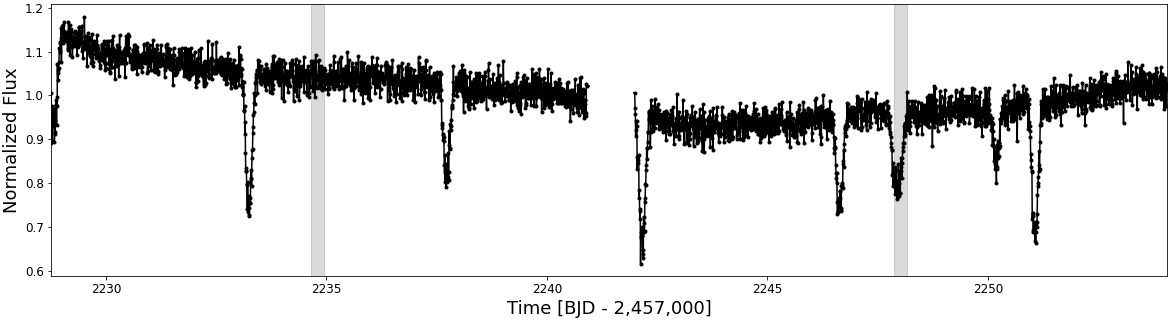}
    \\
    \hfill
    \subfloat{
        \includegraphics[width=0.36\textwidth]{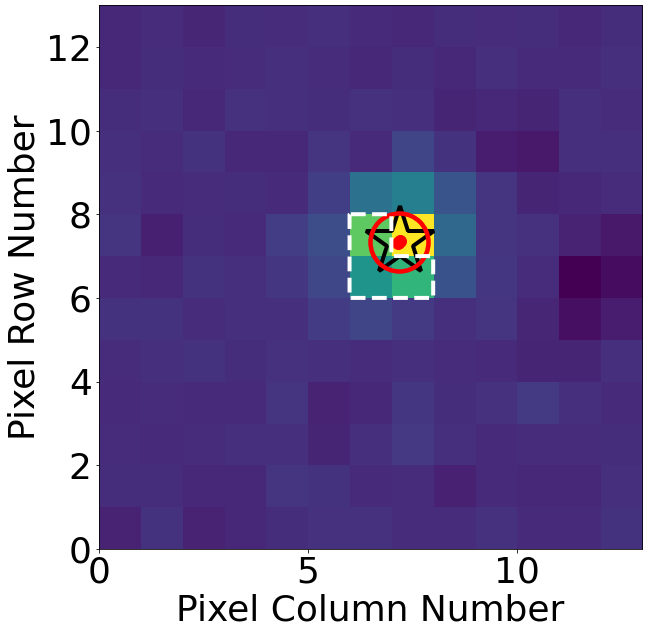}
    }
    \hfill
    \subfloat{
        \includegraphics[width=0.36\textwidth]{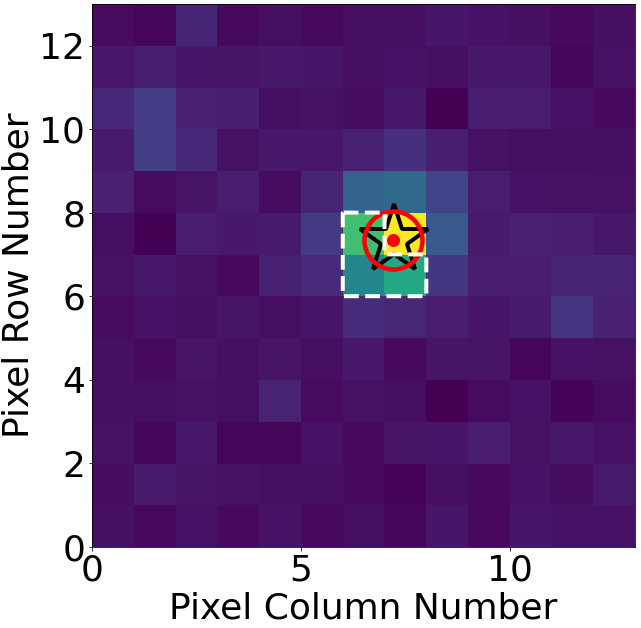}
    }
    \hfill
    \null
    \caption{First and second row: Long-cadence {\em TESS} {\tt eleanor} lightcurve of TIC 53201931, a triply-eclipsing triple star producing a pair of tertiary elipses in Sector 9 (first row, 30-min cadence) and in Sector 34 (second row, 10-min cadence). Third row, left and right panels: photocenter measurements for the inner binary (left, with a period of 8.9-days) and for the third star (right, with a period that is $\sim$700/N, where N is an integer). Note how the pair of tertiary eclipses change shape and order -- the ``smoking gun'' signature of a third star orbiting the EB. This is an example of on-target false positive quadruple due to a triple star.}
    \label{fig:triple_}
\end{figure}

\begin{figure}[h]
    \centering
    \includegraphics[width=0.95\textwidth]{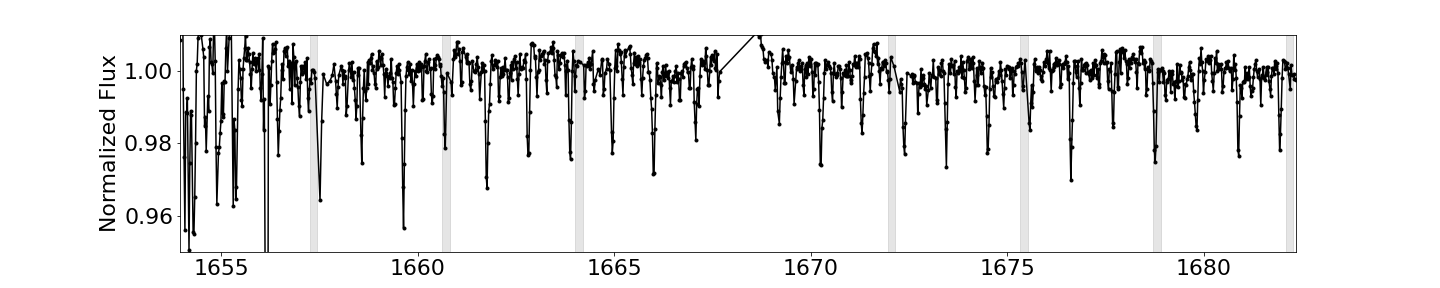}
    \includegraphics[width=0.45\textwidth]{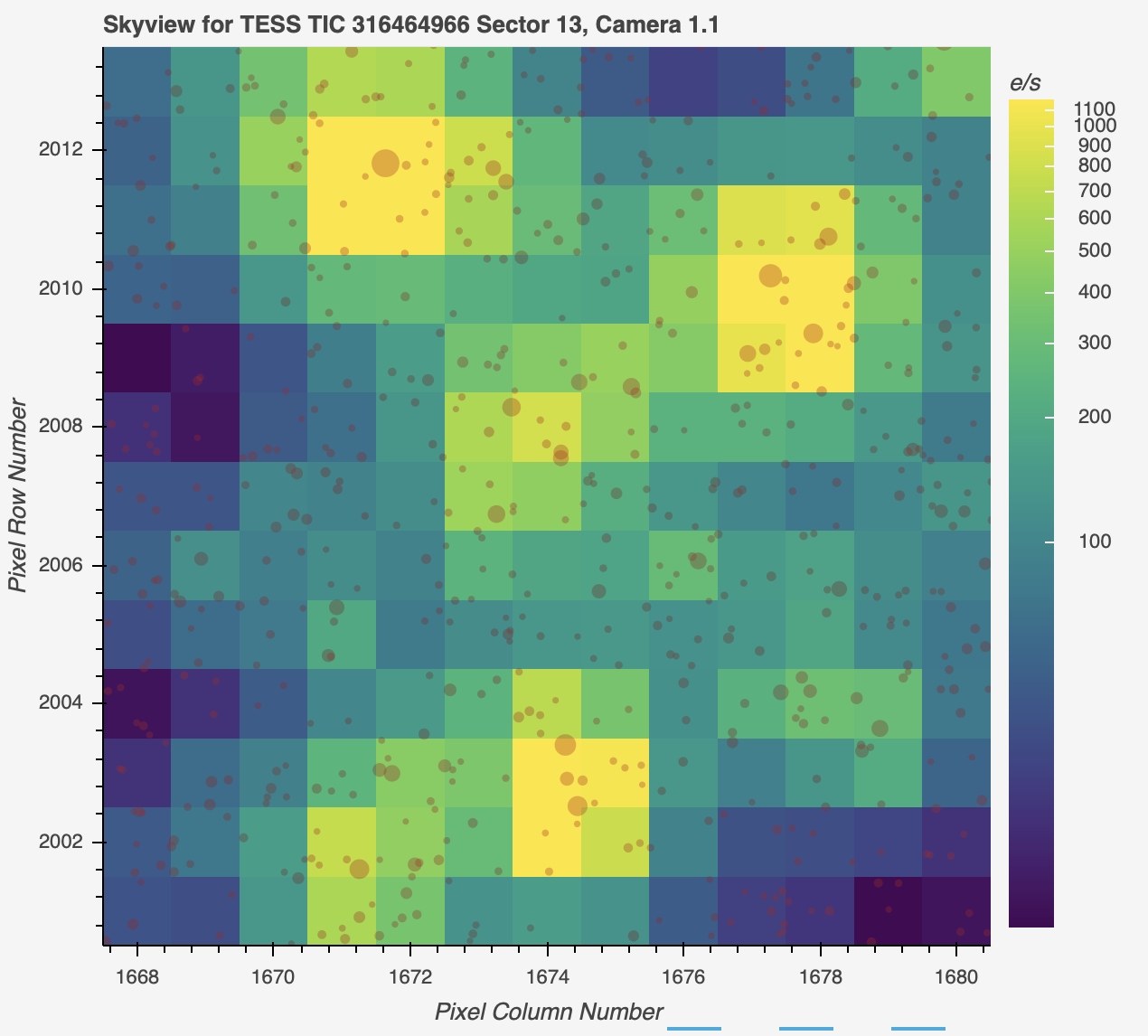}
    \caption{Upper panel: Long-cadence {\em TESS} {\tt eleanor} lightcurve of PTIC 316464966 showing two sets of eclipses, with periods of P1 = 1.06 days and P2 = 0.24 days. The {\em TESS} Input Catalog lists two stars separated by about 1.11 arcsec, TIC 1821336783 and TIC 1821336772, with respective magnitudes of T = 13.83 mag and T = 14.27 mag. Either star can produce either set of eclipses, and the separation between them is too small for reliable photocenter measurements. This is an example of a potential false positive due to blended stars that are too far apart to be of interest for this study (in this case $\sim$ 1000 AU).}
    \label{fig:blend_}
\end{figure}

In general, we encountered the following five false positive scenarios:

\begin{description}
  \item[$\bullet$ Target EB + Field EB] This represents a scenario where the photocenter analysis shows that one of the detected EBs is the target itself and the other is a nearby field star whose signal bleeds into the target's aperture (see Fig. \ref{fig:on_off_target}).
  \item[$\bullet$ Field EB (star 1) + Field EB (star 1)] A scenario where the photocenter analysis shows that both EBs originate from a nearby quadruple candidate which bleeds into the aperture of the target star (see Fig. \ref{fig:off_target_quad}). We note that here the off-target quadruple is a genuine candidate, but the target itself is considered in this work to be a false positive. 
  \item[$\bullet$ Field EB (star 1) + Field EB (star 2)] A scenario where the photocenter analysis shows that two EBs from two field stars bleed into the target's aperture (see Fig. \ref{fig:off_off_target}). Star 1 and star 2 may or may not be in a wide quadruple system. 
  \item[$\bullet$ Target triple star] A scenario where a triply-eclipsing triple star produces a single pair of tertiary eclipses that mimic a second highly-eccentric eclipsing binary with a period longer than the duration of the observations; the photocenter analysis shows that tertiary eclipses originate from the target. If there are more pairs of these eclipses in additional sectors of data, the target can be immediately marked as a triply-eclipsing triple star as the pairs of tertiary eclipses will (usually) vary in shape and order between consecutive conjunctions (see Fig. \ref{fig:triple_})
  \item[$\bullet$ Potential false positive] A case where one or more field stars nearly overlap with the target and are bright enough to produce a second set of eclipses in the target's lightcurve. The angular separation between the target and the field stars (sub-arcsec separation) is too small for reliable photocenter measurements, and is likely beyond the capabilities of dedicated follow-up as well (see Fig. \ref{fig:blend_}). The contaminating field star may or may not form a physical quadruple with the target.
\end{description}

For completeness, we perform a preliminary comparison between each set of ephemerides for each quadruple candidate presented here to those of a sample of 31,154 EB candidates (unvetted) from the GSFC EB Catalog (Kruse et al. in prep). 

Restricting the ephemerides match to $10^{-3}$ fractional difference in orbital period and $10^{-3}$ in orbital phase, 3 of the quadruple candidates show close matches with targets from the GSFC {\em TESS} EB catalog (see Table \ref{tab:quads_vs_ebs}). These are (i) TIC 63459761 vs TIC 63459765/63459804/63459811, which is due to contamination as the coordinates are very close; (ii) TIC 283940788 (ra = 8.85, dec = 62.90) vs TIC 285609529/285609535 (ra = 296.25/296.25, dec = 26.14/26.14); and (iii) TIC 370440624 (ra = 143.23, dec = -68.68) vs TIC 451982722/451982756 (ra = 296.58/296.58, dec = 27.06/27.06). Assuming no cross-talk between distant {\em TESS} pixels, (ii) and (iii) are due to coincidence as the corresponding sky coordinates are quite different. While a comprehensive cross-match is beyond the scope of this work, it could be performed once the GSFC FFI EB Catalog is released and fully vetted (Kruse et al. in prep).

\begin{table}[]
    \centering
    \caption{Ephemerides cross-match between the quadruple candidates presented here and $\sim30,000$ unvetted EB candidates from the GSFC {\em TESS} EB Catalog (Kruse et al. in prep)}
    \begin{tabular}{c|ccccccc}
\hline
TIC & Period & T0 & RA & Dec & Sectors Obs & Comments \\

\hline
{\bf 63459761} & 4.3621 & 1683.8108 & 308.5251 & 41.1359 & 14-15 & contamination \\
63459765 & 4.3621 & 1683.8113 & 308.5181 & 41.1355 & 14-15, 41 \\
63459804 & 4.3621 & 1683.8125 & 308.5183 & 41.1309 & 14-15, 41 \\
63459811 & 4.3620 & 1683.8125 & 308.5300 & 41.1297 & 14-15, 41 \\
\hline
{\bf 283940788} & 0.8768 & 1765.3137 & 8.8514 & 62.9015 & 17-18, 24 & coincidence \\
285609529 & 0.8767 & 1683.7647 & 296.2538 & 26.1412 & 14, 40-41 \\
285609535 & 0.8767 & 1683.7647 & 296.2518 & 26.14063 & 14, 40-41 \\
\hline
{\bf 370440624} & 2.2350 & 1572.7416 & 143.2320 & -68.6811 &  9-11, 36-38 & coincidence \\
451982722 & 2.2334 & 1684.4948 & 296.5768 & 27.0600 & 14, 41 \\
451982756 & 2.2334 & 1684.4950 & 296.5829 & 27.0637 & 14, 41 \\
\hline
\hline
    \end{tabular}
    \label{tab:quads_vs_ebs}
\end{table}

\section{Ephemeris Determination}

To determine the ephemerides of the two EBs for each quadruple candidate in this catalog, we perform the following steps. First, we run a preliminary Box-Least Square \citep[BLS,][]{2002A&A...391..369K} analysis of the lightcurve of each target for each available sector. Where needed, we clean the lightcurve by removing sections with partially- or fully-blended eclipses, as well as sections exhibiting known systematic effects. Additionally, for targets exhibiting prominent out-of-eclipse modulations we detrend the lightcurve by masking out the eclipses, applying a Savitsky-Golay filter to remove the variability, and adding the eclipses back in. For completeness, we include in our catalog relevant comments for any targets exhibiting prominent lightcurve variability due to either potential ellipsoidal modulations or general variability (e.g., TIC 271186951, TIC 357810643, TIC 367448265). Next, we extract small sections of the lightcurve centered on each eclipse for each binary, with a typical length of 2-3 eclipse durations as measured by BLS. Finally, using these sections we measure the eclipse times, depths and durations using four different functions -- a trapezoid, a Gaussian, a generalized Gaussian of the form

\begin{equation}
{\rm F(t) = A - B e^{-(\frac{|t-t_o|}{\omega})^{\beta}} + C(t-t_o)}
\label{eq:GG}
\end{equation}

{\noindent and a generalized hyperbolic secant of the form} 

\begin{equation}
{\rm F(t) = A - \frac{2B}{e^{-(\frac{|t-t_o|}{\omega})^{\beta}} + e^{(\frac{|t-t_o|}{\omega})^{\beta}}} + C(t-t_o)}
\label{eq:GHS}
\end{equation}

{\noindent where A, B, C, $t_o$, $\omega$, and ${\beta}$ are free parameters.} For $\beta = 2$, Eqn. \ref{eq:GG} becomes a standard Gaussian except for an additional factor of 2 in the denominator of the exponential (which can be absorbed into $\omega$). We note that (i) ${\beta}$ can take non-integer values; and (ii) the ${\rm C(t-t_o)}$ term in Equations \ref{eq:GG} and \ref{eq:GHS} helps minimize the effects of in-eclipse lightcurve variability by accounting for a residual linear trend.

\begin{figure}
    \centering
    \includegraphics[width=0.95\textwidth]{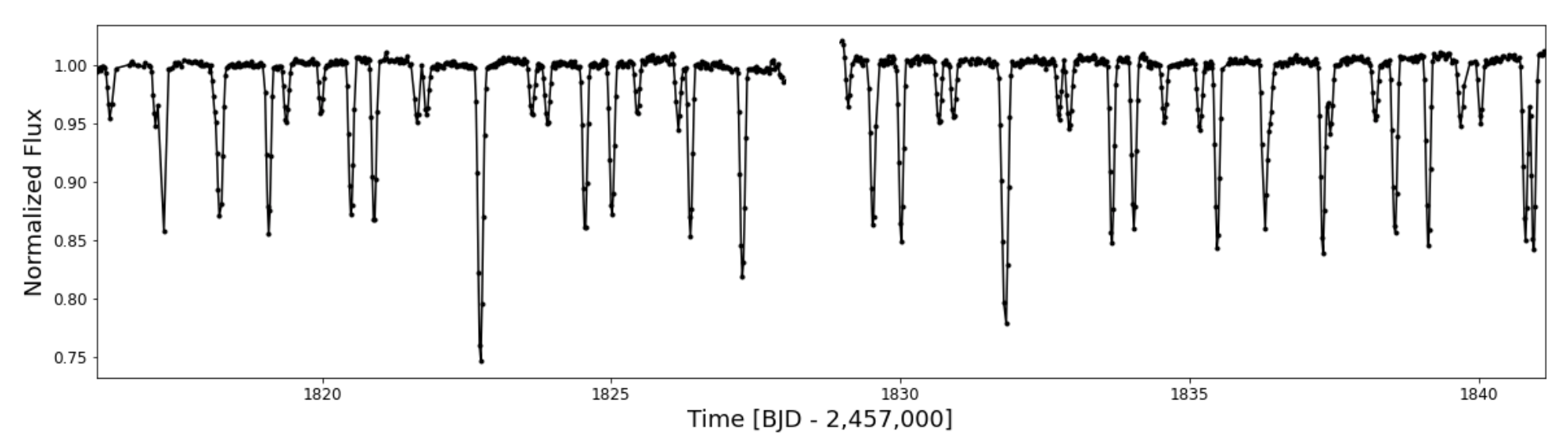}
    \includegraphics[width=0.95\textwidth]{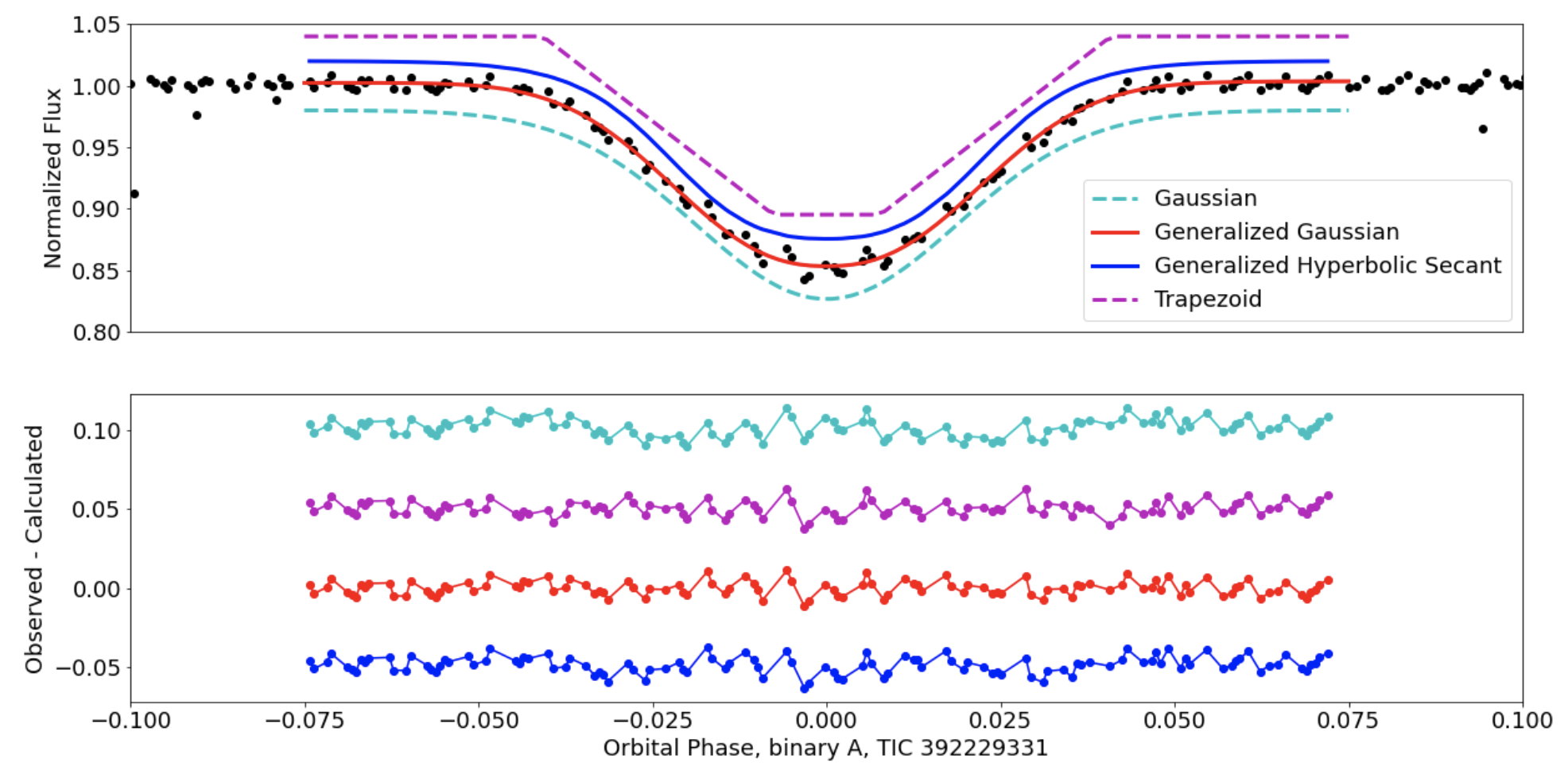}
    \includegraphics[width=0.95\textwidth]{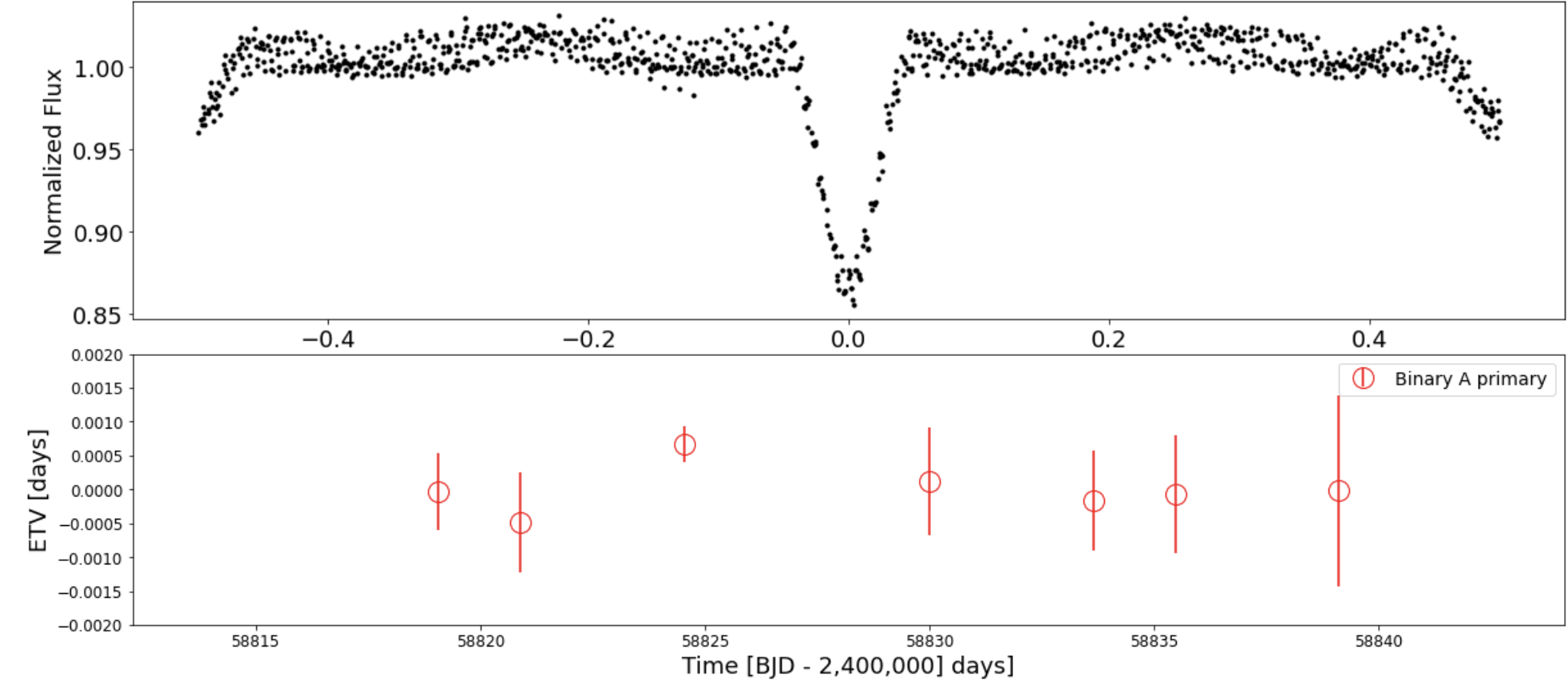}
    \caption{Upper panel: lightcurve for quadruple candidate TIC 392229331 exhibiting four sets of eclipses with two different periods. Second panel from top: Trapezoid (magenta), gaussian (cyan), generalized gaussian (red) and generalized hyperbolic secant (blue) fits to the phase-folded, primary eclipses of Binary A (zoomed-in on phase between -0.1 and 0.1). The blended eclipses have been removed and the lightcurve has been detrended. The magenta, blue, and cyan curves are vertically offset for clarity. Third panel from top: Residuals, in terms of (observed-calculated), for the four respective functions. Fourth panel from top: the undetrended, phase-folded lightcurve of binary A using the ephemeris determined by the fits to the individual eclipses. The secondary eclipse is clearly-distinguished. The eclipses of Binary B have been removed for clarity; Last panel: measured eclipse timing variations for the primary eclipses of Binary A, showing no significant trends. }
    \label{fig:TIC_392229331}
\end{figure}

Because eclipse depths can vary between sectors (due to genuine changes caused by dynamical interactions or simply because of systematics), we fit each eclipse individually and then adopt the measurements from the corresponding function which provides the smallest chi-square. An example is shown in Fig. \ref{fig:TIC_392229331} for the case of TIC 392229331. Here, although the four functions look similar to the eye, the generalized Gaussian provides the best fit: the chi-square ratios between the generalized Gaussian function and the Gaussian, generalized hyperbolic secant, and trapezoid are ${\approx0.77, 0.84, 0.95}$, respectively. 

\begin{figure}
    \centering
    \includegraphics[width=0.95\textwidth]{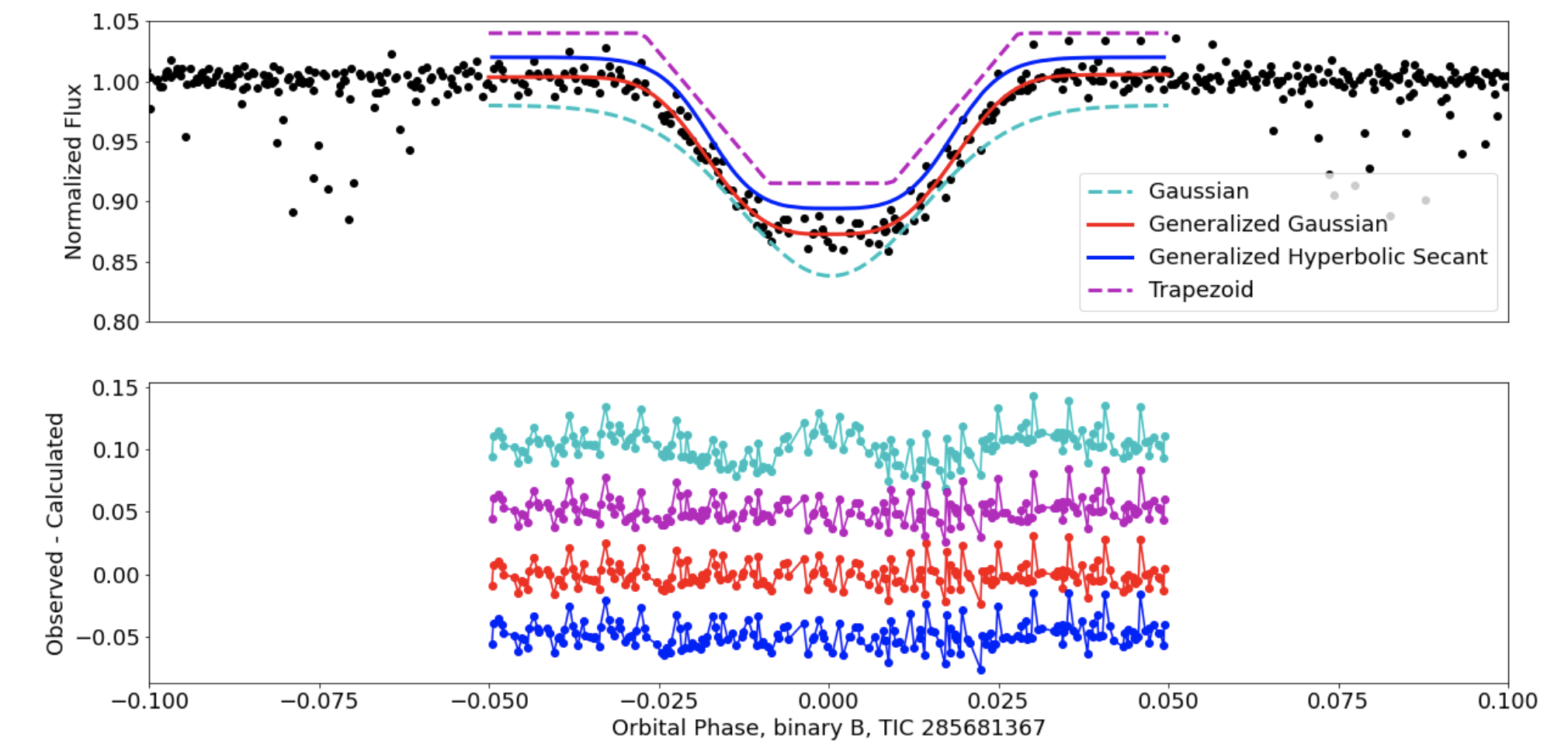}
    \caption{Same as second and third panels in Figure \ref{fig:TIC_392229331} but for a primary eclipse of binary B of TIC 285681367 (see Fig. see \ref{fig:on_on_target}). Here the generalized Gaussian and hyperbolic secant functions are clearly the better representation of the eclipse compared to a Gaussian. The flatter/sharper the eclipse bottom is, the better the former two fit the data.}
    \label{fig:GG_vs_Gauss}
\end{figure}

The differences between the Gaussian and the generalized Gaussian and secant functions -- and the benefit of using the latter two -- become more pronounced the flatter/sharper the eclipse bottom is. This is highlighted in Fig. \ref{fig:GG_vs_Gauss} for the primary eclipses of binary B of TIC 285681367 (see also Fig. \ref{fig:on_on_target}). Here, the eclipse shape clearly deviates from a Gaussian but is well-represented by the generalized Gaussian and secant functions (as well as the trapezoid function). 

This approach allows us to keep track of potential eclipse-time variations that might indicate dynamical interactions between the two EBs. Indeed, several systems exhibit such variations as discussed below.

\section{The Catalog}

\subsection{Contents of the Catalog}

The catalog presents the TIC ID, periods, eclipse times, depths, durations, secondary phases of the potential quadruple systems detected in the GSFC FFI lightcurves (Powell et al. in prep, Kruse et al. in prep), as well as additional comments including important features, issues, caveats, etc. The results are summarized in Table \ref{tbl:fitpar}. For completeness, the table also provides the estimated composite effective temperature and the TESS magnitude from the TESS Input Catalog, as well as the Gaia EDR3 distance and identifier; for convenience, we label the targets as e.g. TGV-1 for ``TESS/Goddard/VSG quadruple candidate-1". 

\begin{figure*}
    \centering
    \includegraphics[width=0.95\textwidth]{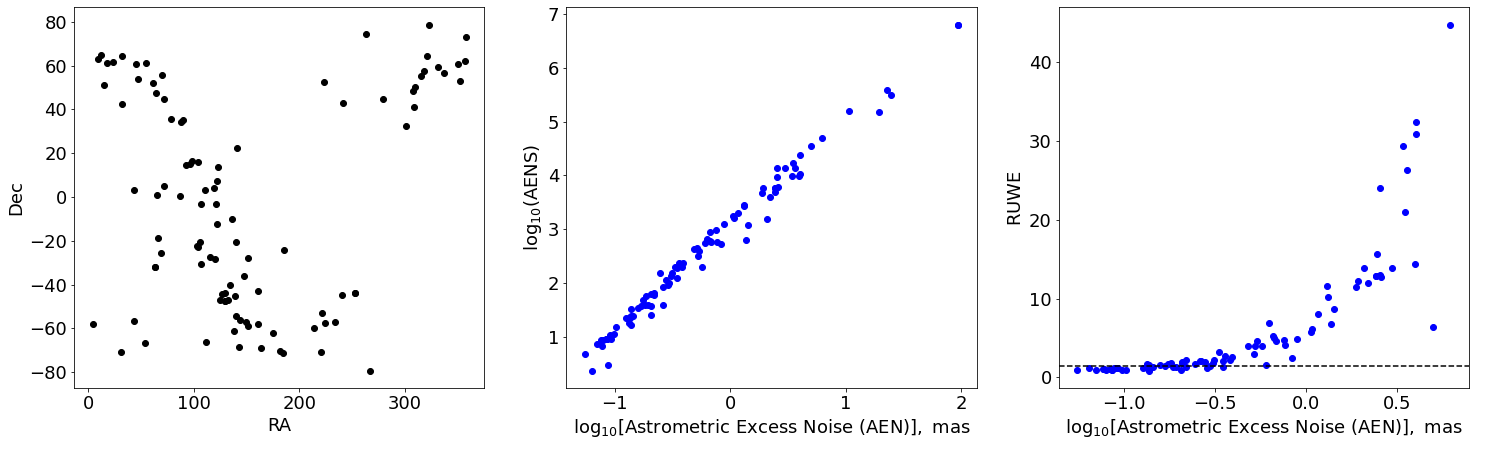}
    \caption{Left panel: RA and Dec of the quadruple candidates in this catalog. Middle panel: The respective $\rm astrometric\_excess\_noise$ (AEN) and $\rm astrometric\_excess\_noise\_sig$ (AENS) as measured by Gaia. We note that both are on a logarhitmic scale. Right panel: Corresponding distribution of the renormalized unit weight error (RUWE) as a function of the $\rm astrometric\_excess\_ noise$. The dashed horizontal line indicates RUWE = 1.5, the value above which the astrometry is likely affected by a presumptive wide companion.}
    \label{fig:RA_vs_Dec}
\end{figure*}

The sky coordinates of the 97 quadruple candidates are shown in Figure \ref{fig:RA_vs_Dec}. Compared to the quadruple candidates listed in \citet{2019A&A...630A.128Z}, our targets are uniformly-spread in declination (see their Figure 12 vs upper right panel of Figure \ref{fig:RA_vs_Dec} here). The apparent lack of targets in some parts of the sky (e.g. southern targets with RA greater than about 250$^\circ$) is likely due to the incompleteness of our catalog -- there are many more potential quadruple candidates in our database awaiting vetting and analysis. 

A powerful tool to infer the potential presence of unresolved companion(s) in a given stellar system is measurements of astrometric noise in excess of that expected from the target's parallax and proper motion. The Gaia EDR3 Catalog \citep{Gaia2021} provides such measurements in terms of an astrometric excess noise above a single star model, i.e. ${\rm astrometric\_excess\_ noise}$ (AEN), with a corresponding significance ${\rm astrometric\_excess\_noise\_sig}$ (AENS), as well as an indicator for the astrometric goodness-of-fit in terms of a renormalized unit weight error (RUWE). The method has been demonstrated for known spectroscopic binaries, used to measure binarity fraction across the HR diagram, detect hierarchical triples, and X-ray binary stars (e.g. \cite{Belokurov2020,Penoyre2020,Stassun2021,Gandhi2022}, and references therein). Figure \ref{fig:RA_vs_Dec} shows Gaia's EDR3 AEN, AENS, and RUWE for those targets in our catalog with measured AEN (91 out of 97). Of these 93, the AEN is greater than 1/5/10 mas for 32/7/6 targets, and can reach up to 93 mas (for TIC 168789840). Most of the targets (90) have AENS greater than 5 and 56 targets have AENS $>$ 100; more than half of the targets (59 out of 91) have RUWE greater than 1.5 (the value above which the astrometry is likely affected by wide companions) and 20 targets have RUWE greater than 10. Altogether, these indicate potentially significant orbital motion between the two unresolved components for a large number of the quadruple candidates presented in this catalog. 

\subsection{Period Distributions}

Recent analysis of quadruple systems consisting of two eclipsing binaries in a 2+2 hierarchical configuration shows strong peaks near period ratios of 1/1 and 3/2, a smaller peak near 5/2, and no significant peak near 2/1 \citep{2019A&A...630A.128Z}. The systems representing the 1/1 peak are close to resonance but not exactly in resonance, and only about a quarter of the systems representing the 3/2 peak are close to an exact period ratio. With the caveat that some of the studied systems may not be genuine quadruples, and also assuming co-planar orbits, it has been suggested that period ratios of 2/1 and 3/2 could be due to resonant capture and should be common \citep{2020MNRAS.493.5583T}. However, a period ratio near or at 1/1 is expected to be rare as the corresponding resonance capture is inefficient \citep{2018MNRAS.475.5215B,2020MNRAS.493.5583T}. This makes makes the origin of the 1/1 peak in the \cite{2019A&A...630A.128Z} sample puzzling. 

\begin{figure*}
    \centering
    \includegraphics[width=0.99\textwidth]{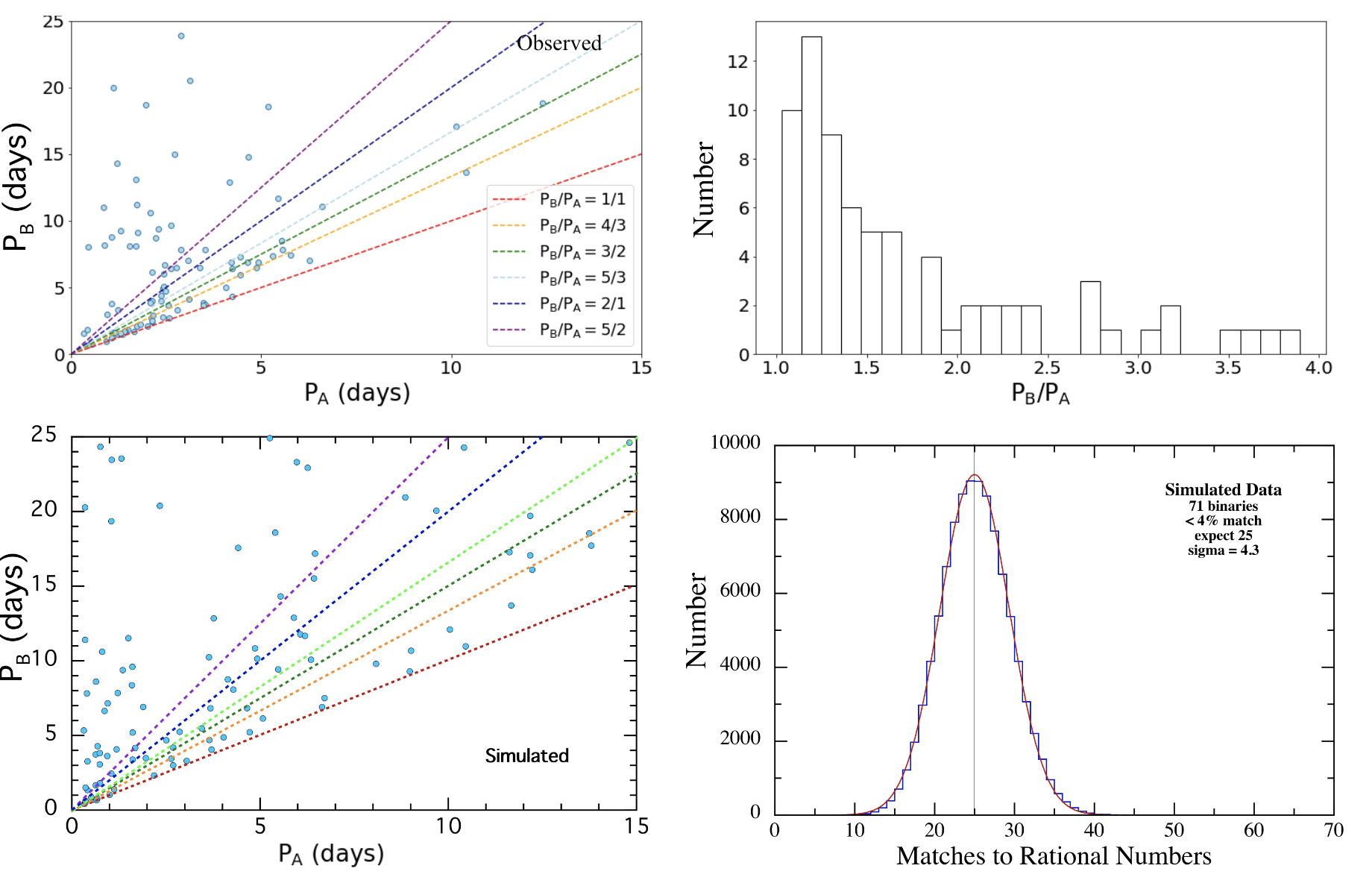}
    \caption{Upper left: Measured period ratios between ${\rm P_B}$ and ${\rm P_A}$ for the 97 quadruple candidates from {\em TESS} that pass vetting tests, compared to select integer ratios. Upper right: Distribution of the number of systems with periods ratio ${\rm P_A/P_B}$ smaller than 4 (72 out of 97 systems). Lower left: Same as upper left but for a simulated set of systems. Lower right: Number of simulated systems with period ratios that match to within 4\% one of the rational numbers 1:1, 5:4, 4:3, 3:2, 5:3, 2:1, 5/2, or 3:1. See text and Table \ref{tab:simulated_period_ratios} for details.}
    \label{fig:PAvPB}
\end{figure*}

Our catalog of uniformly-vetted quadruple candidates presents a new opportunity to study period distributions. These are shown in Fig. \ref{fig:PAvPB} for the 97 candidates that pass our vetting tests measurements. As a comparison to Figure 13 of \citet{2019A&A...630A.128Z}, 72 of the 97 systems have period ratios ${\rm P_B/P_A}$ smaller than 4. The distribution of these 72 systems is shown as a histogram in the upper right panel of Fig. \ref{fig:PAvPB} presented here, using the same number of bins as \citet{2019A&A...630A.128Z} (26). Of these 72 systems, 24(7) have period ratios of 1:1, 5:4, 4:3, 3:2, 5:3, 2:1, 5:2, and 3:1 to within 4\%(1\%), respectively; the value of 4\% was chosen as twice the difference between the 5:4 and 4:3 period ratios.  

To evaluate the significance of the measured period ratios, we computed the matches to rational numbers 1:1, 5:4, 4:3, 3:2, 5:3, 2:1, 5:2, and 3:1 to within 8\%, 4\%, 2\%, and 1\% for a simulated distribution of period ratios as follows. First, we selected $P_A$ periods measured randomly according to a $dp/dP_A \propto P_A^{-1}$ probability distribution covering the range of 0.3 to 16 days (in line with the 72 systems outlined above). Next, for any given $P_A$ we selected $P_B$ randomly according to $dp/dP_B \propto P_B^{-4/3}$ such that $P_B$ ranges from $P_A$ to 25 days; we experimented with various empirical power-law indices and found that the value of 4/3 yielded approximately the observed fraction of systems with $P_B/P_A (< 4)$ vs with $P_B/P_A (> 4)$. This was done 72 times to make a complete simulation of one {\it TESS} dataset. For each simulated period ratio we checked whether this ratio was within a certain percentage of a rational number. We then stored the number of matches within the set of 72 ratios. Finally, the entire process was repeated $10^5$ times and distributions of matches to rational numbers were computed. The mean numbers of expected matches vs those found in the data, as a function of the percentage match requirement, are listed in Table \ref{tab:simulated_period_ratios}. 

Overall, these simulations show that the numbers of accidental matches with rational numbers agrees to within the statistical uncertainties with the observed numbers for each of the four percentage match requirements. From this, we conclude that there is no evidence in our data set for an enhancement of period ratios in quadruples at the rational number values. In turn, this indicates that either (i) we have insufficient statistics to conclude that there is an enhancement at rational number period ratios, or (ii) if Nature prefers special period ratios they are not sufficiently close to rational numbers for us to measure them.

Finally, once the quadruple orbital period and eccentricity are known, a dynamical stability study on these systems would provide further proof that they are not just coincident but gravitationally bound.

\begin{table}[]
    \centering
    \begin{tabular}{c|c|c}
    \hline
    \hline
    Percentage & Observed & Simulated \\
    \hline
    1\% & 7 & $6\pm2$ \\
    2\% & 13 & $12\pm3$ \\
    4\% & 24 & $25\pm4$ \\
    8\% & 47 & $50\pm6$ \\
    \hline
    \end{tabular}
    \caption{Observed vs simulated numbers of systems with period ratios ${\rm P_B/P_A}$ within the listed percentage of 1:1, 5:4, 4:3, 3:2, 5:3, 2:1, 5:2, and 3:1.}
    \label{tab:simulated_period_ratios}
\end{table}

\subsection{Secondary Eclipses, Eclipse Depth and Duration Distributions}

\begin{figure}
    \centering
    \includegraphics[width=0.99\textwidth]{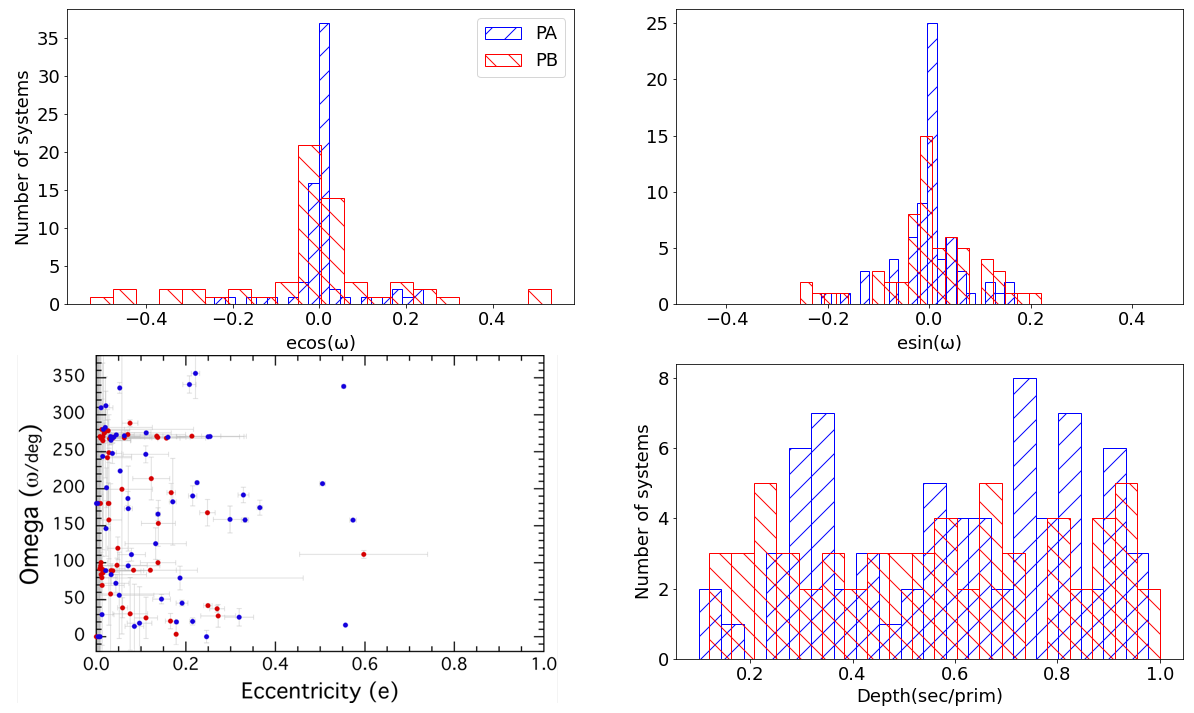}
    \caption{Upper panels: Measured ${\rm e\cos(\omega)}$ (left panel) and ${\rm e\sin(\omega)}$ (right panel) for the corresponding binary A (blue) and B (red) for each quadruple candidate exhibiting both primary and secondary eclipses. Most ${\rm e\cos(\omega)}$ and ${\rm e\sin(\omega)}$ ratios cluster around zero. Lower left panel: Corresponding eccentricity and omega values with the associated measured uncertainties, calculated by propagating the uncertainties in ${\rm e\cos(\omega)}$ and ${\rm e\sin(\omega)}$ into those for e and $\omega$. Lower right panel: Eclipse depth ratios, showing no clear preference.}
    \label{fig:phase_depth_ratios}
\end{figure}

The phase difference between primary and secondary eclipses directly constrains ${\rm e\cos(\omega)}$. As the eclipse times can be measured reasonably well from the data, even at relatively low SNR, ${\rm e\cos(\omega)}$ can be readily estimated. The other component of the orbital eccentricity, ${\rm e\sin(\omega)}$, is constrained by the difference between the primary and secondary eclipses durations (e.g. Prsa et al. 2011). The eclipse durations are more difficult to measure compared to the eclipse times and thus ${\rm e\sin(\omega)}$ is less-well constrained. With this in mind, the distributions of the measured ${\rm e\cos(\omega)}$ and ${\rm e\sin(\omega)}$ for the quadruple candidates presented in this catalog are shown in Figure \ref{fig:phase_depth_ratios}. Both distributions are strongly clustered around 0.0, suggesting a tendency for circular orbits. Given the relatively short orbital periods ($<15-25$ days, see Fig. \ref{fig:PAvPB}), this is not unexpected. The measured duration ratios between the primary and secondary eclipses are smaller than 1.5 with the exception of TIC 1337279468, binary C, where it is $\approx4$.

We note that the individual eclipse depths reported in Table \ref{tbl:fitpar} are guaranteed to be different from the true depths due to the mandatory ``contamination'' produced by the contribution of the other binary to the total light of each A+B system. However, the ratio between the primary and secondary eclipse depths -- an indicator of the relative brightness of the two stars in each binary, as well as of the orbital eccentricity -- are less affected by said contamination. The measured eclipse depth ratios are presented in Figure \ref{fig:phase_depth_ratios}, showing no clear preference towards a specific value. 

\subsection{Discussion}

A uniformly-vetted catalog of eclipsing quadruple systems provides the opportunity to examine in further detail both individual systems of particular interest, as well as study broader questions relevant to their formation and evolution. 

\subsubsection{Individual Systems}

Below we list several potentially interesting systems detected as part of this work. 

\begin{description}
\item[$\bullet$ TIC 168789840]
TIC 168789840 (TGV-96) is a confirmed (2+2)+2 hierarchical sextuple system consisting of an inner quadruple composed of two EBs and an outer EB \citep{2021AJ....161..162P}. The two EBs of the quadruple have orbital periods of ${\rm P_A}$ = 1.31 days and ${\rm P_B}$ = 1.57 days, and a mutual orbital period of about 4 years. The third EB has an orbital period of ${\rm P_C}$ = 8.22 days. The outer orbit of the sextuple has a period of about 2,000 years. The six stars have very similar masses (1.23-1.3  $M_\odot$ for the primaries, 0.56–0.66 $M_\odot$ for the secondaries), sizes (1.46–1.69 $R_\odot$ for the primaries, 0.52–0.62 $R_\odot$ for the secondaries), and effective temperatures (${\rm T_{eff}}$ = 6350–6400 K for the primaries, ${\rm T_{eff}}$ = 3923–4290 K for the secondaries). 

\item[$\bullet$ TIC 454140642]
TIC 454140642 (TGV-89) is a confirmed 2+2 hierarchical quadruple system composed of two EBs that exhibit strong dynamical interactions and eclipse timing variations \citep{2021ApJ...917...93K}. The two EBs have orbital periods of ${\rm P_A}$ = 10.3928 days and ${\rm P_B}$ = 13.6239 days, and a mutual orbital period of about 432 days. The entire system is practically co-planar, with mutual inclinations smaller than 0.5 degrees. The four stars have very similar masses (1.11-1.2 $M_\odot$), sizes (1.1–1.26 $R_\odot$), and effective temperatures (${\rm T_{eff}}$ = 6188–6434 K). 

\item[$\bullet$ TIC 52856877]
TIC 52856877 (TGV-6) is a candidate 2+2 hierarchical quadruple system composed of two EBs with orbital periods of ${\rm P_A}$ = 5.1868 days and ${\rm P_B}$ = 18.5864 days. For simplicity, throughout this manuscript we label the periods of the two EBs as sorted in ascending order. The target was observed in Sectors 18 and 24; the {\em TESS} lightcurve of the system is shown in \ref{fig:tic52856877}. The system exhibits strong eclipse timing variations, as seen from Figure \ref{fig:tic52856877}. Gaia EDR3 shows AEN = 0.31 mas with an AESN of 150.57, and RUWE = 2.19. Altogether, these considerations indicate kinematic motions between the two EBs and suggest that the system is gravitationally-bound. The analysis of the system, including follow-up observations, is in progress. 

\begin{figure}
    \centering
    \includegraphics[width=0.95\textwidth]{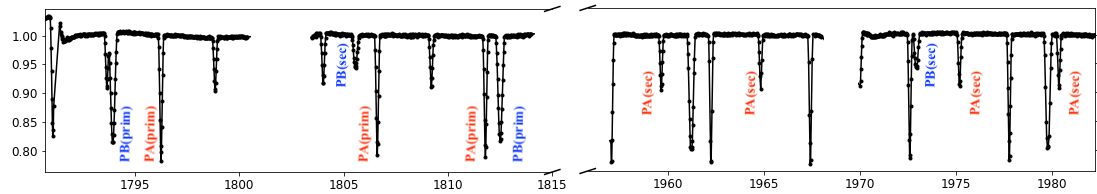}
    \includegraphics[width=0.45\textwidth]{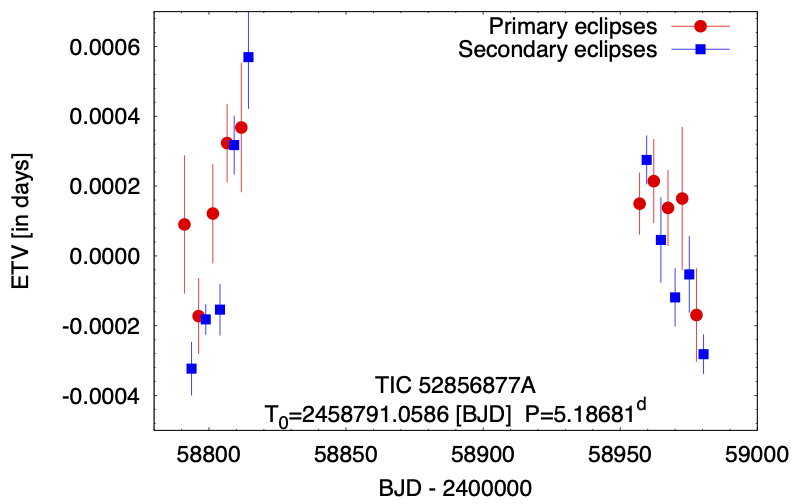}
    \includegraphics[width=0.45\textwidth]{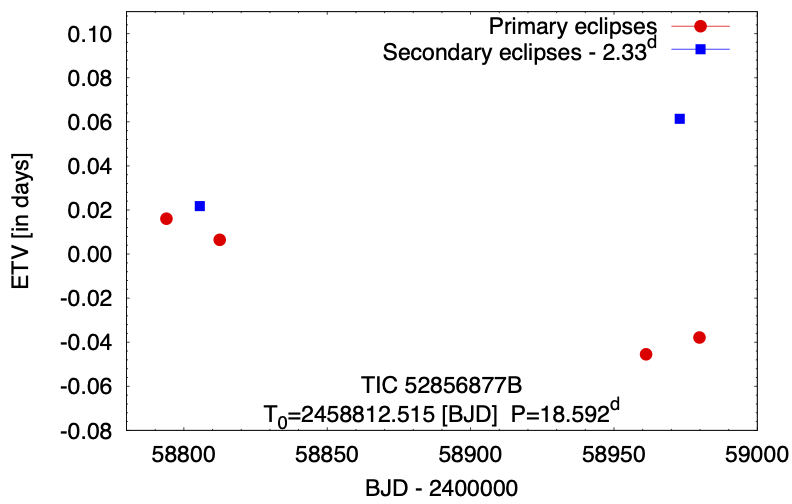}
    \caption{Upper panel: FFI lightcurve of quadruple candidate TIC 52856877. The system consists of two EBs with ${\rm P_A}$ = 5.1868 and ${\rm P_B}$ = 18.5864 days, highlighted in the figure. Lower panels: Measured eclipse-time variations for ${\rm P_A}$ (left panel) and ${\rm P_B}$ (right panel), indicating dynamical interactions between the two binaries and suggesting that the system is gravitationally-bound.}
    \label{fig:tic52856877}
\end{figure}

\item[$\bullet$ TIC 45160946]
TIC 45160946 (TGV-5) is a candidate 2+2 hierarchical quadruple system composed of two EBs with orbital periods of ${\rm P_A}$ = 3.5163 days and ${\rm P_B}$ = 7.8462 days. The target was observed in Sectors 9, 35 and 36; the {\em TESS} lightcurve for the latter two is shown in Fig. \ref{fig:tic45160946}. As in the case for TIC 52856877, TIC 45160946 also exhibits strong eclipse timing variations (see Fig. \ref{fig:tic45160946}). Gaia EDR3 shows AEN = 3.6 mas, with an AESN of 13472, and RUWE = 26.33. This indicates kinematic motions between the two EBs and suggests that the system is gravitationally-bound with a relatively short outer orbital period. A detailed analysis of the system is in progress. 

\begin{figure}
    \centering
    \includegraphics[width=0.95\textwidth]{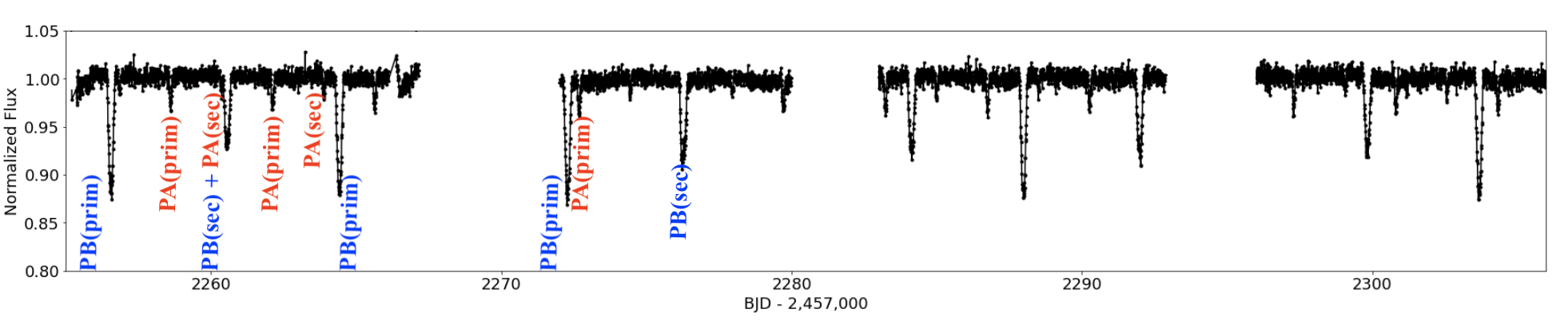}
    \includegraphics[width=0.45\textwidth]{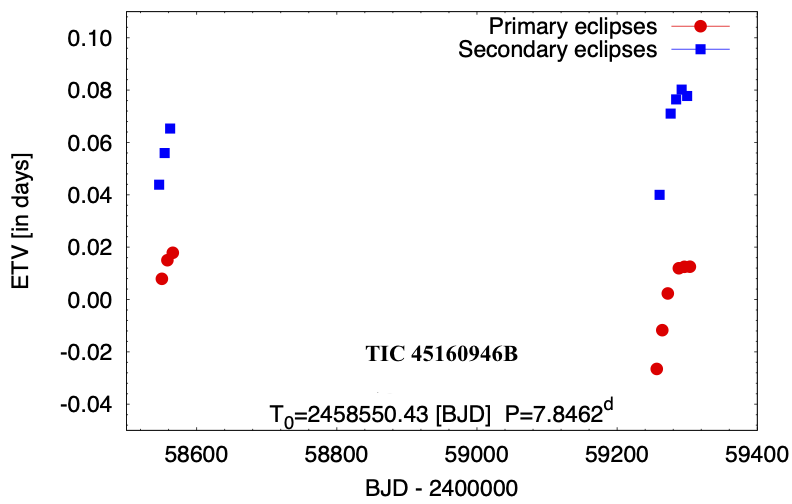}
    \includegraphics[width=0.45\textwidth]{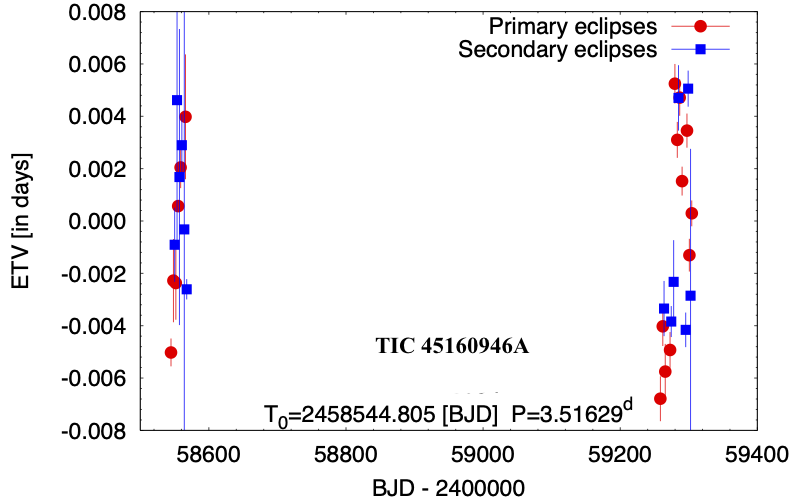}
    \caption{Same as Fig. \ref{fig:tic52856877} but for quadruple candidate TIC 45160946. The system consists of two EBs with ${\rm P_A}$ = 3.5163 and ${\rm P_B}$ = 7.8462 days, highlighted in the figure. Lower panels: Measured eclipse-time variations for ${\rm P_A}$ (left panel) and ${\rm P_B}$ (right panel), indicating dynamical interactions between the two binaries and suggesting that the system is gravitationally-bound.}
    \label{fig:tic45160946}
\end{figure}

\item[$\bullet$ TIC 256158466]
TIC 256158466 (TGV-37) is a candidate 2+2 hierarchical quadruple system composed of two EBs with orbital periods of ${\rm P_A}$ = 5.7745 days and ${\rm P_B}$ = 7.4544 days, with nearly-identical primary and secondary eclipses for ${\rm P_A}$ (depths of 131 parts-per-thousand (ppt) and 128 ppt, respectively). The target was observed in Sectors 12, 13 and 39; the {\em TESS} lightcurve for the latter is shown in Fig. \ref{fig:tic256158466}. Our analysis shows potential sinusoidal variations for the times of the ${\rm P_A}$ eclipses, although the variations only appear if we use relatively narrow sections of the lightcurve centered on the corresponding eclipses. If we use somewhat wider sections of the eclipses, the variations disappear in the scatter, as demonstrated in Figure \ref{fig:tic256158466}. Thus we label this target as showing potential ETVs. 

We also note that there is a nearby field star, TIC 1508756606, with a separation of 5.65 arcsec, coordinates of RA = 17:47:35.21 and Dec = -79:22:40.94, and magnitude difference ${\rm \Delta T\approx 4.8\ mag}$. Our analysis rules out TIC 1508756606 as a source of either ${\rm P_A}$ or PB. The coordinates of TIC 1508756606, along with Gaia's parallax ($1.26\pm0.32$ arcsec vs $1.4\pm0.02$ arcsec for TIC 256158466) and proper motion (pmRA = $0.54\pm0.3$ mas/yr, pmDec = $-29.57\pm0.36$ mas/yr vs pmRA = $-0.92\pm0.02$ mas/yr, pmDec = $-29.18\pm0.02$ mas/yr for TIC 256158466), suggest that it might in fact form a co-moving quintuple system with TIC 256158466. Gaia's EDR3 AEN for both targets is zero, and the corresponding RUWE is about 0.96. 

\begin{figure}
    \centering
    \includegraphics[width=0.99\textwidth]{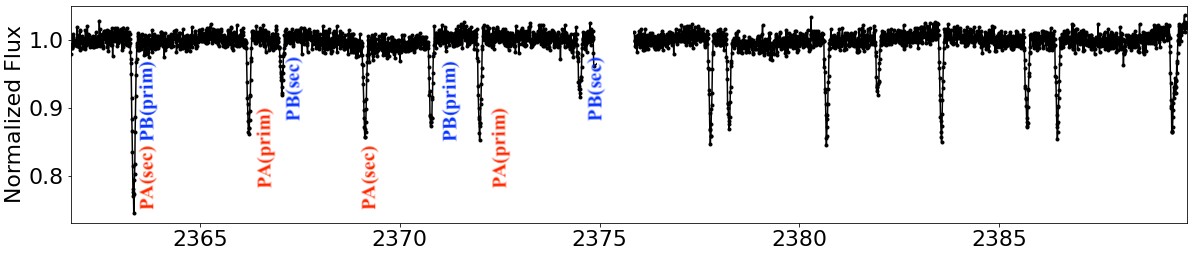}
    \includegraphics[width=0.44\textwidth]{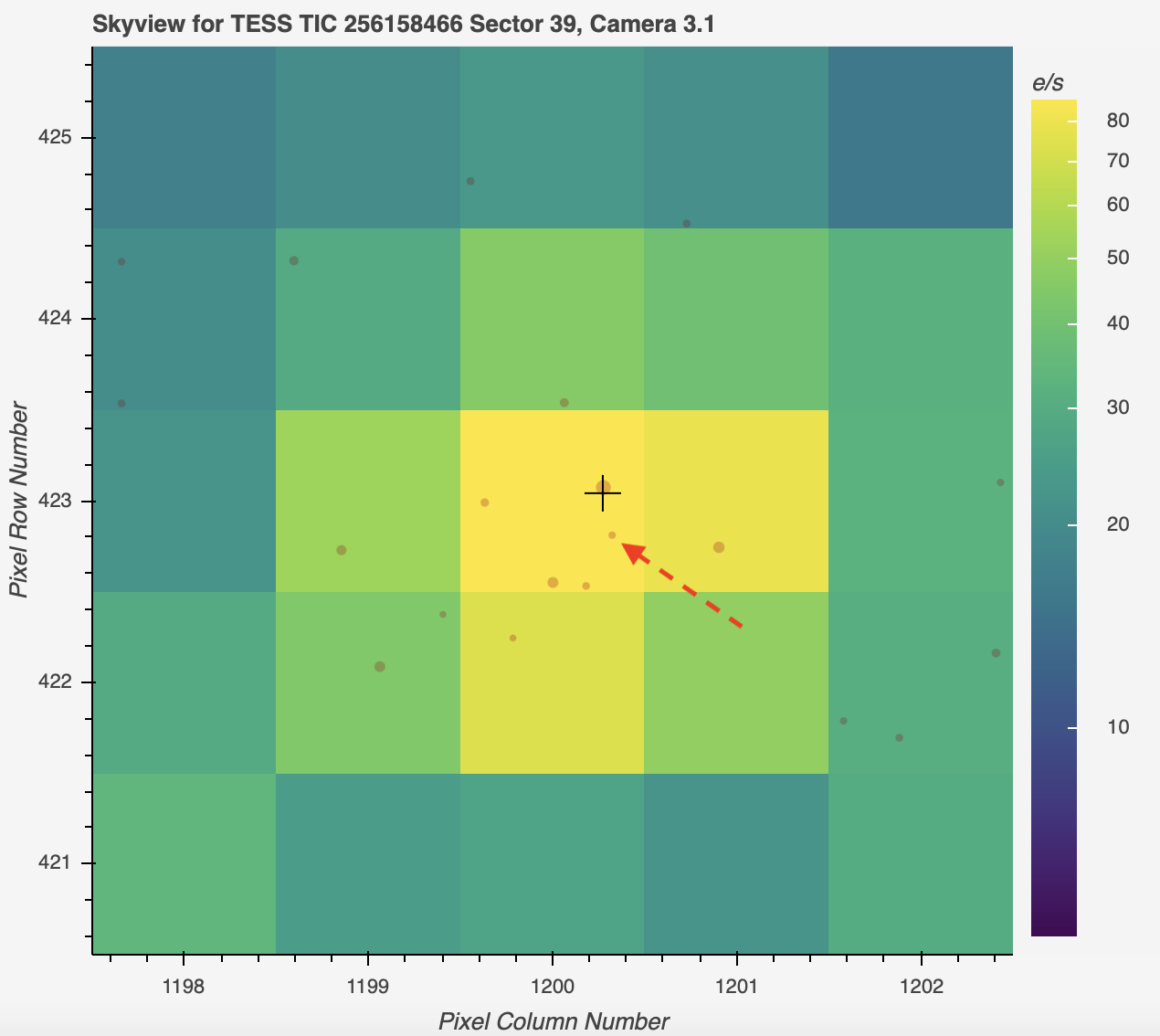}
    \includegraphics[width=0.54\textwidth]{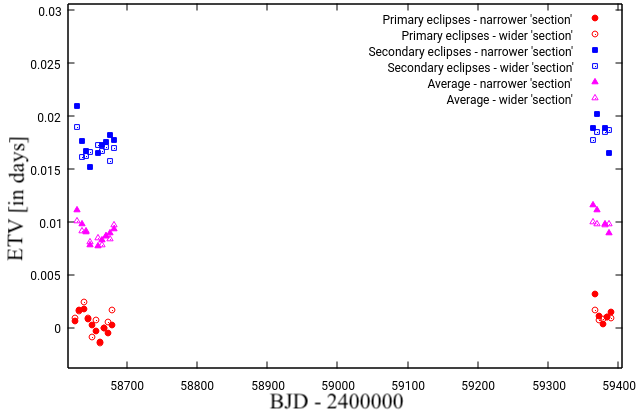}
    \caption{Upper panel: Same as Fig. \ref{fig:tic52856877} but for quadruple candidate TIC 256158466. The system consists of two EBs with ${\rm P_A}$ = 5.7745 and ${\rm P_B}$ = 7.4544 days, highlighted in the figure, both on-target. Lower left panel: Skyview image of the target for Sector 39, showing all Gaia sources down to G = 21 mag. We note that compared to Skyview images shown above, here the image is zoomed-in to a size of $5\times5$ pixels in order to highlight the position of the nearby field star TIC 1508756606 (5.65 arcsec separation, ${\rm \Delta T\approx4.8}$ mag fainter, marked with dashed red arrow). TIC 1508756606 has parallax and proper motion within one sigma of those of the target star and thus might be in a co-moving quintuple system with the quadruple TIC 256158466. Third panel: Measured eclipse-time variations for ${\rm P_A}$ using two lightcurve sections centered on each eclipse -- narrow and wide -- show potential sine-like variations only in the former case. }
    \label{fig:tic256158466}
\end{figure}

\item[$\bullet$ TIC 141733685]
The lightcurve of TIC 141733685 (TGV-95) exhibits three EBs. The target was observed in Sectors 8, 9, 35 and 36. Two of the EBs have orbital periods of ${\rm P_A}$ = 5.29 days and ${\rm P_B}$ = 7.37 days. We note that the former is seen in {\tt eleanor} data only in Sectors 35 and 36, where it is unclear whether the period is 5.29 days or half of that. To account for this, we extract the target's lightcurve using the {\tt FITSH} pipeline \citep{Pal2012} which shows the primary and secondary eclipses of ${\rm P_A}$ in all 4 sectors of data. The third EB shows three eclipses but it is unclear from {\em TESS} data whether its period, PC, is $\sim$21.8 days or $\sim$43.6 days. This is because one of three eclipses coincides with a momentum dump in S35 and is blended with a ${\rm P_A}$ eclipse, and another (in Sector 36) is blended with a ${\rm P_A}$ primary and a ${\rm P_B}$ secondary. The {\tt eleanor} and {\tt FITSH} lightcurves for Sectors 35 and 36 are shown in Fig. \ref{fig:tic141733685}. Analysis of archival ASAS-SN data for TIC 141733685 shows a clear periodicity at ${\rm P_C}$ = 43.63 days with primary and secondary eclipses, and a slight eccentricity. We note that TIC 141733685 contaminates the {\em TESS} {\tt eleanor} lightcurve of TIC 141733683 and TIC 141733701. 

Closer inspection of the field around TIC 141733685 shows that there is a nearby field star, TIC 141733688, with a separation of 4.53 arcsec, coordinates of RA = 08:39:24.8 and Dec = -47:21:39.85, and magnitude difference ${\rm \Delta T\approx 3.15\ mag}$ (see Fig. \ref{fig:tic141733685_skyview}). This field star is not bright enough to produce the ${\rm P_B}$ or ${\rm P_C}$ eclipses as the needed ${\rm \Delta T\ < 0.9\ mag}$ and ${\rm \Delta T\ < 2.06\ mag}$, respectively. TIC 141733688 is bright enough to produce the ${\rm P_A}$ eclipses (needed ${\rm \Delta T\ < 4.05\ mag}$). Another field star, TIC 141733701, has a separation of 12.6 arcsec, magnitude difference of (${\rm \Delta T\approx 2.03\ mag}$), and is nominally bright enough to produce all three sets of eclipses. 

As seen from Figure \ref{fig:tic141733685_centAB}, our photocenter analysis clearly rules out the nearby TIC 141733701 as a potential source of the detected eclipses, and shows that all three EBs originate from the position of TIC 141733685. However, the measurement is not precise enough to distinguish between TIC 141733685 and TIC 141733688 for ${\rm P_A}$ as the eclipses are very shallow and the separation very small. Thus the source of ${\rm P_A}$ can be either TIC 141733685 or TIC 141733688. With that said, the ${\rm P_A}$ photocenters for TIC 141733685 ``pull'' away from TIC 141733688 (first row of panels, Fig. \ref{fig:tic141733685_centAB}) indicating that the latter is not a likely source. Overall, these considerations suggest that there are two possibilities for the structure of the system: (i) quadruple (PB+PC) on TIC 141733685, ${\rm P_A}$ on TIC 141733688; or (ii) sextuple (PA+PB+PC) on TIC 141733685. Furthermore, the measured ETVs for both ${\rm P_A}$ and ${\rm P_B}$ imply non-linear effects (more prominent for PB, see Fig. \ref{fig:141733685_ETVs}) which strengthens the sextuple interpretation. 

Gaia EDR3 measurements show AEN = 0.085 mas, with an AENS of 9.19, and RUWE = 0.95 for TIC 141733685. The target's parallax is within one sigma of that for TIC 141733688, $0.55\pm0.01$ mas for the former vs $0.58\pm0.03$ mas for the latter. The corresponding proper motions are comparable, 7.25 mas vs 7.45 mas, respectively, although the proper motions in declination are nominally different at a greater than 3 sigma level: (pmRA = $-6.12\pm0.03$ mas/yr, pmDec = $3.89\pm0.04$ mas/yr) vs (pmRA = $-6.08\pm0.02$ mas/yr, pmDec = $4.3\pm0.02$ mas/yr). Thus TIC 141733685 and TIC 141733688 might be a co-moving system, in which case possibility (ii) above would represent a septuple at a projected separation of $\sim$8,000 AU. Given the magnitude and early spectral type of both sources, the prospect of spectroscopic follow-up might be poor.

Finally, we note that the coordinates of TIC 141733685 (RA = 08:39:25.2, Dec = -47:21:38.36) also coincide with those of the open Galactic cluster [KPS2012] MWSC 1523 (RA = 08:39:27, Dec = -47:21.4, Monteiro et al. 2020, MNRAS, 499, 1874). The proper motion of two are, however, different: (pmRA = $-6.08\pm0.01$ mas/yr, pmDec = $4.29\pm0.01$ mas/yr) for TIC 141733685 vs (average pmRA = $-4.12\pm0.77$ mas/yr, pmDec = $8.37\pm0.77$ mas/yr) for MWSC 1523.
  
\begin{figure}
    \centering
    \includegraphics[width=0.95\textwidth]{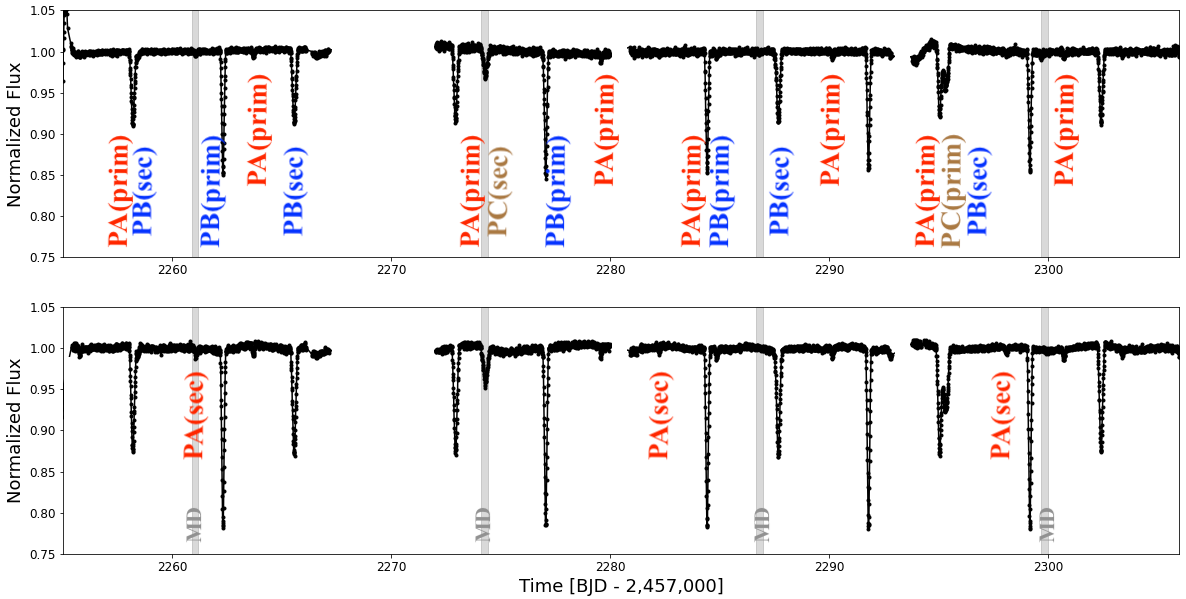}
    \includegraphics[width=0.48\textwidth]{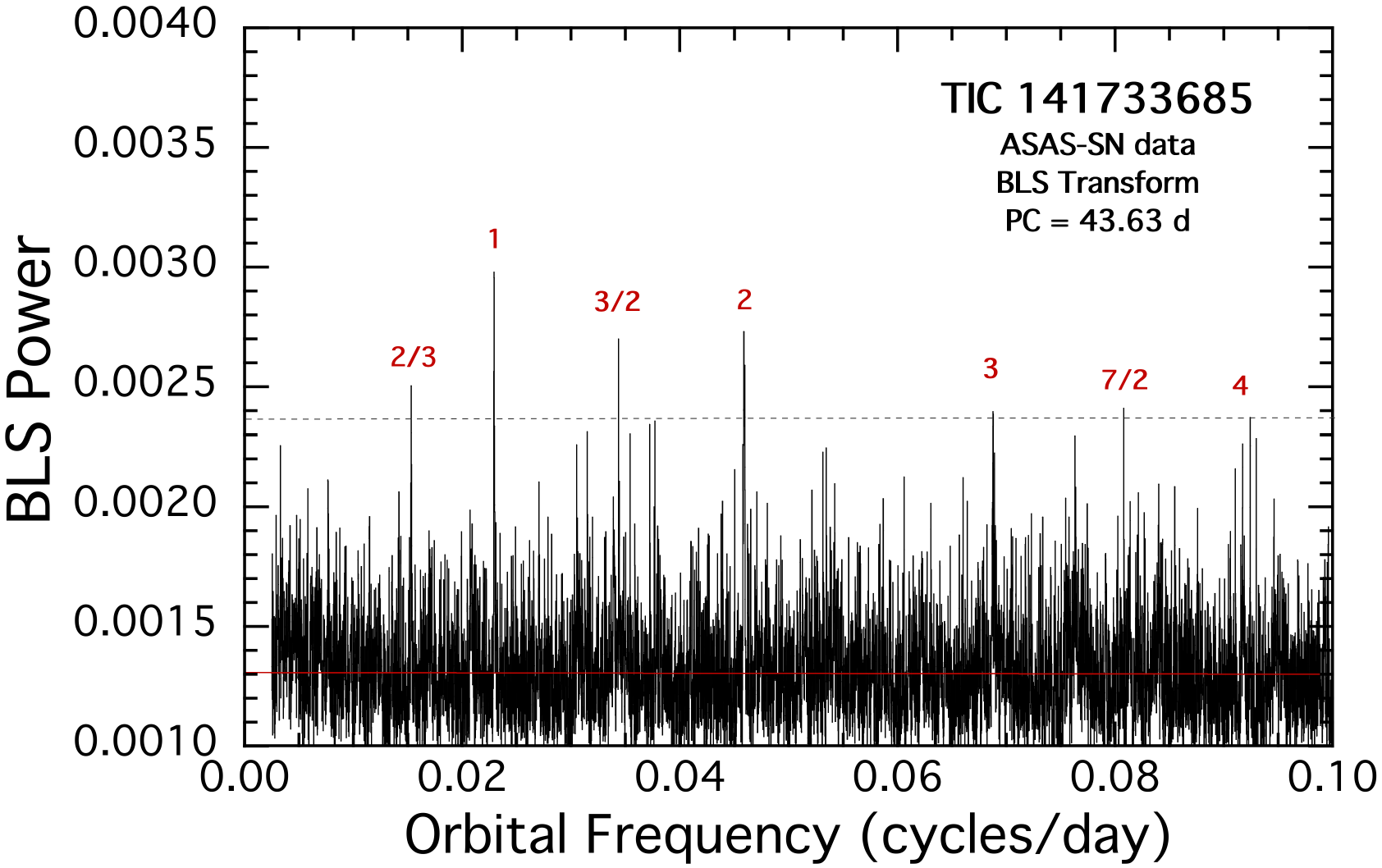}
    \includegraphics[width=0.48\textwidth]{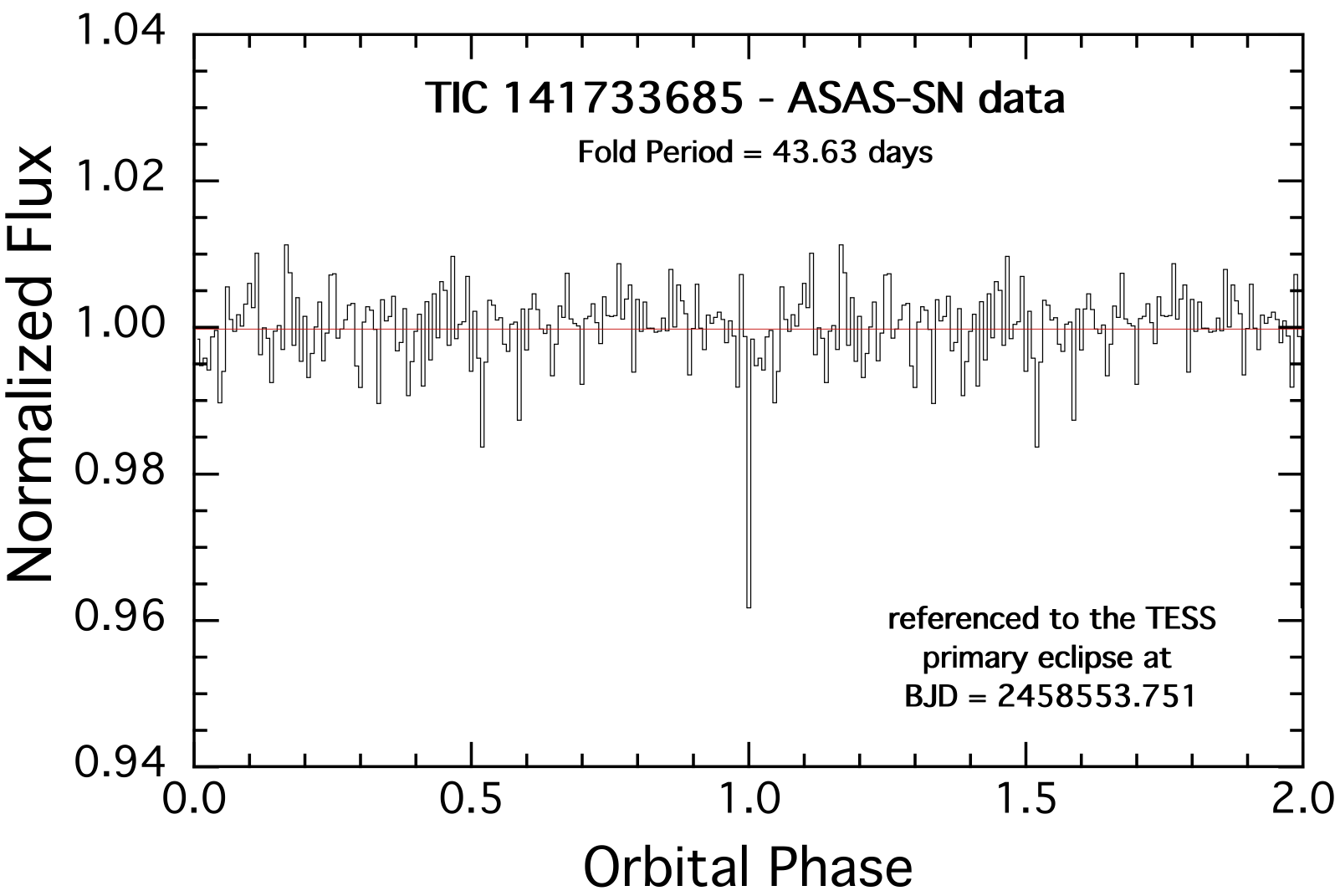}
    \caption{First panel from top: Same as Fig. \ref{fig:tic256158466} but for quadruple candidate TIC 141733685, showing the Sectors 35 and 36 {\tt eleanor} data. The target exhibits three EBs with ${\rm P_A}$ = 5.29 days, ${\rm P_B}$ = 7.37 days, and ${\rm P_C}$ = 43.63 days, as highlighted in the figure. Second panel: same as first panel but showing the {\tt FITSH} lightcurve; Lower left panel: BLS transform of the All-Sky Automated Survey for Supernovae (ASAS-SN) archival data for TIC 141733685 after removing binary A from the data train. The transform is plotted as a function of frequency in order to better identify harmonics.  The orbital period of binary C is found to be 43.63 days ($\nu = 0.02292$ cycles/day).  This peak is marked as ``1'' for the first harmonic.  Other harmonics and subharmonics are also marked in red. Nearly all of the highest peaks (i.e., above the dotted line) are due to this period or its associated harmonics. Lower right panel: Folded, binned, and averaged lightcurve from the ASAS-SN archival data. There are 150 phase bins in the plot, each corresponding to $\sim$0.3 days. The phasing is referenced to one of the {\em TESS} detected primary eclipses. There is also a likely secondary eclipse near phase 0.5, but its statistical significance is much less than that of the primary eclipse.}
    \label{fig:tic141733685}
\end{figure}

\begin{figure}
    \centering
    \includegraphics[width=0.6\textwidth]{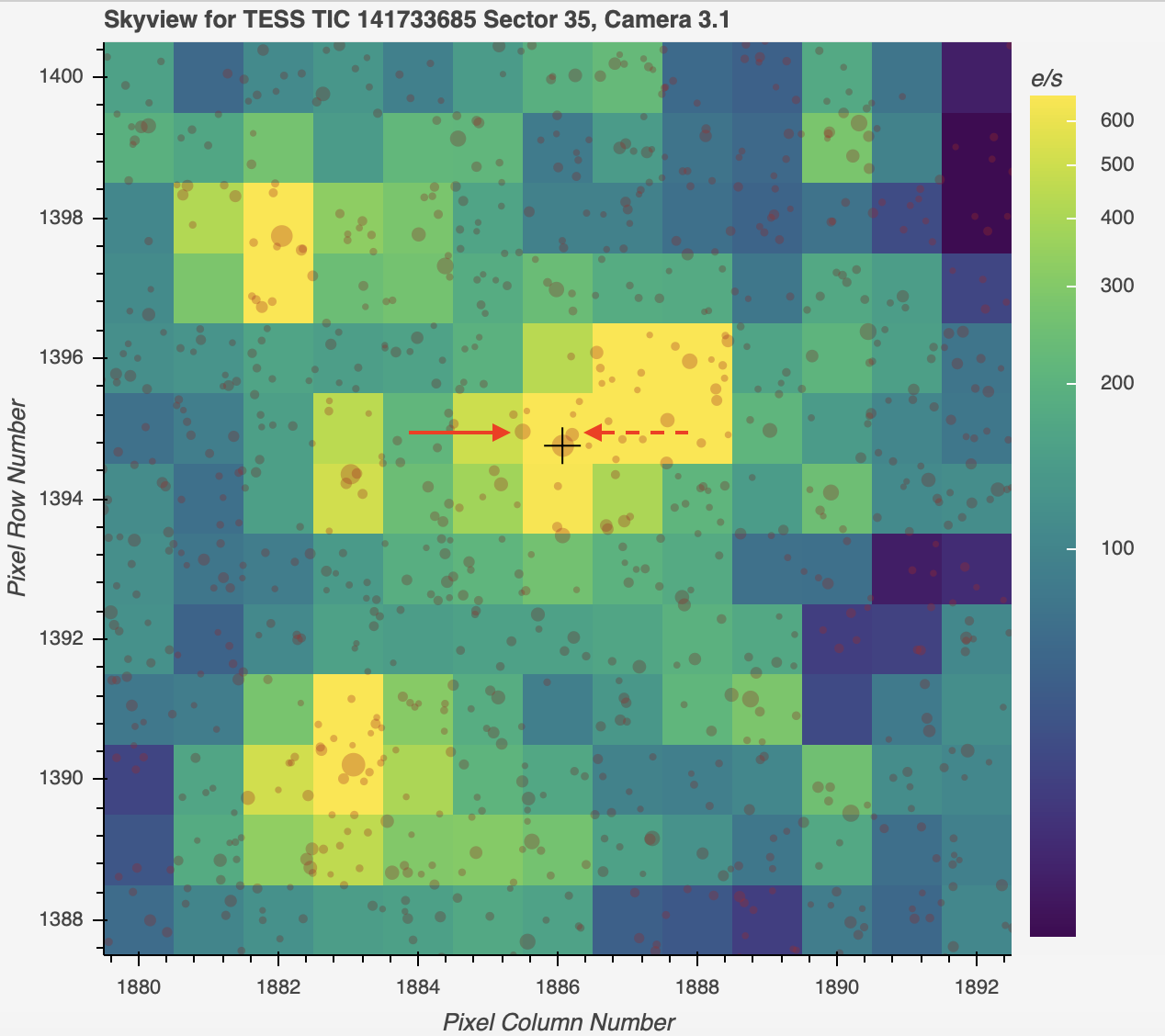}
    \caption{Skyview image of TIC 141733685 (black plus symbol) for Sector 35, showing all Gaia sources brighter than G = 21 mag. There is a nearby field star, TIC 141733688 (dashed arrow), separated from the target by 4.53 arcsec, and with magnitude difference ${\rm \Delta T\approx 3.15\ mag}$. Another field star, TIC 141733701 (solid arrow), has a separation of 12.6 arcsec and magnitude difference ${\rm \Delta T\approx 2.03\ mag}$. TIC 141733685 and TIC 141733688 have comparable parallax and, to a lesser degree, proper motion. See text for details.
    }
    \label{fig:tic141733685_skyview}
\end{figure}

\begin{figure*}
    \centering
    \includegraphics[width=0.38\textwidth]{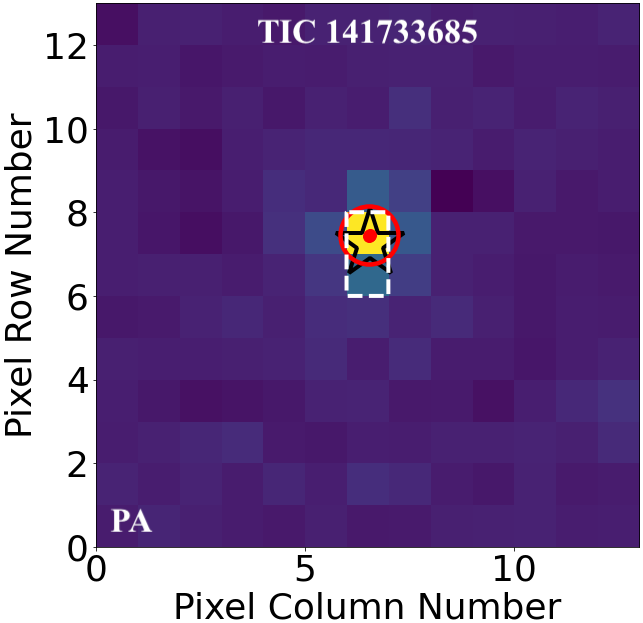}
    \includegraphics[width=0.38\textwidth]{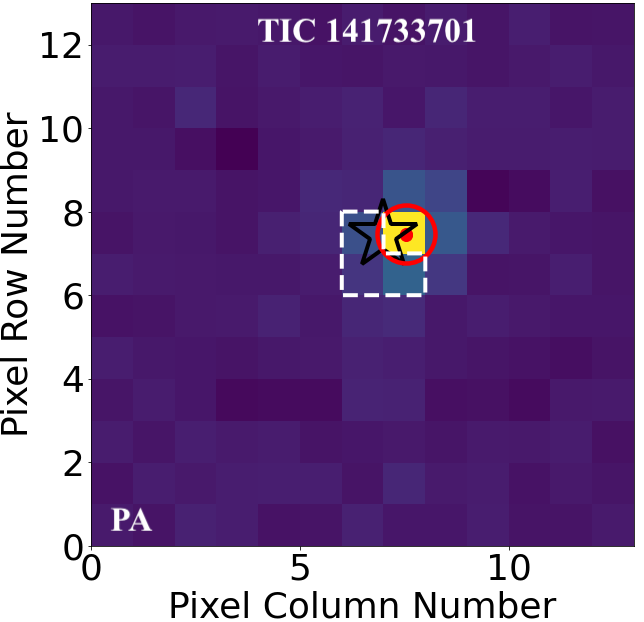}
    \includegraphics[width=0.38\textwidth]{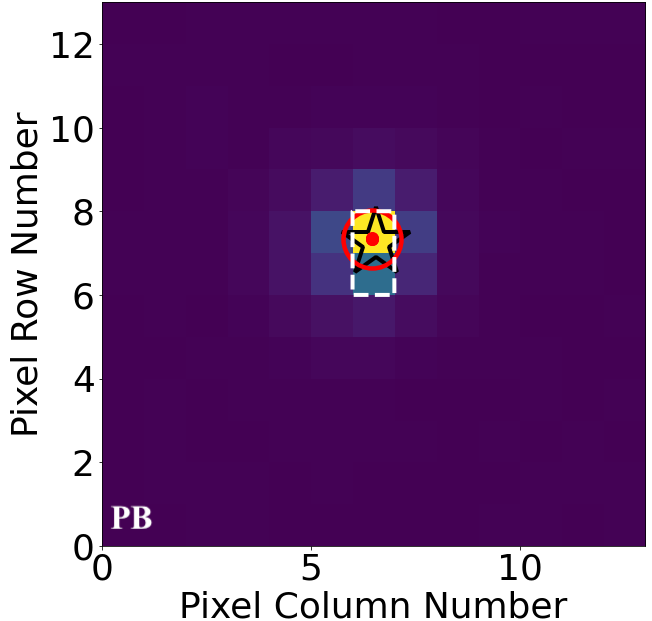}
    \includegraphics[width=0.38\textwidth]{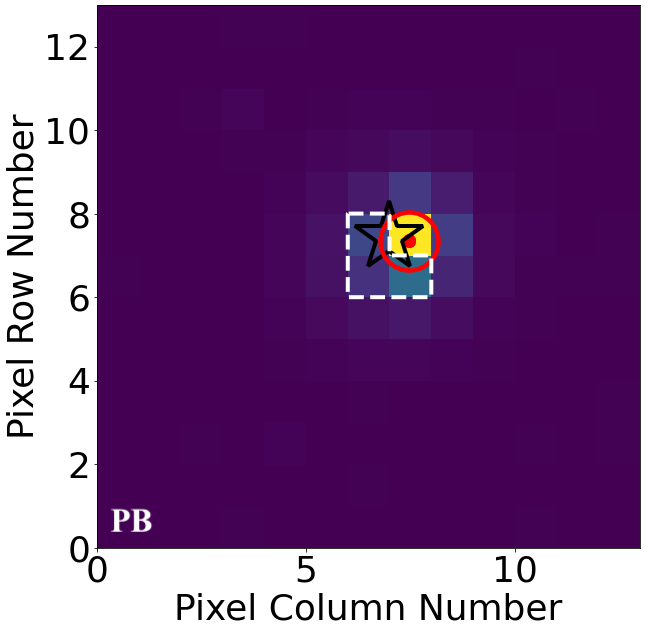}
    \includegraphics[width=0.38\textwidth]{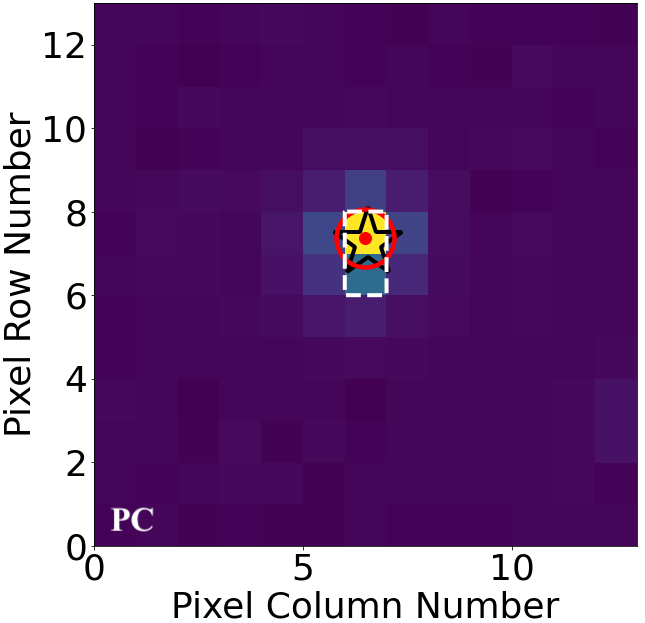}
    \includegraphics[width=0.38\textwidth]{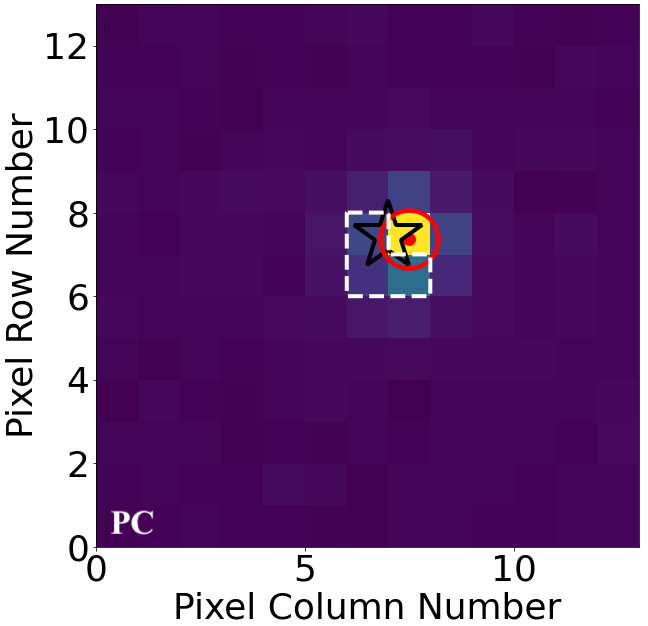}
    \caption{Photocenter analysis for ${\rm P_A}$ = 5.29 d (first row from top), ${\rm P_B}$ = 7.37 d (second row), and ${\rm P_C}$ = 43.63 d (third row) for TIC 141733685 (left panels), and the nearby field star TIC 141733701 (right panels) using Sector 35 data. The panels show the corresponding difference images along with the location of the respective target on the detector (black star, label ``TIC''), the measured per-eclipse photocenters (small red symbols, label ``Indiv Cent'') and the average photocenters (large red symbols, label ``Avg Cent''). The dashed white contours indicate the aperture used by {\tt eleanor} for the respective target. For PC, there is a single eclipse detected in this sector (see Fig. \ref{fig:tic141733685}), corresponding to a single photocenter measurement. Our analysis clearly rules out TIC 141733701 as a potential source of PA, PB, and ${\rm P_C}$ (right panels), and confirms that all three EBs are associated with TIC 141733685 (left panels). We note that the sky orientation of Fig. \ref{fig:tic141733685_skyview} is flipped along the x-axis (i.e. x $\rightarrow$ -x) with respect to the images shown here. }
    \label{fig:tic141733685_centAB}
\end{figure*}

\begin{figure*}
    \centering
    \includegraphics[width=0.45\textwidth]{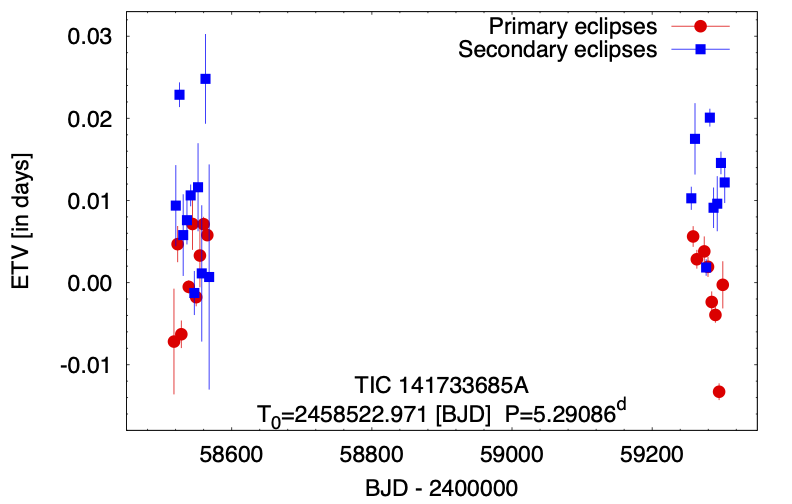}
    \includegraphics[width=0.45\textwidth]{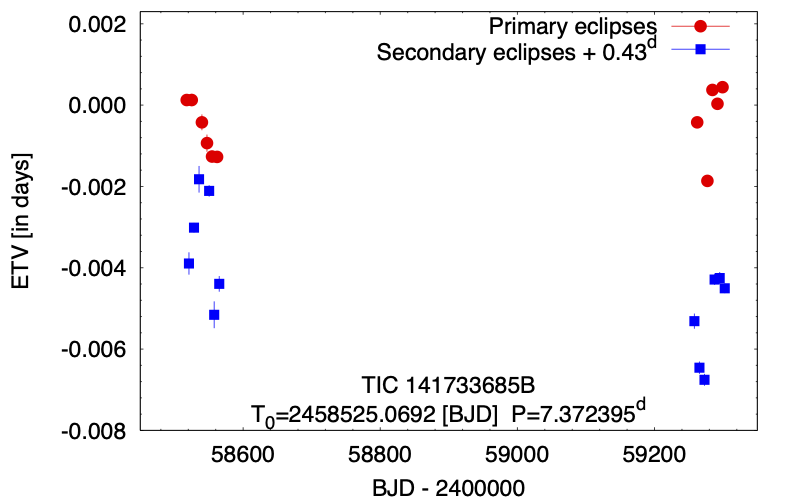}
    \caption{Measured ETVs for ${\rm P_A}$ (left) and ${\rm P_B}$ (right) for TIC 141733685, showing non-linear effects and indicating that the that the two are likely to be gravitationally-bound.}
    \label{fig:141733685_ETVs}
\end{figure*}

\item[$\bullet$ TIC 1337279468]
The lightcurve of TIC 1337279468 (TGV-97) exhibits three EBs with orbital periods ${\rm P_A}$ = 4.45 days, ${\rm P_B}$ = 5.94 days, and ${\rm P_C}$ = 10.57 days. The target was observed in Sectors 12 and 39. The {\em TESS} lightcurve for the latter is shown in Fig. \ref{fig:tic1337279468}. We note that the TIC lists a source at a separation of 0.59 arcsec (TIC 246039685) and another at a separation of 4.84 arcsec (TIC 246039695). Neither is present in the Gaia EDR3 data. 

There are three nearby field stars (TIC 1337279457, TIC 1337279471, and TIC 1337279458,) with separations of $\sim4-6$ arcsec and magnitude differences ${\rm \Delta T\ \sim5.5-6\ mag}$ (see Fig. \ref{fig:tic1337279468_skyview}). None of these is bright enough to produce the eclipses seen in the lightcurve as the shallowest primary eclipses (PB, depth ${\sim 25}$ parts-per-thousand, ppt) require a magnitude difference of ${\rm \Delta T\ < 3.2\ mag}$. Our photocenter analysis indicates that all three sets of eclipses are on-target (Fig. \ref{fig:tic1337279468}) and, accordingly, we consider the system to be a likely sextuple. At the time of writing, there is insufficient data to evaluate its hierarchy and structure. 

Gaia EDR3 shows AEN = 2.43 mas, with an AENS of 5891, and RUWE = 12.82 for TIC 1337279468. The target's proper motion is within ${\rm \sim1 \sigma}$ of those for TIC 1337279471, and the two might be potential co-moving septuple. However, the proper motions are small and the parallax accuracy is relatively low (the EDR3 parallax for TIC 1337279471 is ${\rm 0.4\pm0.5}$mas) so it is currently unclear whether this is the case. 

Finally, we note that there is another EB in the $13 \times 13$ {\em TESS} pixels surrounding TIC 1337279468 -- TIC 246039698 -- separated by about 4 pixels. The ephemeris of TIC 246039698 is unrelated to any of the three EBs of TIC 1337279468. 

\begin{figure}
    \centering
    \includegraphics[width=0.95\textwidth]{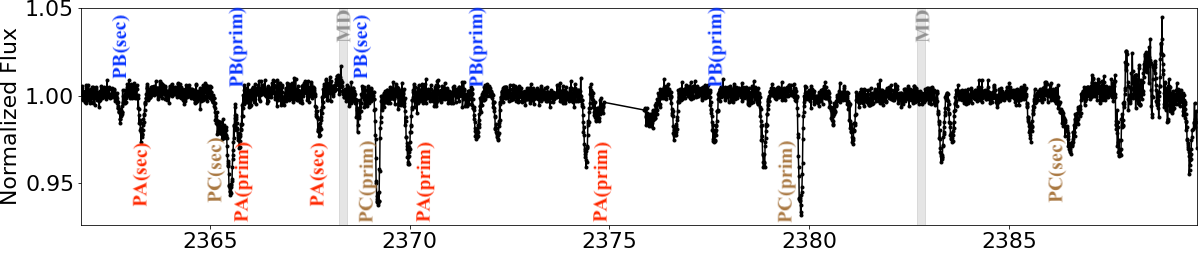}
    \includegraphics[width=0.85\textwidth]{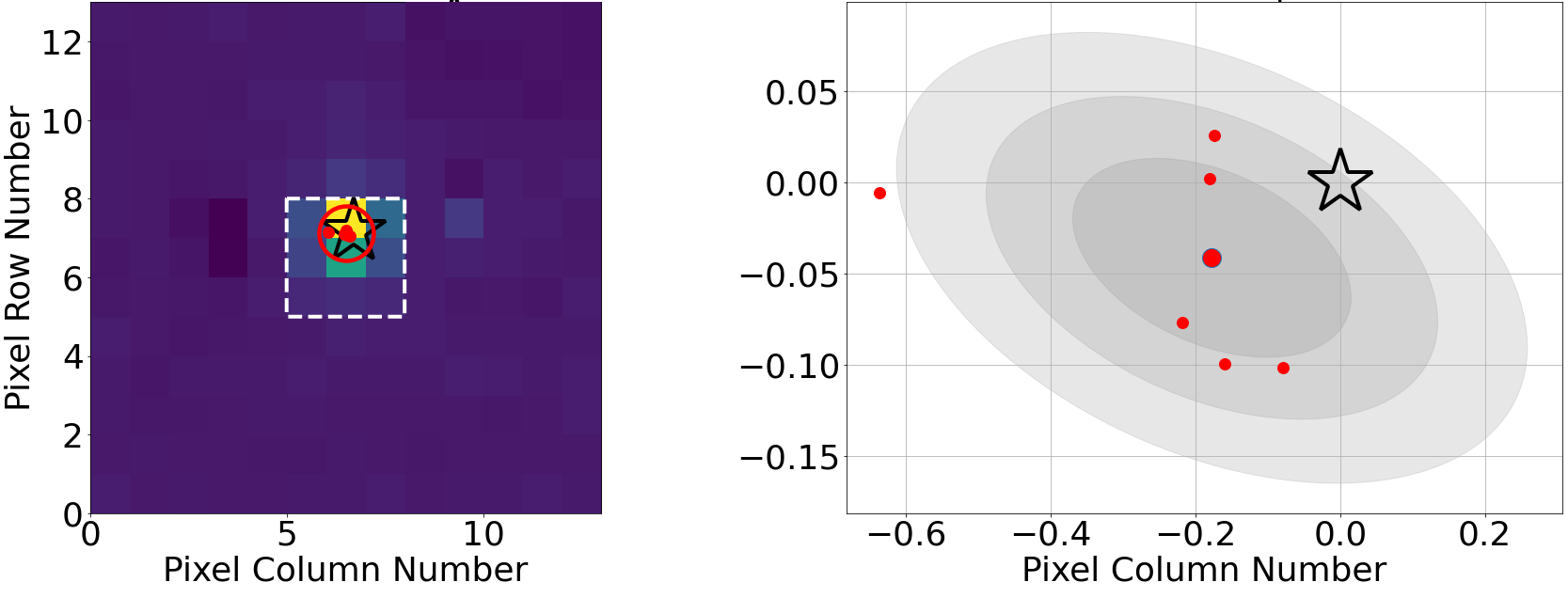}
    \includegraphics[width=0.85\textwidth]{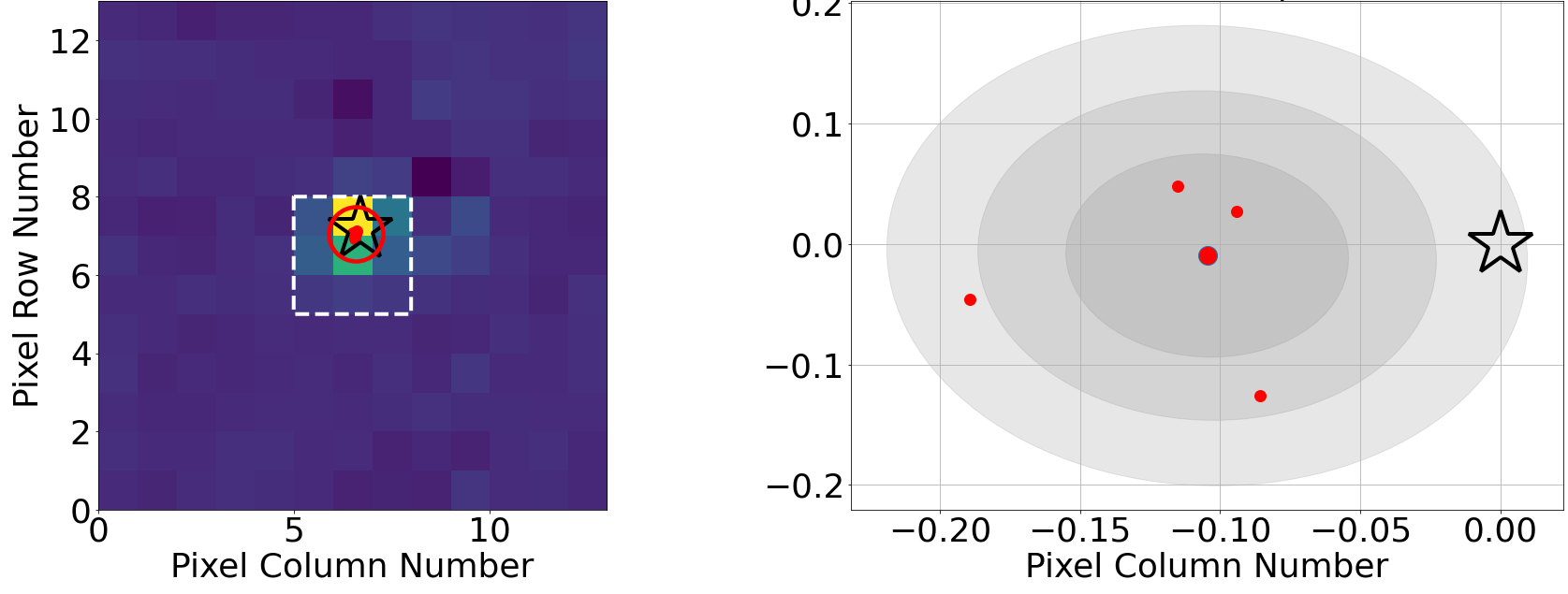}
    \includegraphics[width=0.85\textwidth]{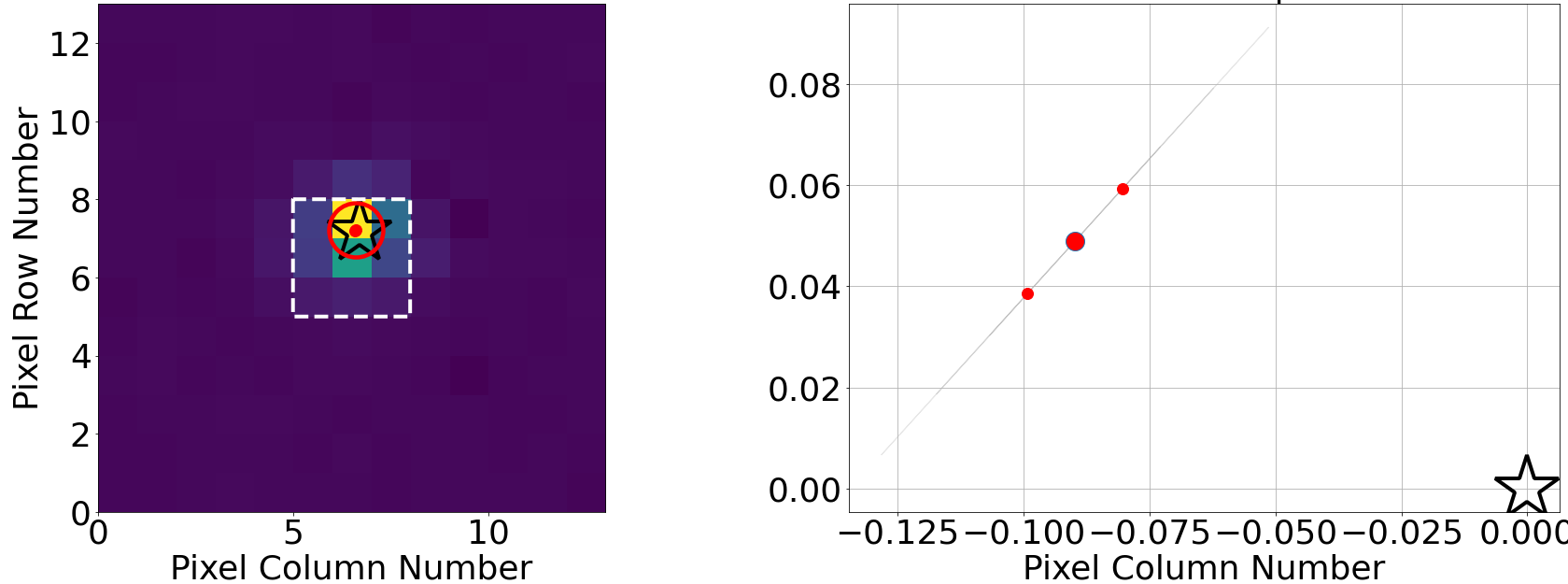}
    \caption{Upper panel: \textsf{eleanor} lightcurve for sextuple candidate TIC 1337279468, showing the Sector 39 data. The target exhibits three EBs with ${\rm P_A}$ = 4.45 days, ${\rm P_B}$ = 5.94 days, and ${\rm P_C}$ = 10.57 days, as highlighted in the panel. Second, third and fourth rows: Photocenter analysis of TIC 1337279468 for ${\rm P_A}$ (second row), ${\rm P_B}$ (third row) and ${\rm P_C}$ (fourth row) for Sector 39. The right panels show zoom-ins on the central pixel along with the corresponding confidence intervals (grey colors, ${\rm 1-, 2-, 3-\sigma}$, respectively) of the scatter in the measured photocenters. We note that there are only two eclipses of binary ${\rm P_C}$ in this sector, corresponding to two measured photocenters and thus the measured offset of 0.1 pixels (4th row, right panel) is not significant. Thus all three EBs originate from TIC 1337279468. }
    \label{fig:tic1337279468}
\end{figure}

\begin{figure}
    \centering
    \includegraphics[width=0.85\textwidth]{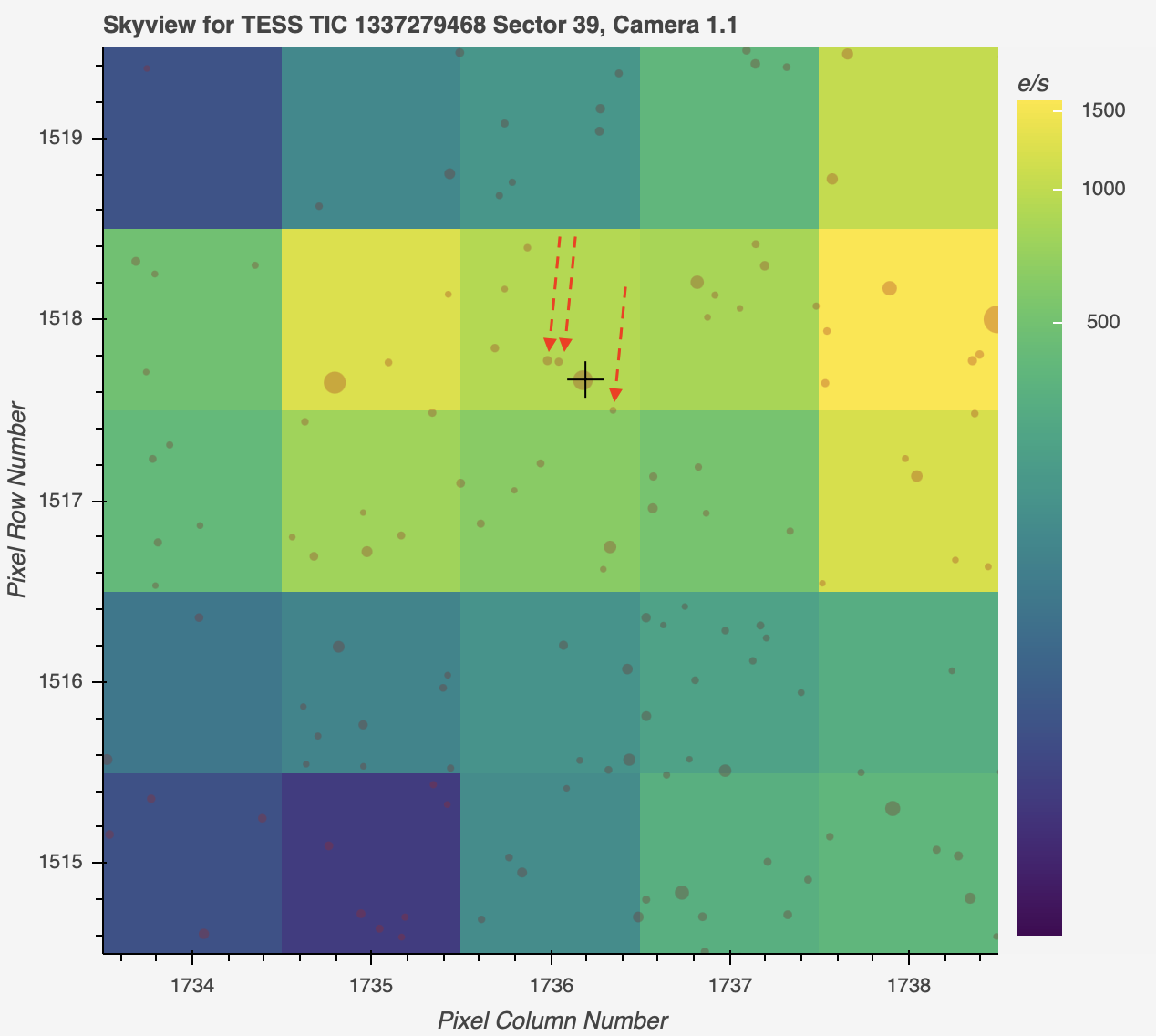}
    \caption{$5\times5$ pixels Skyview image of TIC 1337279468 for Sector 39, showing all Gaia sources brighter than G = 21 mag. There are three nearby field stars (TIC 1337279457, TIC 1337279471, and TIC 1337279458), all with separations of $\sim4-5$ arcsec and magnitude differences ${\rm \Delta T = 5.5-6\ mag}$ each (marked with dashed red arrows). None of them is bright enough to produce any of the eclipses seen in the target's lightcurve. The coordinates, parallax and proper motion for TIC 1337279468 and TIC 1337279471 are within ${\rm \sim1 \sigma}$, suggesting a potential co-moving septuple system.}
    \label{fig:tic1337279468_skyview}
\end{figure}

\item[$\bullet$ TIC 438226195]

TIC 438226195 (TGV-85) produced a single extra transit-like event in Sector 6 near time t = 1488.8598 (see Fig. \ref{fig:tic_438226195}). Our analysis indicates that the feature is on-target and, as we suspected this to be a candidate for a circumbinary planet, we added the target for 2-min cadence observations with {\em TESS} through the DDT program (\#27). It was observed again in Sector 33 and indeed produced another clear extra transit-like event, strengthening the CBP hypothesis. However, closer inspection of the Sector 33 lightcurve showed that the first secondary eclipse of the main EB, about 11.7 days before the clear extra event, is deeper than the rest---the trademark signature of a blend between two events. This indicated that the extra events are due to a second, on-target, EB making this a quadruple candidate. 

\begin{figure}
    \centering
    \includegraphics[width=1.05\textwidth]{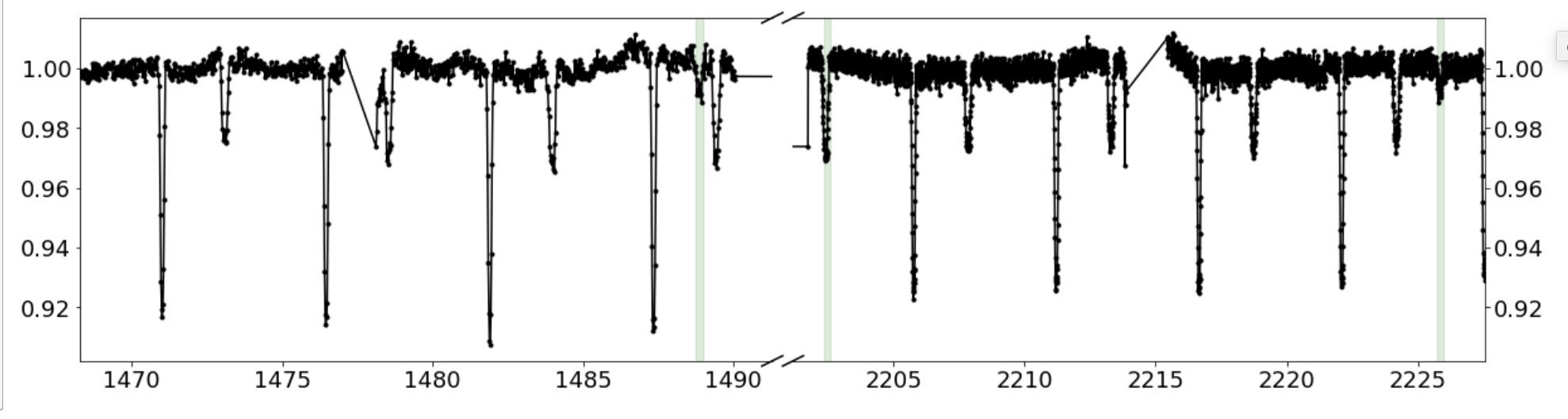}
    \caption{Lightcurve of TIC 438226195, highlighting the extra events in Sectors 6 and 33 (first and last green vertical bands) that mimic a circumbinary planet. Close inspection shows that the two events are due to a second EB with a period of ${\sim 11.7}$ days, with a third event blended with the first secondary eclipse in Sector 33 (second vertical green band), slightly deeper compared to the other secondaries in Sector 33. }
    \label{fig:tic_438226195}
\end{figure}

\end{description}

\subsubsection{Cross-match with archival data}

The results presented here can be further used to compare the measured ephemerides from TESS, especially for targets with multi-sector observations, to available archival photometric data from e.g. ASAS-SN \citep{2014ApJ...788...48S,2017PASP..129j4502K}, Digital Access to a Sky Century @ Harvard \citep[DASCH,][]{2012IAUS..285...29G}, Hungarian-made Automated Telescope Network \citep[HATNet,][]{2004PASP..116..266B}, Kilodegree Extremely Little Telescope \citep[KELT,][]{2007PASP..119..923P}, Kepler/K2 \citep{2010Sci...327..977B}, Northern Sky Variability Survey \citep[NSVS,][]{2004AJ....127.2436W}, Optical Gravitational Lensing Experiment \citep[OGLE,][]{1992AcA....42..253U}, Wide Angle Search for Planets \citep[WASP,][]{2006PASP..118.1407P}. Such a comparison could also allow testing for potential apsidal motion which, if large enough to be readily detected, would strengthen the case for a genuine quadruple system. 

\begin{figure}
    \centering
    \includegraphics[width=0.9\textwidth]{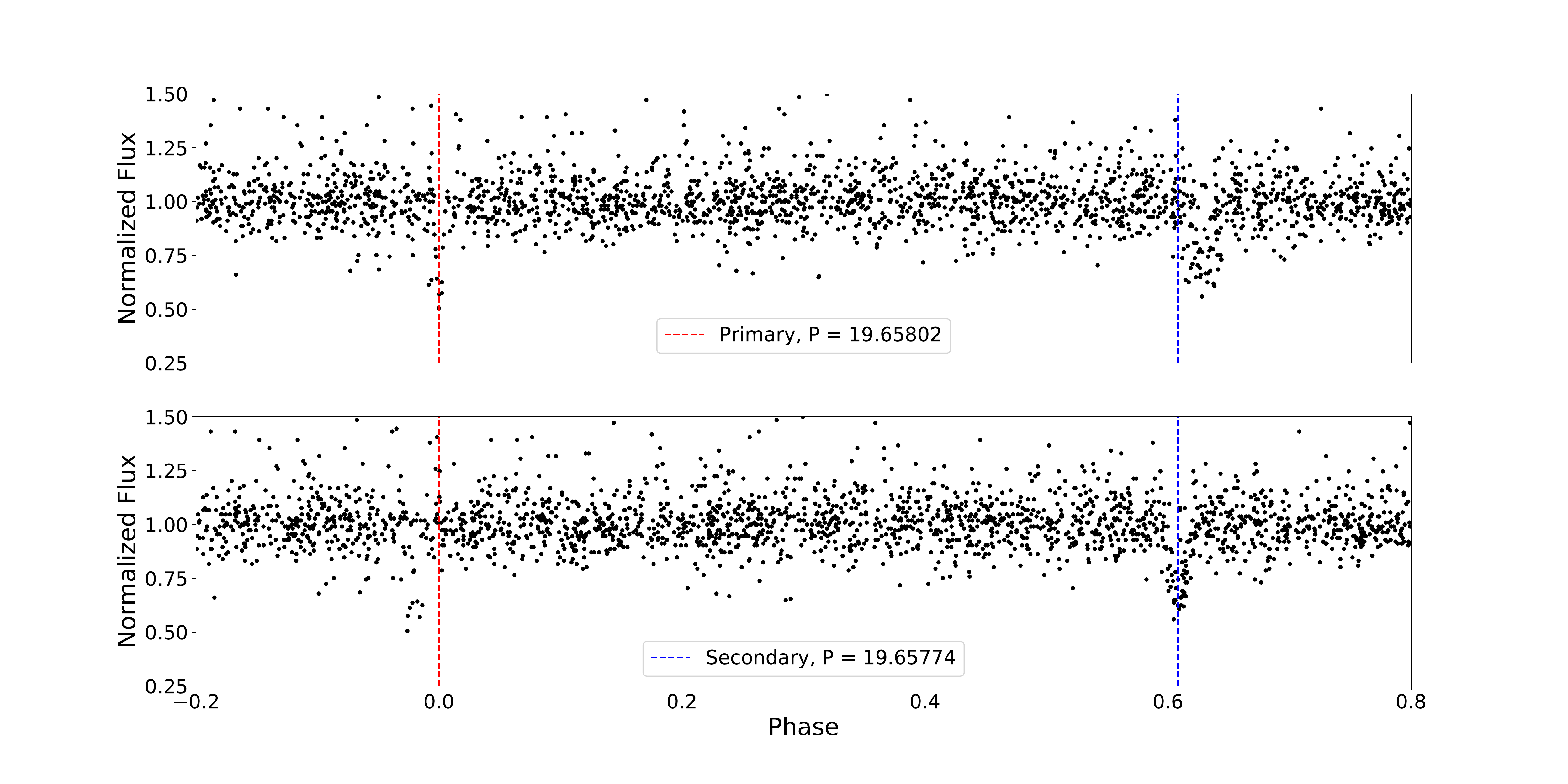}
    \caption{Phase-folded DASCH data of TIC 172900988 -- an eclipsing binary hosting a transiting circumbinary planet. The upper panel shows the data folded on the primary eclipse (dashed red line) and the lower panel shows the data folded on the secondary eclipse (dashed blue line). The two eclipses follow slightly different periods due to the apsidal motion of the binary caused by dynamical perturbation from the planet. }
    \label{fig:TIC_172900988_DASCH}
\end{figure}

As an example of this approach, we show on Fig. \ref{fig:TIC_172900988_DASCH} the DASCH data of TIC 172900988. This is an eclipsing binary system hosting a transiting circumbinary planet \citep{2021arXiv210508614K} and exhibiting clear apsidal motion. While the cadence of the DASCH observations is low and there are very few in-eclipse datapoints, the phase change of the secondary eclipse relative to the primary is clearly seen in the phase-folded DASCH data, and the two eclipses follow slightly different periods. However, checking the archival data for the bulk of our 97 targets is beyond the scope of this paper.

Another potential comparison can be with spectroscopic archives. For example, \citet{2021AJ....162..184K} examined the APOGEE DR16 and D17 for double-lined spectroscopic binaries. They detected several SB2 with corresponding eclipses in {\em TESS} data -- demonstrating the cross-match potential -- and uncovered 813 SB3 and 19 SB4 systems. One of the targets in our catalog, TIC 219469945, is listed as an SB3 system in the \citet{2021AJ....162..184K} database. None is listed as SB4. 

Finally, the targets in our catalog can be compared to astrometric binary stars detected by Gaia, especially those with significant astrometric excess noise (e.g. Belokurov et al. 2020, Peynore et al. 2021).

\subsubsection{Different datasets, different lightcurves}

During our visual examination of {\tt eleanor} and QLP lightcurves we have noticed that sometimes there are distinct differences between the two datasets. As another potential source of false positives, it is important to keep track of such differences and investigate their source. 

\begin{figure}
    \centering
    \includegraphics[width=0.9\textwidth]{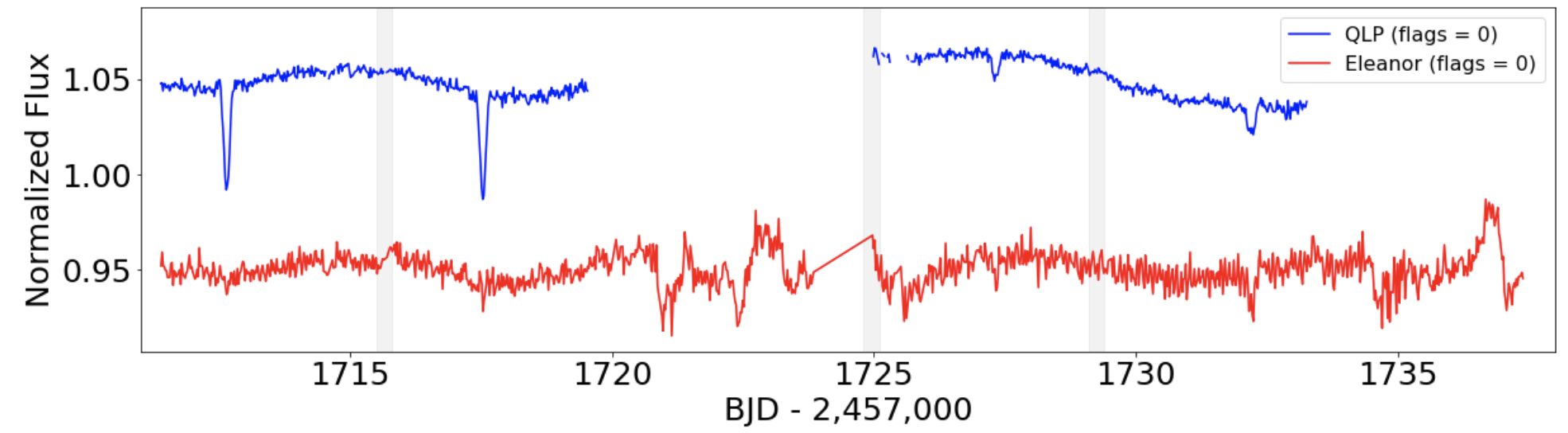}
    \includegraphics[width=0.9\textwidth]{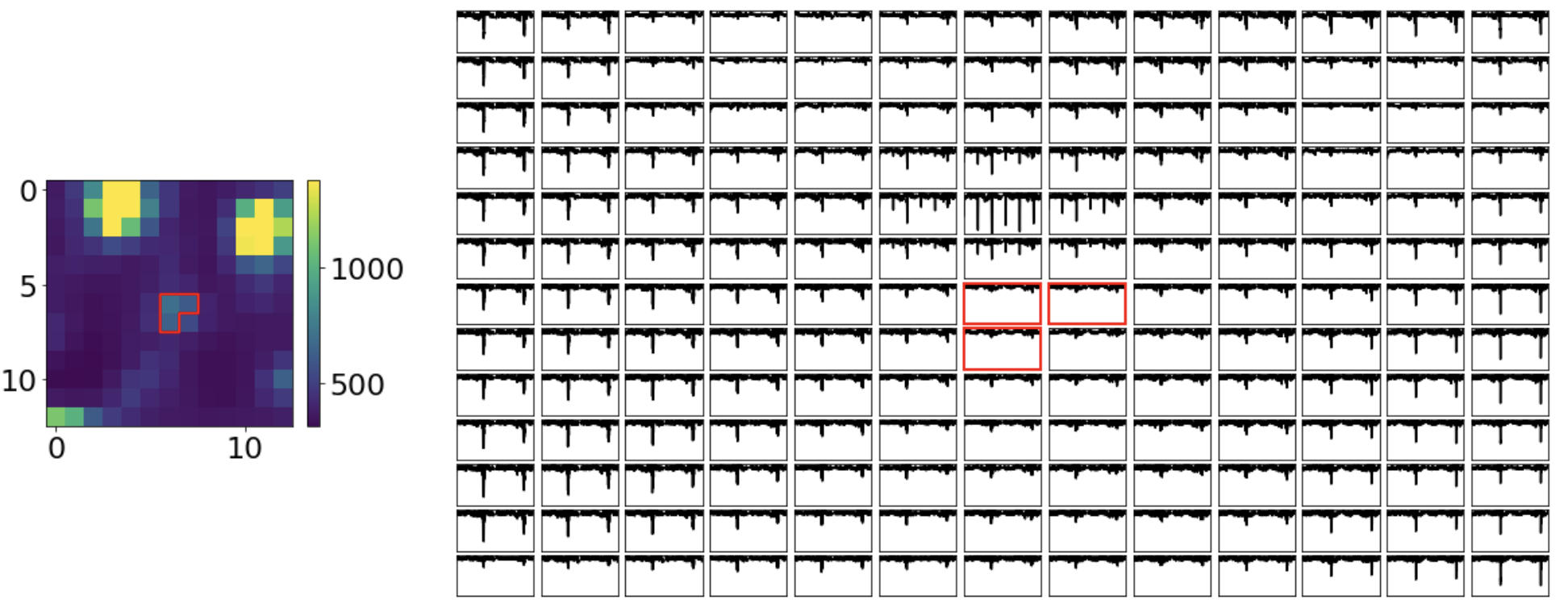}
    \caption{Upper panel: {\tt eleanor} (red) and QLP (blue) lightcurves for TIC 13120007, showing clear eclipses in the latter but no eclipses in the former. The vertical grey bands indicate momentum dumps. Lower left panel: TESS' $13 \times 13$ pixels field of view centered on the target. The red contour indicates the aperture used by {\tt eleanor} to extract the lightcurve. Lower right panel. Pixel-by-pixel {\tt eleanor} data for TIC 13120007 showing that the source of the EB is 2 pixels above the target.}
    \label{fig:eleanor_vs_QLP_1}
\end{figure}

An example is shown in Fig. \ref{fig:eleanor_vs_QLP_1} for the case of TIC 13120007 where there are clear eclipses in QLP data but no discernible features in {\tt eleanor} data. Note that the Sector 15 eclipses in QLP data have notably different depths before and after the data gap, which is a potential indicator for a false positive. The photocenter vetting analysis discussed above cannot be applied to {\tt eleanor} data as there are no eclipses. However, in cases like this we use the pixel-by-pixel {\tt eleanor} data to vet the target. Here, the data show that the target is not the source of the signal and the EB is in fact off-target, about 2 pixels away. 

\begin{figure}
    \centering
    \includegraphics[width=0.9\textwidth]{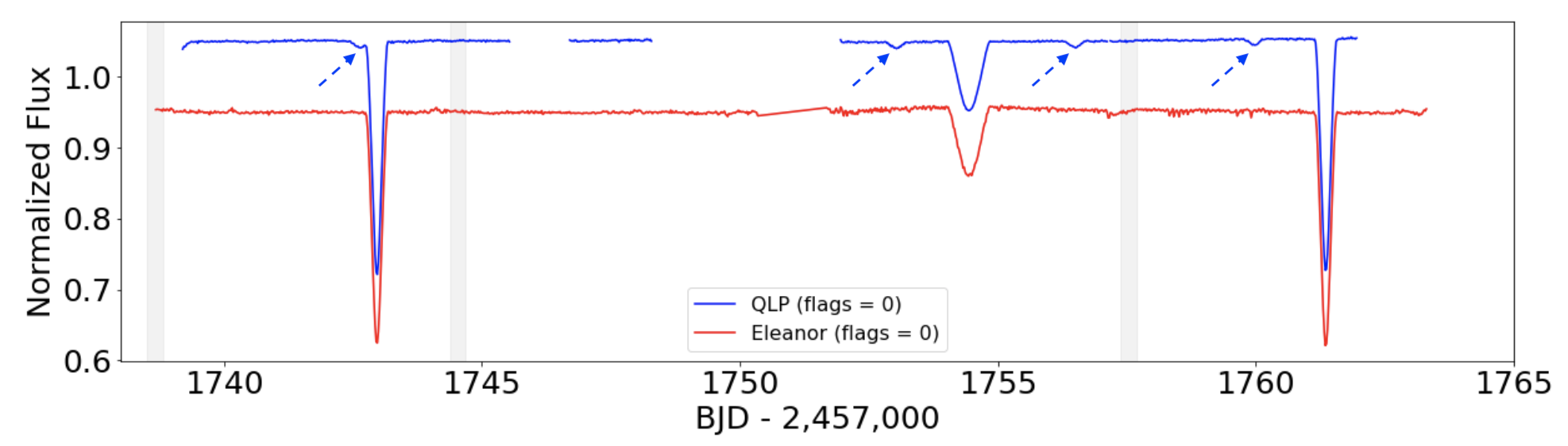}
    \includegraphics[width=0.9\textwidth]{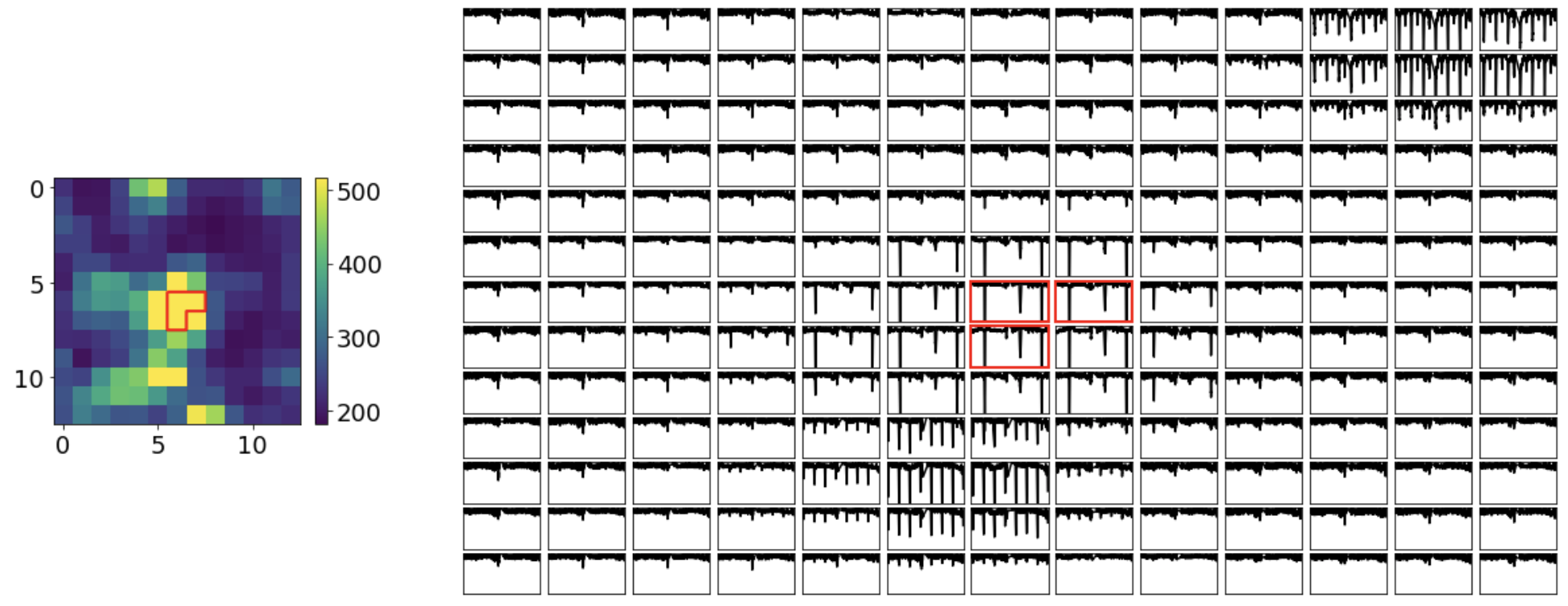}
    \caption{Same as Fig. \ref{fig:eleanor_vs_QLP_1} but for TIC 63708251. A potential quadruple candidate in QLP data (see blue curve in upper panel, and blue arrows) which is a false positive due to a nearby EB located about 5 pixels below the target. The vertical grey bands indicate momentum dumps. We note that there is another EB in the upper right corner of the pixel field of view. }
    \label{fig:eleanor_vs_QLP_2}
\end{figure}

Another example is shown in Fig. \ref{fig:eleanor_vs_QLP_2} for the case of TIC 63708251. Here, the {\tt eleanor} lightcurve shows one set of eclipses whereas QLP shows two sets of eclipses. As in the case of TIC 13120007, we cannot perform photocenter analysis of the {\tt eleanor} data for the shallow eclipses. However, as seen from Fig. \ref{fig:eleanor_vs_QLP_2}, the pixel-by-pixel {\tt eleanor} data show that only the deeper eclipses seen in both datasets are on-target whereas the second EB seen in QLP data is off-target, about 5 pixels away. Interestingly, there is a third EB in the target's $13 \times 13$ pixels {\em TESS} aperture, near the upper right corner. 

\subsubsection{Quadruple stars from TESS}

Our group has visually inspected millions of {\em TESS} lightcurves produced by several different pipelines. These include (i) $\approx2.2$ million {\tt eleanor} lightcurves (Powell et al. in prep) for Sectors 1-40; (ii) all {\em TESS} CTL \citep{2019AJ....158..138S} SC lightcurves for Sectors 1-42; (iii) QLP lightcurves for Sectors 2-4, 9, 13-27, 35 \citep{2020RNAAS...4..204H}; (iv) \citep{2018AJ....156..132O} lightcurves for Sectors 1-5; (v) CDIPS FFI lightcurves for Sectors 6-13 \citep{2020yCat..22450013B}; (vi) PATHOS Sector 4-14 lightcurves \citep{2019MNRAS.490.3806N}. Altogether, the different lightcurve sets include targets as faint as T = 15 mag and thus represent a significant portion of all available {\em TESS} data. 

Using these lightcurves, at the time of submission we have detected 2311 candidates for multiple stellar systems (triple and higher-order). Of these, 1319 (${\sim57\%}$) have been already vetted such that (i) 10\% passed all vetting tests; (ii) 9\% passed preliminary vetting tests, including the analysis of the pixel-by-pixel data; and (iii) 81\% were ruled out as false positives. The catalog of 97 quadruple candidates presented here represents all such fully-vetted systems except the handful that need further analysis. 905 candidates ($\sim39\%$) still need to be vetted, and the nature of 87 candidates (4\%) is currently unclear. Overall, the detection, vetting and analysis of our candidates is a continuous process, and we plan to present the results in a series of papers. 

The number of false positives we have encountered is more than an order of magnitude larger than the number of fully-vetted quadruple candidates presented here. Thus while completeness analysis is beyond the scope of this work, given the large number of targets inspected and assuming many of the additional candidates turn out to be real, {\em TESS} has the potential to increase the number of known eclipsing quadruple systems by more than a factor of two.

\section{Summary}
\label{sec:summary}

We have presented a catalog of 97 eclipsing quadruple star candidates detected in {\em TESS} Full Frame Images. The target stars have been identified through visual inspection and exhibit two sets of eclipses with two distinct periods, each with primary and, in most cases, secondary eclipses. All targets have been uniformly-vetted and passed a series of tests, including pixel-by-pixel and photocenter motion analysis. We outlined the procedures for determining orbital periods, eclipse depths and durations, and discussed the statistical properties of the sample.

\acknowledgments
This paper includes data collected by the \emph{TESS} mission, which are publicly available from the Mikulski Archive for Space Telescopes (MAST). Funding for the \emph{TESS} mission is provided by NASA's Science Mission directorate.

Resources supporting this work were provided by the NASA High-End Computing (HEC) Program through the NASA Center for Climate Simulation (NCCS) at Goddard Space Flight Center.  Personnel directly supporting this effort were Mark L. Carroll, Laura E. Carriere, Ellen M. Salmon, Nicko D. Acks, Matthew J. Stroud, Bruce E. Pfaff, Lyn E. Gerner, Timothy M. Burch, and Savannah L. Strong.

This research has made use of the Exoplanet Follow-up Observation Program website, which is operated by the California Institute of Technology, under contract with the National Aeronautics and Space Administration under the Exoplanet Exploration Program. 

This research is based on observations made with the Galaxy Evolution Explorer, obtained from the MAST data archive at the Space Telescope Science Institute, which is operated by the Association of Universities for Research in Astronomy, Inc., under NASA contract NAS5-26555.

TB acknowledges the financial support of the Hungarian National Research, Development and Innovation Office -- NKFIH Grant KH-130372.

This work has made use of data from the European Space Agency (ESA) mission {\it Gaia} (\url{https://www.cosmos.esa.int/gaia}), processed by the {\it Gaia} Data Processing and Analysis Consortium (DPAC, \url{https://www.cosmos.esa.int/web/gaia/dpac/consortium}). Funding for the DPAC has been provided by national institutions, in particular the institutions participating in the {\it Gaia} Multilateral Agreement.

Resources supporting this work were provided by the NASA High-End Computing (HEC) Program through the NASA Advanced Supercomputing (NAS) Division at Ames Research Center for the production of the SPOC data products.

This work makes use of observations from the LCOGT network.

\facilities{
\emph{Gaia},
MAST,
TESS,
WASP,
ASAS-SN,
NCCS,
FRAM,
PEST,
CHIRON,
TRES,
SOAR,
LCOGT}

\software{
{\tt Astrocut} \citep{astrocut},
{\tt AstroImageJ} \citep{Collins:2017},
{\tt Astropy} \citep{astropy2013,astropy2018}, 
{\tt Eleanor} \citep{eleanor},
{\tt IPython} \citep{ipython},
{\tt Keras} \citep{keras},
{\tt LcTools} \citep{2019arXiv191008034S,2021arXiv210310285S},
{\tt Lightcurvefactory} \citep{2013MNRAS.428.1656B,2017MNRAS.467.2160R,2018MNRAS.478.5135B},
{\tt Lightkurve} \citep{lightkurve},
{\tt Matplotlib} \citep{matplotlib},
{\tt Mpi4py} \citep{mpi4py2008},
{\tt NumPy} \citep{numpy}, 
{\tt Pandas} \citep{pandas},
{\tt PHOEBE} \citep{2011ascl.soft06002P},
{\tt Scikit-learn} \citep{scikit-learn},
{\tt SciPy} \citep{scipy},
{\tt Tensorflow} \citep{tensorflow},
{\tt Tess-point} \citep{tess-point}
{\tt wotan} \citep{2019AJ....158..143H}
}

\bibliography{refs}{}
\bibliographystyle{aasjournal}


\newpage
\newpage
\newpage

{\onecolumngrid
\begin{longtable}{llllllllrrrr}

\hline
\hline
TIC ID & RA & Dec & Binary & Period & T${_0}$ & Phase$_{s}$ & Dep$_{p}$ & Dep$_{s}$ & Dur$_{p}$ & Dur$_{s}$ \\
- & degrees & degrees & - & d & BJD-2457000 & - & ppt & ppt & hr & hr\\
\hline
\endhead

\multicolumn{11}{c}{{------------------------------------------------------------------}}\\
\multicolumn{11}{c}{{Continued on next page}}\\
\multicolumn{11}{c}{{TGV-N = TESS/Goddard/VSG quadruple candidate -N, Phase$_{s}$ = Secondary phase, Dep$_{n}$ = Depth of eclipse $n$,}}\\
\multicolumn{11}{c}{{Dur$_{n}$ = Duration of eclipse $n$, Teff = Composite effective temperature, ppt = parts-per-thousand}}\\
\endfoot

\hline \hline
\endlastfoot

9493888 & 69.510209 & 55.731524 & A &  2.098992 & 1816.2345 & 0.4999 & 146 & 117 & 2.9 & 2.7 \\
& & & B & 2.706156 & 1818.6919 & 0.5018 & 96 & 90 & 3.7 & 2.2 \\
\multicolumn{11}{l}{Additional information: TGV-1, Gaia EDR3 277142591660752128, Tmag: 11.98, Teff: 5112 K, Dist: 374.29 pc } \\
\multicolumn{11}{l}{Comments A: -- } \\
\multicolumn{11}{l}{Comments B: -- } \\
\hline
25818450 & 352.743444 & 53.069150 & A &  10.132402 & 1769.9109 & 0.6396 & 12 & 9 & -- & -- \\
& & & B & 17.101657 & 1765.8009 & -- & 80 & -- & -- & -- \\
\multicolumn{11}{l}{Additional information: TGV-2, Gaia EDR3 1992486494566143744, Tmag: 11.14, Teff: 7172 K, Dist: 838.61 pc } \\
\multicolumn{11}{l}{Comments A: depth difference between sectors } \\
\multicolumn{11}{l}{Comments B: depth difference between sectors } \\
\hline
27543409 & 122.702004 & 13.567217 & A &  2.122862 & 1493.1001 & 0.4964 & 50 & 15 & -- & -- \\
& & & B & 4.013356 & 1494.513 & -- & 75 & -- & -- & -- \\
\multicolumn{11}{l}{Additional information: TGV-3, Gaia EDR3 653620084592824960, Tmag: 13.22, Teff: 6421 K, Dist: 1878.88 pc } \\
\multicolumn{11}{l}{Comments A: -- } \\
\multicolumn{11}{l}{Comments B: SNR too low for secondary measurements } \\
\hline
31928452 & 53.969191 & -66.936899 & A &  2.8823 & 1337.9129 & 0.5014 & 30 & 25 & 2.1 & 1.6 \\
& & & B & 7.829944 & 1326.0719 & 0.5584 & 89 & 72 & 3.2 & 3.4 \\
\multicolumn{11}{l}{Additional information: TGV-4, Gaia EDR3 4670910529358997888, Tmag: 13.28, Teff: -- K, Dist: 574.18 pc } \\
\multicolumn{11}{l}{Comments A: Ellipsoidal variations; potential depth differences between sectors } \\
\multicolumn{11}{l}{Comments B: potential depth differences between sectors } \\
\hline
45160946 & 147.614561 & -36.191917 & A &  3.516299 & 1544.8002 & 0.4989 & 35 & 20 & 2.9 & 2.2 \\
& & & B & 7.846200 & 1550.438 & 0.4954 & 125 & 75 & 5.9 & 6.8 \\
\multicolumn{11}{l}{Additional information: TGV-5, Gaia EDR3 5434831348413276160, Tmag: 13.21, Teff: -- K, Dist: 446.90 pc } \\
\multicolumn{11}{l}{Comments A: Prominent ETVs; compact 2+2 quadruple system; Contaminator for TIC 45160944;}\\
\multicolumn{11}{l}{~~~~~~~~~~~~~~~~~~~Nearly-blended with TIC 872919203;  } \\
\multicolumn{11}{l}{Comments A: Prominent ETVs; compact 2+2 quadruple system; Contaminator for TIC 45160944;}\\
\multicolumn{11}{l}{~~~~~~~~~~~~~~~~~~~Nearly-blended with TIC 872919203;  } \\
\hline
52856877 & 17.334288 & 61.041245 & A &  5.186818 & 1791.059 & 0.5001 & 220 & 90 & 5.5 & 5.5 \\
& & & B & 18.586410 & 1812.5146 & 0.3736 & 200 & 55 & 9.4 & 7.6 \\
\multicolumn{11}{l}{Additional information: TGV-6, Gaia EDR3 522450134113950336, Tmag: 10.62, Teff: 8886 K, Dist: 847.17 pc } \\
\multicolumn{11}{l}{Comments A: -- } \\
\multicolumn{11}{l}{Comments B: Prominent ETVs } \\
\hline
63459761 & 308.525065 & 41.135869 & A &  4.244072 & 1715.1118 & 0.4846 & 15 & 10 & 7.0 & 9.2 \\
& & & B & 4.362293 & 1683.8128 & 0.6822 & 70 & 45 & 6.7 & 8.9 \\
\multicolumn{11}{l}{Additional information: TGV-7, Gaia EDR3 2067766791544561920, Tmag: 10.93, Teff: 3960 K, Dist: 2005.35 pc } \\
\multicolumn{11}{l}{Comments A: heavily-blended eclipses; ephemeris might be slightly off } \\
\multicolumn{11}{l}{Comments B: Potential ETVs; heavily-blended eclipses; ephemeris might be slightly off; }\\
\multicolumn{11}{l}{~~~~~~~~~~~~~~~~~~~False positive for TIC 63459765 } \\
\hline
73296637 & 121.017527 & -3.380218 & A &  1.483742 & 1493.5234 & -- & 11 & -- & -- & -- \\
& & & B & 1.844061 & 1494.0154 & -- & 27 & -- & -- & -- \\
\multicolumn{11}{l}{Additional information: TGV-8, Gaia EDR3 3069066742193077760, Tmag: 10.57, Teff: 7617 K, Dist: 592.68 pc } \\
\multicolumn{11}{l}{Comments A: Potential ETVs; heavily-blended eclipses; bump after eclipse? } \\
\multicolumn{11}{l}{Comments B: heavily-blended eclipses } \\
\hline
75740921 & 139.330597 & -45.038900 & A &  0.93308 & 1519.6949 & 0.5012 & 87 & 59 & 2.7 & 2.6 \\
& & & B & 0.986341 & 1520.8984 & 0.5067 & 81 & 26 & 2.5 & 2.0 \\
\multicolumn{11}{l}{Additional information: TGV-9, Gaia EDR3 5423787166437660032, Tmag: 12.40, Teff: -- K, Dist: 780.64 pc } \\
\multicolumn{11}{l}{Comments A: heavily-blended eclipses } \\
\multicolumn{11}{l}{Comments B: heavily-blended eclipses } \\
\hline
78568780 & 102.848953 & -22.167204 & A &  2.88838 & 1468.5374 & 0.4935 & 57 & 20 & 3.9 & 3.7 \\
& & & B & 23.903000 & 1500.8335 & -- & 37 & -- & -- & -- \\
\multicolumn{11}{l}{Additional information: TGV-10, Gaia EDR3 2925879645017239552, Tmag: 11.05, Teff: 7199 K, Dist: 2777.22 pc } \\
\multicolumn{11}{l}{Comments A: -- } \\
\multicolumn{11}{l}{Comments B: Period might be an integer of the listed value } \\
\hline
79140936 & 103.846452 & -22.623862 & A &  3.54389 & 1468.3795 & 0.3969 & 23 & 7 & 3.7 & 3.4 \\
& & & B & 30.913745 & 1479.6195 & 0.6070 & 400 & 325 & 11.5 & 13.0 \\
\multicolumn{11}{l}{Additional information: TGV-11, Gaia EDR3 2922782286399030912, Tmag: 10.90, Teff: 9022 K, Dist: 1084.70 pc } \\
\multicolumn{11}{l}{Comments A: -- } \\
\multicolumn{11}{l}{Comments B: -- } \\
\hline
80914862 & 106.116464 & -20.563763 & A &  1.967319 & 1492.6062 & -- & 95 & -- & -- & -- \\
& & & B & 18.666628 & 1495.0672 & -- & 158 & -- & -- & -- \\
\multicolumn{11}{l}{Additional information: TGV-12, Gaia EDR3 2929421068892142336, Tmag: 12.20, Teff: 8746 K, Dist: 2443.33 pc } \\
\multicolumn{11}{l}{Comments A: -- } \\
\multicolumn{11}{l}{Comments B: Period might be half of the listed value } \\
\hline
82818966 & 124.931906 & -47.096644 & A &  2.417501 & 1522.0949 & -- & 15 & -- & -- & -- \\
& & & B & 4.930024 & 1521.5166 & 0.3652 & 25 & 10 & 2.7 & 2.5 \\
\multicolumn{11}{l}{Additional information: TGV-13, Gaia EDR3 5516638037184056960, Tmag: 13.63, Teff: 8231 K, Dist: 1533.25 pc } \\
\multicolumn{11}{l}{Comments A: False positive for 82818975; low SNR; ephemeris might be slightly off } \\
\multicolumn{11}{l}{Comments B: -- } \\
\hline
89278612 & 301.219498 & 32.643051 & A &  2.557052 & 1684.2273 & 0.5008 & 53 & 32 & 4.5 & 4.6 \\
& & & B & 3.641763 & 1685.1175 & 0.4547 & 111 & 60 & 5.8 & 5.9 \\
\multicolumn{11}{l}{Additional information: TGV-14, Gaia EDR3 2055125259707080832, Tmag: 10.88, Teff: 8740 K, Dist: 20073.90 pc } \\
\multicolumn{11}{l}{Comments A: Potential ETVs } \\
\multicolumn{11}{l}{Comments B: -- } \\
\hline
95928255 & 136.380566 & -10.058331 & A &  2.36543 & 1518.123 & 0.5003 & 150 & 120 & 2.5 & 2.5 \\
& & & B & 4.426586 & 1518.7943 & -- & 40 & -- & -- & -- \\
\multicolumn{11}{l}{Additional information: TGV-15, Gaia EDR3 5743326067457954944, Tmag: 13.39, Teff: 4316 K, Dist: 391.73 pc } \\
\multicolumn{11}{l}{Comments A: -- } \\
\multicolumn{11}{l}{Comments B: -- } \\
\hline
97356407 & 106.502031 & -30.655710 & A &  1.533535 & 1494.5954 & 0.5009 & 15 & 5 & 4.0 & 4.0 \\
& & & B & 8.098533 & 1493.0889 & 0.6567 & 140 & 20 & 7.7 & 7.7 \\
\multicolumn{11}{l}{Additional information: TGV-16, Gaia EDR3 5604686034976764928, Tmag: 6.46, Teff: -- K, Dist: 300.37 pc } \\
\multicolumn{11}{l}{Comments A: Coherent lightcurve modulations (likely spots) with a period of about 1.2 days } \\
\multicolumn{11}{l}{Comments B: -- } \\
\hline
123098844 & 279.572843 & 44.698600 & A &  1.730707 & 1685.852 & -- & 35 & -- & -- & -- \\
& & & B & 11.210254 & 1685.345 & -- & 83 & -- & -- & -- \\
\multicolumn{11}{l}{Additional information: TGV-17, Gaia EDR3 2117607962866960256, Tmag: 10.79, Teff: 6761 K, Dist: 778.60 pc } \\
\multicolumn{11}{l}{Comments A: Ellipsoidal variations } \\
\multicolumn{11}{l}{Comments B: -- } \\
\hline
125952257 & 115.426817 & -27.582624 & A &  2.161915 & 1492.877 & 0.6134 & 140 & 100 & 4.8 & 4.9 \\
& & & B & 2.898585 & 1494.5706 & -- & 30 & -- & -- & -- \\
\multicolumn{11}{l}{Additional information: TGV-18, Gaia EDR3 5600205937417955968, Tmag: 11.36, Teff: 9578 K, Dist: 2851.53 pc } \\
\multicolumn{11}{l}{Comments A: -- } \\
\multicolumn{11}{l}{Comments B: -- } \\
\hline
130276377 & 119.825676 & -28.378980 & A &  2.757776 & 1495.0112 & 0.4833 & 74 & 48 & 4.6 & 4.7 \\
& & & B & 6.457989 & 1497.1837 & -- & 40 & -- & -- & -- \\
\multicolumn{11}{l}{Additional information: TGV-19, Gaia EDR3 5597797624687631232, Tmag: 11.93, Teff: 10329 K, Dist: 3549.38 pc } \\
\multicolumn{11}{l}{Comments A: Potential ETVs; ``ringing'' in folded LC } \\
\multicolumn{11}{l}{Comments B: Potential ETVs } \\
\hline
139650665 & 65.602471 & -18.916383 & A &  2.091887 & 1439.7823 & 0.4992 & 110 & 31 & 3.4 & 3.0 \\
& & & B & 10.631474 & 1438.907 & -- & 30 & -- & -- & -- \\
\multicolumn{11}{l}{Additional information: TGV-20, Gaia EDR3 5092393365381610880, Tmag: 10.74, Teff: 5515 K, Dist: 257.83 pc } \\
\multicolumn{11}{l}{Comments A: -- } \\
\multicolumn{11}{l}{Comments B: S32 data low SNR; ephemeris might be slightly off } \\
\hline
139944266 & 127.035827 & -44.334557 & A &  1.443586 & 1518.7658 & -- & 10 & -- & -- & -- \\
& & & B & 27.065312 & 1560.4232 & -- & 45 & -- & -- & -- \\
\multicolumn{11}{l}{Additional information: TGV-21, Gaia EDR3 5523049701799164160, Tmag: 10.36, Teff: -- K, Dist: 1041.59 pc } \\
\multicolumn{11}{l}{Comments A: -- } \\
\multicolumn{11}{l}{Comments B: -- } \\
\hline
146810480 & 160.527028 & -42.877977 & A &  0.544981 & 1545.7233 & -- & 25 & -- & -- & -- \\
& & & B & 0.734297 & 1544.2573 & -- & 131 & -- & -- & -- \\
\multicolumn{11}{l}{Additional information: TGV-22, Gaia EDR3 5391302370263824768, Tmag: 8.67, Teff: 6990 K, Dist: 368.84 pc } \\
\multicolumn{11}{l}{Comments A: heavily-blended eclipses; ephemeris might be slightly off } \\
\multicolumn{11}{l}{Comments B: heavily-blended eclipses; ephemeris might be slightly off } \\
\hline
161043618 & 223.425163 & 52.715848 & A &  1.350249 & 1744.7203 & 0.4995 & 75 & 20 & 4.0 & 4.0 \\
& & & B & 1.488497 & 1739.1594 & 0.5029 & 50 & 20 & 2.6 & 2.5 \\
\multicolumn{11}{l}{Additional information: TGV-23, Gaia EDR3 1594082407606370176, Tmag: 11.91, Teff: 5860 K, Dist: 1822.53 pc } \\
\multicolumn{11}{l}{Comments A: Ellipsoidal variations; heavily-blended eclipses; ephemeris might be slightly off } \\
\multicolumn{11}{l}{Comments B: heavily-blended eclipses; ephemeris might be slightly off } \\
\hline
177810207 & 106.524515 & -3.007379 & A &  1.422857 & 1493.1115 & 0.5001 & 125 & 90 & 3.4 & 3.3 \\
& & & B & 1.737808 & 1492.9002 & -- & 60 & -- & -- & -- \\
\multicolumn{11}{l}{Additional information: TGV-24, Gaia EDR3 3107984987053417728, Tmag: 13.47, Teff: -- K, Dist: 2485.24 pc } \\
\multicolumn{11}{l}{Comments A: blended eclipses } \\
\multicolumn{11}{l}{Comments B: blended eclipses } \\
\hline
178953404 & 69.096433 & -25.587820 & A &  3.182144 & 1438.3095 & 0.5001 & 40 & 25 & 2.2 & 2.6 \\
& & & B & 28.005157 & 1454.5725 & 0.8268 & 110 & 75 & 5.9 & 3.9 \\
\multicolumn{11}{l}{Additional information: TGV-25, Gaia EDR3 4894302326864856064, Tmag: 11.62, Teff: 5393 K, Dist: 332.55 pc } \\
\multicolumn{11}{l}{Comments A: -- } \\
\multicolumn{11}{l}{Comments B: Period may be an integer of the listed value } \\
\hline
190895730 & 134.131178 & -40.236827 & A &  0.459147 & 1519.026 & -- & 85 & -- & -- & -- \\
& & & B & 0.658889 & 1522.0179 & -- & 40 & -- & -- & -- \\
\multicolumn{11}{l}{Additional information: TGV-26, Gaia EDR3 5620856616906378880, Tmag: 11.63, Teff: 6059 K, Dist: 943.64 pc } \\
\multicolumn{11}{l}{Comments A: -- } \\
\multicolumn{11}{l}{Comments B: heavily-blended eclipses; ephemeris might be slightly off } \\
\hline
200094011 & 86.782967 & 0.298943 & A &  2.135567 & 1473.4704 & 0.5016 & 180 & 55 & 6.3 & 4.8 \\
& & & B & 2.437293 & 1475.0203 & 0.4993 & 230 & 60 & 8.7 & 6.3 \\
\multicolumn{11}{l}{Additional information: TGV-27, Gaia EDR3 3219205978378160512, Tmag: 8.66, Teff: -- K, Dist: 486.92 pc } \\
\multicolumn{11}{l}{Comments A: heavily-blended blends; ephemeris might be slightly off} \\
\multicolumn{11}{l}{Comments B: Ellipsoidal variations; heavily-blended blends; ephemeris might be slightly off } \\
\hline
201310151 & 4.111881 & -58.141759 & A &  5.538208 & 1335.888 & 0.4655 & 59 & 52 & 2.7 & 2.6 \\
& & & B & 8.485997 & 1346.3409 & 0.6407 & 120 & 37 & 3.2 & 3.1 \\
\multicolumn{11}{l}{Additional information: TGV-28, Gaia EDR3 4918400117051127424, Tmag: 14.29, Teff: -- K, Dist: 959.11 pc } \\
\multicolumn{11}{l}{Comments A: -- } \\
\multicolumn{11}{l}{Comments B: -- } \\
\hline
204698586 & 185.551846 & -24.224846 & A &  0.84381 & 1571.127 & 0.5109 & 40 & 13 & 1.8 & 1.9 \\
& & & B & 11.006707 & 1575.2285 & -- & 125 & -- & -- & -- \\
\multicolumn{11}{l}{Additional information: TGV-29, Gaia EDR3 3512835476313571072, Tmag: 10.84, Teff: 6392 K, Dist: 557.44 pc } \\
\multicolumn{11}{l}{Comments A: -- } \\
\multicolumn{11}{l}{Comments B: -- } \\
\hline
207137124 & 43.702325 & -56.810784 & A &  0.995134 & 1354.9037 & 0.4994 & 50 & 45 & 1.4 & 1.3 \\
& & & B & 32.155225 & 1379.1547 & 0.5011 & 180 & 30 & 6.8 & 6.0 \\
\multicolumn{11}{l}{Additional information: TGV-30, Gaia EDR3 4728114305421574528, Tmag: 13.49, Teff: 5451 K, Dist: 626.58 pc } \\
\multicolumn{11}{l}{Comments A: Period might be half of the quote value; depth differences between sectors; }\\
\multicolumn{11}{l}{~~~~~~~~~~~~~~~~~~~Potential apsidal motion } \\
\multicolumn{11}{l}{Comments B: depth differences between sectors; secondary heavily-blended; potential apsidal motion } \\
\hline
219469945 & 241.047908 & 43.030301 & A &  2.717596 & 1959.7065 & 0.5014 & 125 & 43 & 4.1 & 4.0 \\
& & & B & 14.965529 & 1957.7099 & 0.4821 & 80 & 20 & 6.3 & 7.3 \\
\multicolumn{11}{l}{Additional information: TGV-31, Gaia EDR3 1383708278019279616, Tmag: 12.04, Teff: 6323 K, Dist: 873.32 pc } \\
\multicolumn{11}{l}{Comments A: -- } \\
\multicolumn{11}{l}{Comments B: Potential ETVs; SB3 in Kounkel et al. (2021) } \\
\hline
232087348 & 30.996867 & -70.737328 & A &  2.614296 & 1325.4831 & 0.4988 & 150 & 40 & 4.7 & 4.8 \\
& & & B & 9.648651 & 2050.0082 & -- & 10 & -- & -- & -- \\
\multicolumn{11}{l}{Additional information: TGV-32, Gaia EDR3 4693178663478574464, Tmag: 12.32, Teff: 6154 K, Dist: 899.26 pc } \\
\multicolumn{11}{l}{Comments A: Ellipsoidal variations } \\
\multicolumn{11}{l}{Comments B: heavily-blended eclipses } \\
\hline
239872462 & 87.514177 & 34.417596 & A &  0.935825 & 1817.9507 & 0.5007 & 55 & 30 & -- & -- \\
& & & B & 2.961924 & 1819.4395 & -- & 60 & -- & -- & -- \\
\multicolumn{11}{l}{Additional information: TGV-33, Gaia EDR3 3454503601324455040, Tmag: 11.14, Teff: -- K, Dist: 1940.21 pc } \\
\multicolumn{11}{l}{Comments A: Ellipsoidal variations } \\
\multicolumn{11}{l}{Comments B: heavily-blended eclipses; ephemeris might be slightly off } \\
\hline
250119205 & 221.669507 & -53.195218 & A &  2.426677 & 1601.8218 & 0.4998 & 50 & 25 & 3.5 & 3.6 \\
& & & B & 5.097049 & 1601.5004 & 0.5002 & 80 & 40 & 4.6 & 4.8 \\
\multicolumn{11}{l}{Additional information: TGV-34, Gaia EDR3 5894766751792531328, Tmag: 10.25, Teff: 9090 K, Dist: 671.86 pc } \\
\multicolumn{11}{l}{Comments A: -- } \\
\multicolumn{11}{l}{Comments B: -- } \\
\hline
251757935 & 46.959379 & 54.066164 & A &  1.203606 & 1790.9094 & -- & 110 & -- & -- & -- \\
& & & B & 1.524962 & 1792.6865 & 0.4984 & 116 & 107 & 4.7 & 4.4 \\
\multicolumn{11}{l}{Additional information: TGV-35, Gaia EDR3 447114998986540672, Tmag: 11.51, Teff: 7736 K, Dist: 1665.22 pc } \\
\multicolumn{11}{l}{Comments A: heavily-blended eclipses } \\
\multicolumn{11}{l}{Comments B: heavily-blended eclipses } \\
\hline
255532033 & 240.549599 & -44.712993 & A &  4.173996 & 1629.2844 & 0.5638 & 140 & 130 & 7.0 & 7.7 \\
& & & B & 12.927714 & 1638.7287 & -- & 80 & -- & -- & -- \\
\multicolumn{11}{l}{Additional information: TGV-36, Gaia EDR3 5991306965083680128, Tmag: 10.34, Teff: 8685 K, Dist: 8414.36 pc } \\
\multicolumn{11}{l}{Comments A: -- } \\
\multicolumn{11}{l}{Comments B: -- } \\
\hline
256158466 & 266.899180 & -79.379329 & A &  5.774547 & 1627.0833 & 0.4966 & 131 & 128 & 3.1 & 3.4 \\
& & & B & 7.454373 & 1647.7103 & 0.5014 & 110 & 75 & 3.5 & 3.2 \\
\multicolumn{11}{l}{Additional information: TGV-37, Gaia EDR3 5776452909696577152, Tmag: 14.08, Teff: 4505 K, Dist: 713.86 pc } \\
\multicolumn{11}{l}{Comments A: Nearly-identical primary and secondary eclipses (secondary very slightly offset from phase 0.5); }\\
\multicolumn{11}{l}{~~~~~~~~~~~~~~~~~~~potential ellipsoidal variations; potential ETVs; potential co-moving quintuple with TIC 1508756606} \\
\multicolumn{11}{l}{Comments B: -- } \\
\hline
257776944 & 220.351703 & -71.048111 & A &  1.225515 & 1627.1031 & -- & 120 & -- & -- & -- \\
& & & B & 3.304494 & 1632.8798 & 0.4985 & 38 & 20 & -- & -- \\
\multicolumn{11}{l}{Additional information: TGV-38, Gaia EDR3 5798003887367982848, Tmag: 9.42, Teff: 9720 K, Dist: 993.91 pc } \\
\multicolumn{11}{l}{Comments A: -- } \\
\multicolumn{11}{l}{Comments B: -- } \\
\hline
260056937 & 63.922207 & 47.422198 & A &  2.446538 & 1819.2142 & 0.5001 & 230 & 190 & 6.0 & 5.9 \\
& & & B & 5.998071 & 1818.4123 & -- & 25 & -- & -- & -- \\
\multicolumn{11}{l}{Additional information: TGV-39, Gaia EDR3 234153100062678400, Tmag: 9.85, Teff: 8030 K, Dist: 742.95 pc } \\
\multicolumn{11}{l}{Comments A: Ellipsoidal variations } \\
\multicolumn{11}{l}{Comments B: heavily-blended eclipses } \\
\hline
264402353 & 323.090421 & 78.695151 & A &  1.697811 & 1765.9711 & -- & 15 & -- & -- & -- \\
& & & B & 8.096013 & 1769.7168 & -- & 17 & -- & -- & -- \\
\multicolumn{11}{l}{Additional information: TGV-40, Gaia EDR3 2284721597003238272, Tmag: 11.65, Teff: 6358 K, Dist: 1025.10 pc } \\
\multicolumn{11}{l}{Comments A: Ellipsoidal variations; secondary unclear } \\
\multicolumn{11}{l}{Comments B: Potential ETVs } \\
\hline
265274458 & 357.694256 & 73.156742 & A &  2.997813 & 1766.3538 & -- & 60 & -- & -- & -- \\
& & & B & 57.333800 & 1781.2291 & -- & 265 & -- & -- & -- \\
\multicolumn{11}{l}{Additional information: TGV-41, Gaia EDR3 2228101077501418240, Tmag: 12.03, Teff: 7972 K, Dist: 827.91 pc } \\
\multicolumn{11}{l}{Comments A: depth differences between sectors } \\
\multicolumn{11}{l}{Comments B: period may be an integer of the listed value; measured depths might be slightly off}\\ 
\multicolumn{11}{l}{~~~~~~~~~~~~~~~~~~~due to blended eclipses } \\
\hline
266657256 & 119.266500 & 4.186727 & A &  4.919396 & 1492.7016 & -- & 40 & -- & -- & -- \\
& & & B & 6.870338 & 1495.591 & 0.5222 & 50 & 44 & 2.0 & 2.9 \\
\multicolumn{11}{l}{Additional information: TGV-42, Gaia EDR3 3095184060360074368, Tmag: 13.17, Teff: 5514 K, Dist: 776.35 pc } \\
\multicolumn{11}{l}{Comments A: -- } \\
\multicolumn{11}{l}{Comments B: -- } \\
\hline
266771301 & 61.462323 & 52.245283 & A &  3.479611 & 1818.1561 & 0.5026 & 11 & 4 & 3.8 & 3.3 \\
& & & B & 3.833457 & 1824.8298 & -- & 5 & -- & -- & -- \\
\multicolumn{11}{l}{Additional information: TGV-43, Gaia EDR3 251000293962850944, Tmag: 11.51, Teff: -- K, Dist: 211.93 pc } \\
\multicolumn{11}{l}{Comments A: Potential ETVs; prominent systematics; depths might be slightly off due to systematics at}\\
\multicolumn{11}{l}{~~~~~~~~~~~~~~~~~~~momentum dumps} \\
\multicolumn{11}{l}{Comments B: prominent systematics; depths might be slightly off due to systematics at }\\
\multicolumn{11}{l}{~~~~~~~~~~~~~~~~~~~momentum dumps } \\
\hline
269811101 & 350.883799 & 60.886918 & A &  1.094351 & 1767.6151 & 0.5032 & 260 & 175 & 4.4 & 4.4 \\
& & & B & 1.660019 & 1778.8075 & -- & 95 & -- & 4.1 & 4.7 \\
\multicolumn{11}{l}{Additional information: TGV-44, Gaia EDR3 2015432374533799424, Tmag: 13.30, Teff: -- K, Dist: -- pc } \\
\multicolumn{11}{l}{Comments A: heavily-blended eclipses; measurements might be slightly off } \\
\multicolumn{11}{l}{Comments B: heavily-blended eclipses; primary completely blended; measurements might be slightly off } \\
\hline
271186951 & 132.355842 & -46.868040 & A &  1.731754 & 1520.1093 & 0.5026 & 80 & 45 & 5.4 & 3.5 \\
& & & B & 2.094425 & 1520.5993 & 0.5018 & 280 & 40 & 6.5 & 6.3 \\
\multicolumn{11}{l}{Additional information: TGV-45, Gaia EDR3 5329659041142860032, Tmag: 11.78, Teff: 9933 K, Dist: 2230.72 pc } \\
\multicolumn{11}{l}{Comments A: heavily-blended eclipses; prominent lightcurve variability; potential ETVs } \\
\multicolumn{11}{l}{Comments B: heavily-blended eclipses; prominent lightcurve variability } \\
\hline
274791367 & 149.395501 & -57.049303 & A &  1.207163 & 1544.2019 & 0.4998 & 110 & 100 & 4.9 & 4.8 \\
& & & B & 14.311675 & 1551.6412 & -- & 115 & -- & -- & -- \\
\multicolumn{11}{l}{Additional information: TGV-46, Gaia EDR3 5259511504953856640, Tmag: 10.97, Teff: -- K, Dist: 3114.95 pc } \\
\multicolumn{11}{l}{Comments A: Period might be half of the listed value; depth differences between sectors } \\
\multicolumn{11}{l}{Comments B: heavily-blended eclipses } \\
\hline
278352276 & 307.503640 & 48.607056 & A &  12.403102 & 1686.0153 & 0.5986 & 125 & 100 & 8.7 & 9.8 \\
& & & B & 18.810761 & 1684.6749 & 0.3913 & 85 & 20 & 7.2 & 7.1 \\
\multicolumn{11}{l}{Additional information: TGV-47, Gaia EDR3 2083748708463645952, Tmag: 10.00, Teff: 7156 K, Dist: 754.23 pc } \\
\multicolumn{11}{l}{Comments A: prominent ETVs; prominent systematics between sectors; secondary period might be slightly different } \\
\multicolumn{11}{l}{Comments B: prominent ETVs; prominent systematics between sectors; measurements might be slightly off } \\
\hline
283940788 & 8.851436 & 62.901596 & A &  0.876867 & 1765.7469 & 0.4940 & 30 & 10 & 6.5 & 6.5 \\
& & & B & 8.167894 & 1769.557 & 0.3044 & 165 & 130 & 6.6 & 8.5 \\
\multicolumn{11}{l}{Additional information: TGV-48, Gaia EDR3 430724888397553280, Tmag: 11.47, Teff: 7384 K, Dist: 4917.71 pc } \\
\multicolumn{11}{l}{Comments A: Ellipsoidal variations } \\
\multicolumn{11}{l}{Comments B: heavily-blended eclipses; potential ETVs } \\
\hline
284814380 & 11.863676 & 64.818018 & A &  4.079268 & 1796.4922 & 0.4998 & 175 & 128 & 6.2 & 6.3 \\
& & & B & 4.986977 & 1799.1701 & 0.2950 & 70 & 37 & 5.7 & 5.0 \\
\multicolumn{11}{l}{Additional information: TGV-49, Gaia EDR3 524420699465597184, Tmag: 11.54, Teff: 8669 K, Dist: 2573.12 pc } \\
\multicolumn{11}{l}{Comments A: -- } \\
\multicolumn{11}{l}{Comments B: -- } \\
\hline
285655079 & 233.992493 & -57.122535 & A &  2.456281 & 1631.341 & 0.5153 & 220 & 200 & 7.5 & 6.5 \\
& & & B & 6.691905 & 1627.9654 & 0.4507 & 110 & 97 & 5.8 & 7.2 \\
\multicolumn{11}{l}{Additional information: TGV-50, Gaia EDR3 5882657654103107072, Tmag: 11.96, Teff: 8463 K, Dist: 1701.55 pc } \\
\multicolumn{11}{l}{Comments A: -- } \\
\multicolumn{11}{l}{Comments B: -- } \\
\hline
285681367 & 31.557025 & 64.580199 & A &  2.366008 & 1791.9779 & 0.3457 & 135 & 110 & 4.4 & 4.9 \\
& & & B & 3.970279 & 1791.6442 & 0.2687 & 130 & 65 & 5.4 & 5.8 \\
\multicolumn{11}{l}{Additional information: TGV-51, Gaia EDR3 515086979621620992, Tmag: 11.80, Teff: 8363 K, Dist: 2416.95 pc } \\
\multicolumn{11}{l}{Comments A: blended eclipses } \\
\multicolumn{11}{l}{Comments B: -- } \\
\hline
286470992 & 45.330717 & 60.572294 & A &  3.110381 & 1820.1502 & -- & 20 & -- & -- & -- \\
& & & B & 4.128543 & 1818.6702 & 0.4151 & 200 & 200 & 9.9 & 10.6 \\
\multicolumn{11}{l}{Additional information: TGV-52, Gaia EDR3 463122720055223168, Tmag: 9.60, Teff: 8693 K, Dist: 2234.99 pc } \\
\multicolumn{11}{l}{Comments A: heavily-blended eclipses; ephemeris might be slightly off; potential ETVs } \\
\multicolumn{11}{l}{Comments B: -- } \\
\hline
292318612 & 31.769613 & 42.338435 & A &  1.665448 & 1792.9663 & -- & 130 & -- & -- & -- \\
& & & B & 1.733857 & 1792.3033 & -- & 140 & -- & -- & -- \\
\multicolumn{11}{l}{Additional information: TGV-53, Gaia EDR3 345585408079704704, Tmag: 13.25, Teff: 3959 K, Dist: 296.69 pc } \\
\multicolumn{11}{l}{Comments A: potential ETVs } \\
\multicolumn{11}{l}{Comments B: -- } \\
\hline
300446218 & 111.594902 & -66.281066 & A &  5.557302 & 1659.78 & 0.4925 & 50 & 30 & 2.3 & 2.2 \\
& & & B & 7.862233 & 1662.1022 & 0.4913 & 85 & 60 & 3.1 & 2.9 \\
\multicolumn{11}{l}{Additional information: TGV-54, Gaia EDR3 5281235758763812224, Tmag: 13.46, Teff: -- K, Dist: 460.87 pc } \\
\multicolumn{11}{l}{Comments A: depth differences between sectors; in CVZ; depths listed are for Sector 26 where }\\
\multicolumn{11}{l}{~~~~~~~~~~~~~~~~~~~the eclipses are the deepest } \\
\multicolumn{11}{l}{Comments B: depth differences between sectors; in CVZ; depths listed are for Sector 26 where }\\
\multicolumn{11}{l}{~~~~~~~~~~~~~~~~~~~the eclipses are the deepest } \\
\hline
306903715 & 98.004532 & 16.585394 & A &  4.86821 & 1471.9531 & -- & 24 & -- & -- & -- \\
& & & B & 6.516885 & 1472.6149 & -- & 32 & -- & -- & -- \\
\multicolumn{11}{l}{Additional information: TGV-55, Gaia EDR3 3369407242490695168, Tmag: 13.36, Teff: -- K, Dist: -- pc } \\
\multicolumn{11}{l}{Comments A: -- } \\
\multicolumn{11}{l}{Comments B: -- } \\
\hline
307119043 & 14.827524 & 51.221643 & A &  2.492484 & 1766.5396 & 0.5002 & 70 & 38 & 3.3 & 3.4 \\
& & & B & 4.742467 & 1766.2003 & 0.4885 & 90 & 63 & 4.1 & 4.2 \\
\multicolumn{11}{l}{Additional information: TGV-56, Gaia EDR3 404537094094988800, Tmag: 9.77, Teff: 7815 K, Dist: 468.08 pc } \\
\multicolumn{11}{l}{Comments A: potential ETVs } \\
\multicolumn{11}{l}{Comments B: blended eclipses } \\
\hline
309025182 & 151.007452 & -27.707244 & A &  1.401379 & 1547.7899 & -- & 75 & -- & -- & -- \\
& & & B & 1.680424 & 1548.1386 & -- & 114 & -- & -- & -- \\
\multicolumn{11}{l}{Additional information: TGV-57, Gaia EDR3 5466403275045876096, Tmag: 10.87, Teff: 6828 K, Dist: 825.89 pc } \\
\multicolumn{11}{l}{Comments A: blended eclipses } \\
\multicolumn{11}{l}{Comments B: blended eclipses } \\
\hline
309262405 & 89.424588 & 34.988493 & A &  4.198947 & 1819.386 & 0.5003 & 40 & 30 & 4.0 & 5.1 \\
& & & B & 6.908918 & 1818.9338 & -- & 35 & -- & -- & -- \\
\multicolumn{11}{l}{Additional information: TGV-58, Gaia EDR3 3451961801022167552, Tmag: 13.03, Teff: 8216 K, Dist: 3713.15 pc } \\
\multicolumn{11}{l}{Comments A: potential ETVs } \\
\multicolumn{11}{l}{Comments B: blended eclipses } \\
\hline
311838200 & 181.861947 & -70.489362 & A &  2.13397 & 1604.3638 & 0.4350 & 150 & 137 & 4.7 & 4.1 \\
& & & B & 2.496180 & 1604.849 & 0.5043 & 50 & 10 & 6.1 & 6.0 \\
\multicolumn{11}{l}{Additional information: TGV-59, Gaia EDR3 5855840775977123456, Tmag: 9.82, Teff: 7034 K, Dist: -- pc } \\
\multicolumn{11}{l}{Comments A: blended eclipses } \\
\multicolumn{11}{l}{Comments B: blended eclipses } \\
\hline
314802266 & 184.932193 & -71.498860 & A &  1.694052 & 1601.969 & 0.5005 & 73 & 60 & 3.7 & 3.7 \\
& & & B & 13.069676 & 1622.694 & 0.2130 & 266 & 224 & 10.5 & 6.6 \\
\multicolumn{11}{l}{Additional information: TGV-60, Gaia EDR3 5842816888938573440, Tmag: 11.18, Teff: 7512 K, Dist: 2499.64 pc } \\
\multicolumn{11}{l}{Comments A: heavily-blended eclipses } \\
\multicolumn{11}{l}{Comments B: heavily-blended eclipses; potential ETVs } \\
\hline
317863971 & 110.567508 & 3.031925 & A &  3.526276 & 1507.8495 & 0.4994 & 135 & 32 & 5.2 & 5.3 \\
& & & B & 3.733625 & 1499.5708 & 0.5857 & 60 & 15 & 3.8 & 5.0 \\
\multicolumn{11}{l}{Additional information: TGV-61, Gaia EDR3 3136148427638162176, Tmag: 10.24, Teff: 8506 K, Dist: 732.24 pc } \\
\multicolumn{11}{l}{Comments A: blended eclipses; depth differences between sectors } \\
\multicolumn{11}{l}{Comments B: blended eclipses; depth differences between sectors } \\
\hline
321471064 & 175.527634 & -62.267268 & A &  0.342094 & 1571.8063 & -- & 75 & -- & -- & -- \\
& & & B & 1.557273 & 1575.3683 & -- & 65 & -- & -- & -- \\
\multicolumn{11}{l}{Additional information: TGV-62, Gaia EDR3 5333374492778342400, Tmag: 11.91, Teff: 7202 K, Dist: 833.28 pc } \\
\multicolumn{11}{l}{Comments A: heavily-blended eclipses} \\
\multicolumn{11}{l}{Comments B: heavily-blended eclipses; period may be twice the listed value;}\\
\multicolumn{11}{l}{~~~~~~~~~~~~~~~~~~~secondary too blended and too low SNR for reliable measurements} \\
\hline
322727163 & 309.716625 & 50.466821 & A &  1.156328 & 1714.2718 & -- & 35 & -- & -- & -- \\
& & & B & 1.640142 & 1713.8436 & 0.4997 & 110 & 75 & 6.2 & 6.2 \\
\multicolumn{11}{l}{Additional information: TGV-63, Gaia EDR3 2180387912158160512, Tmag: 10.65, Teff: 7877 K, Dist: 826.96 pc } \\
\multicolumn{11}{l}{Comments A: heavily-blended eclipses; ephemeris might be slightly off; secondary too low SNR and too blended } \\
\multicolumn{11}{l}{Comments B: heavily-blended eclipses; ephemeris, depth and duration might be slightly off } \\
\hline
327885074 & 331.328300 & 59.445019 & A &  2.79561 & 1742.8408 & 0.4849 & 120 & 100 & 8.0 & 8.7 \\
& & & B & 3.345351 & 1743.5301 & -- & 15 & -- & -- & -- \\
\multicolumn{11}{l}{Additional information: TGV-64, Gaia EDR3 2199860842895906176, Tmag: 12.75, Teff: 9444 K, Dist: 8892.61 pc } \\
\multicolumn{11}{l}{Comments A: depth differences between sectors } \\
\multicolumn{11}{l}{Comments B: heavily-blended eclipses; measurements might be slightly off; potential ETVs } \\
\hline
328181241 & 43.153958 & 3.347882 & A &  22.625534 & 1430.1497 & 0.6359 & 150 & 60 & 12.9 & 18.0 \\
& & & B & 26.418506 & 1436.3582 & 0.8410 & 170 & 140 & 6.2 & 8.4 \\
\multicolumn{11}{l}{Additional information: TGV-65, Gaia EDR3 4657501191249536, Tmag: 10.57, Teff: 5183 K, Dist: 618.86 pc } \\
\multicolumn{11}{l}{Comments A: heavily-blended eclipses; flare in S31? extra event near 1427 } \\
\multicolumn{11}{l}{Comments B: secondary heavily-blended Sector 31  } \\
\hline
336882813 & 92.390615 & 14.628986 & A &  2.625028 & 1468.7932 & -- & 10 & -- & -- & -- \\
& & & B & 6.422862 & 1470.9941 & 0.6251 & 65 & 30 & 4.7 & 4.1 \\
\multicolumn{11}{l}{Additional information: TGV-66, Gaia EDR3 3345256224768675584, Tmag: 11.40, Teff: 6121 K, Dist: 2288.43 pc } \\
\multicolumn{11}{l}{Comments A: -- } \\
\multicolumn{11}{l}{Comments B: potential ETVs } \\
\hline
344541836 & 317.850729 & 57.620410 & A &  2.409932 & 1713.8429 & 0.6529 & 45 & 40 & 3.4 & 4.4 \\
& & & B & 2.755276 & 1713.3358 & 0.5010 & 30 & 11 & 5.0 & 3.0 \\
\multicolumn{11}{l}{Additional information: TGV-67, Gaia EDR3 2189576290314119936, Tmag: 7.77, Teff: -- K, Dist: 436.32 pc } \\
\multicolumn{11}{l}{Comments A: False positive for 344541871 } \\
\multicolumn{11}{l}{Comments B: -- } \\
\hline
348651800 & 121.724336 & -12.435678 & A &  2.077867 & 1496.1994 & 0.5030 & 10 & 5 & 4.0 & 4.1 \\
& & & B & 3.836246 & 1493.8673 & 0.4756 & 60 & 15 & 4.2 & 3.9 \\
\multicolumn{11}{l}{Additional information: TGV-68, Gaia EDR3 5727738565990859008, Tmag: 11.40, Teff: -- K, Dist: 675.02 pc } \\
\multicolumn{11}{l}{Comments A: Bright field star in central pixel } \\
\multicolumn{11}{l}{Comments B: Bright field star in central pixel } \\
\hline
357810643 & 138.466380 & -61.279782 & A &  3.120829 & 1545.9917 & -- & 15 & -- & -- & -- \\
& & & B & 20.528963 & 1563.6067 & 0.1628 & 80 & 15 & 7.3 & 11.4 \\
\multicolumn{11}{l}{Additional information: TGV-69, Gaia EDR3 5298906628619404672, Tmag: 7.03, Teff: 13910 K, Dist: 555.82 pc } \\
\multicolumn{11}{l}{Comments A: False positive for TIC 357810555; Potential slight centroid offset ($<$0.1 pixels) but no known source} \\
\multicolumn{11}{l}{~~~~~~~~~~~~~~~~~~~at the location; offset potentially due to the strong stellar variability on timescales}\\
\multicolumn{11}{l}{~~~~~~~~~~~~~~~~~~~comparable to the eclipse duration; Member of Cl Platais 8 open (galactic) Cluster}\\
\multicolumn{11}{l}{Comments B: False positive for TIC 357810555; Potential slight centroid offset ($<$0.1 pixels) but no known source} \\
\multicolumn{11}{l}{~~~~~~~~~~~~~~~~~~~at the location; offset potentially due to the strong stellar variability on timescales}\\
\multicolumn{11}{l}{~~~~~~~~~~~~~~~~~~~comparable to the eclipse duration; Member of Cl Platais 8 open (galactic) Cluster}\\
\hline
367448265 & 78.382438 & 35.653053 & A &  0.418238 & 1816.2930 & -- & 20 & -- & -- & -- \\
& & & B & 1.865520 & 1816.8633 & 0.5044 & 205 & 180 & 4.3 & 4.3 \\
\multicolumn{11}{l}{Additional information: TGV-70, Gaia EDR3 185395080832205440, Tmag: 7.83, Teff: 9212 K, Dist: 330.94 pc } \\
\multicolumn{11}{l}{Comments A: heavily blended eclipses; ellipsoidal variations} \\
\multicolumn{11}{l}{Comments B: heavily blended eclipses; potential ETVs; ellipsoidal variations} \\
\hline
370440624 & 143.232035 & -68.681123 & A &  2.235057 & 1572.7416 & 0.5004 & 50 & 15 & 2.6 & 2.8 \\
& & & B & 8.704980 & 1589.1513 & 0.5183 & 55 & 40 & 4.5 & 4.9 \\
\multicolumn{11}{l}{Additional information: TGV-71, Gaia EDR3 5243829651635260544, Tmag: 11.34, Teff: 6542 K, Dist: 672.99 pc } \\
\multicolumn{11}{l}{Comments A: -- } \\
\multicolumn{11}{l}{Comments B: heavily-blended eclipses } \\
\hline
375325607 & 315.793095 & 55.469314 & A &  1.311984 & 1711.9648 & 0.4972 & 45 & 25 & 4.8 & 3.5 \\
& & & B & 9.223201 & 1719.9224 & 0.3231 & 75 & 50 & 6.9 & 8.6 \\
\multicolumn{11}{l}{Additional information: TGV-72, Gaia EDR3 2188911081472114304, Tmag: 12.00, Teff: -- K, Dist: 1359.01 pc } \\
\multicolumn{11}{l}{Comments A: heavily-blended eclipses; measurements might be slightly off } \\
\multicolumn{11}{l}{Comments B: heavily-blended eclipses; measurements might be slightly off } \\
\hline
386202029 & 104.069404 & 15.809043 & A &  1.05668 & 1470.0427 & 0.5010 & 110 & 90 & 6.0 & 6.0 \\
& & & B & 1.252423 & 1469.0895 & 0.5054 & 240 & 150 & 5.5 & 5.5 \\
\multicolumn{11}{l}{Additional information: TGV-73, Gaia EDR3 3355042672829484032, Tmag: 13.24, Teff: 7096 K, Dist: 3995.38 pc } \\
\multicolumn{11}{l}{Comments A: heavily-blended eclipses; measurements might be slightly off; Ellipsoidal variations } \\
\multicolumn{11}{l}{Comments B: heavily-blended eclipses; measurements might be slightly off; Ellipsoidal variations } \\
\hline
387096013 & 140.130575 & -54.410580 & A &  2.134084 & 1545.8437 & -- & 75 & -- & -- & -- \\
& & & B & 6.144240 & 1544.9975 & 0.5527 & 40 & 35 & 4.8 & 5.0 \\
\multicolumn{11}{l}{Additional information: TGV-74, Gaia EDR3 5310631717561703168, Tmag: 12.02, Teff: -- K, Dist: 3920.34 pc } \\
\multicolumn{11}{l}{Comments A: -- } \\
\multicolumn{11}{l}{Comments B: secondary heavily blended } \\
\hline
389836747 & 23.295470 & 61.585307 & A &  2.56703 & 1792.2945 & 0.4970 & 180 & 130 & 8.2 & 8.2 \\
& & & B & 2.730402 & 1794.4352 & -- & 20 & -- & -- & -- \\
\multicolumn{11}{l}{Additional information: TGV-75, Gaia EDR3 510748306738071680, Tmag: 10.36, Teff: 8874 K, Dist: -- pc } \\
\multicolumn{11}{l}{Comments A: -- } \\
\multicolumn{11}{l}{Comments B: -- } \\
\hline
391620600 & 71.640389 & 44.753662 & A &  3.381354 & 1817.713 & 0.5003 & 90 & 40 & 2.9 & 3.1 \\
& & & B & 6.473460 & 1835.8181 & -- & 20 & -- & -- & -- \\
\multicolumn{11}{l}{Additional information: TGV-76, Gaia EDR3 204904785092921216, Tmag: 11.95, Teff: 5950 K, Dist: 460.63 pc } \\
\multicolumn{11}{l}{Comments A: -- } \\
\multicolumn{11}{l}{Comments B: potential slight centroid offset but no known source at the location; {\tt eleanor} aperture off-target; }\\
\multicolumn{11}{l}{~~~~~~~~~~~~~~~~~~~secondary eclipses clear but SNR too low for reliable measurements; heavily-blended eclipses; }\\
\multicolumn{11}{l}{~~~~~~~~~~~~~~~~~~~ephemeris might be slightly off} \\
\hline
392229331 & 54.767927 & 61.064204 & A &  1.822309 & 1819.0783 & 0.5040 & 148 & 47 & 3.6 & 3.6 \\
& & & B & 2.255905 & 1820.5001 & 0.4959 & 141 & 52 & 3.9 & 3.8 \\
\multicolumn{11}{l}{Additional information: TGV-77, Gaia EDR3 486430957815961344, Tmag: 10.34, Teff: 8028 K, Dist: 644.45 pc } \\
\multicolumn{11}{l}{Comments A: -- } \\
\multicolumn{11}{l}{Comments B: -- } \\
\hline
399492452 & 163.985711 & -69.196354 & A &  1.75514 & 1571.7531 & -- & 5 & -- & -- & -- \\
& & & B & 9.152187 & 1577.7638 & 0.6281 & 215 & 210 & 11.0 & 12.8 \\
\multicolumn{11}{l}{Additional information: TGV-78, Gaia EDR3 5231925239249670912, Tmag: 10.39, Teff: 6216 K, Dist: 923.85 pc } \\
\multicolumn{11}{l}{Comments A: ephemeris might be slightly off } \\
\multicolumn{11}{l}{Comments B: depth differences between sectors due to systematics; depth measurements might be slightly off } \\
\hline
408147984 & 213.482501 & -59.643098 & A &  1.072756 & 1601.2922 & -- & 65 & -- & -- & -- \\
& & & B & 3.804398 & 1604.1065 & 0.5025 & 23 & 10 & 5.5 & 5.5 \\
\multicolumn{11}{l}{Additional information: TGV-79, Gaia EDR3 5867020537173073024, Tmag: 11.10, Teff: -- K, Dist: 1458.91 pc } \\
\multicolumn{11}{l}{Comments A: very-short period ``ringing'' } \\
\multicolumn{11}{l}{Comments B: heavily-blended eclipses; ephemeris might be slightly off } \\
\hline
414026507 & 336.837717 & 56.740362 & A &  4.229981 & 1746.723 & 0.4212 & 70 & 55 & 7.5 & 8.5 \\
& & & B & 6.455288 & 1768.2955 & 0.5305 & 135 & 114 & 9.7 & 9.3 \\
\multicolumn{11}{l}{Additional information: TGV-80, Gaia EDR3 2007566227834657280, Tmag: 10.61, Teff: 9187 K, Dist: 2993.35 pc } \\
\multicolumn{11}{l}{Comments A: potential ETVs; heavily-blended eclipses } \\
\multicolumn{11}{l}{Comments B: heavily-blended eclipses } \\
\hline
414475823 & 224.211179 & -57.685879 & A &  3.478022 & 1602.0353 & 0.5010 & 140 & 90 & 5.8 & 5.6 \\
& & & B & 3.657410 & 1602.5871 & 0.4862 & 215 & 205 & 6.0 & 5.9 \\
\multicolumn{11}{l}{Additional information: TGV-81, Gaia EDR3 5880732619728164736, Tmag: 11.79, Teff: 7796 K, Dist: 1539.20 pc } \\
\multicolumn{11}{l}{Comments A: -- } \\
\multicolumn{11}{l}{Comments B: -- } \\
\hline
414969157 & 141.176162 & 22.200757 & A &  4.630508 & 1873.9443 & 0.5006 & 160 & 120 & 3.3 & 3.4 \\
& & & B & 6.928951 & 1880.0717 & 0.4718 & 110 & 50 & 3.8 & 3.1 \\
\multicolumn{11}{l}{Additional information: TGV-82, Gaia EDR3 638215205129955584, Tmag: 13.52, Teff: 5358 K, Dist: 854.30 pc } \\
\multicolumn{11}{l}{Comments A: potential ETVs } \\
\multicolumn{11}{l}{Comments B: -- } \\
\hline
427092089 & 321.112009 & 64.380683 & A &  2.00191 & 1739.3903 & -- & 13 & -- & 2.8 & -- \\
& & & B & 2.086000 & 1740.3478 & -- & 10 & -- & 3.0 & -- \\
\multicolumn{11}{l}{Additional information: TGV-83, Gaia EDR3 2220339079463012480, Tmag: 11.98, Teff: 8760 K, Dist: 1170.05 pc } \\
\multicolumn{11}{l}{Comments A: heavy blends } \\
\multicolumn{11}{l}{Comments B: heavy blends } \\
\hline
434452777 & 139.740446 & -20.557157 & A &  0.449904 & 1518.8906 & -- & 70 & -- & -- & -- \\
& & & B & 8.040997 & 1524.9251 & 0.4979 & 70 & 20 & 8.0 & 8.0 \\
\multicolumn{11}{l}{Additional information: TGV-84, Gaia EDR3 5676629108002652416, Tmag: 10.56, Teff: 7272 K, Dist: 668.13 pc } \\
\multicolumn{11}{l}{Comments A: blended star too faint to be source (${\rm \Delta T\sim8}$ mag) } \\
\multicolumn{11}{l}{Comments B: completely blended eclipses; ephemeris might be off } \\
\hline
438226195 & 96.502751 & 15.216367 & A &  5.442645 & 1471.014 & 0.6180 & 70 & 25 & 5.0 & 7.0 \\
& & & B & 11.697926 & 1488.8598 & -- & 10 & -- & -- & -- \\
\multicolumn{11}{l}{Additional information: TGV-85, Gaia EDR3 3356826458648806784, Tmag: 12.43, Teff: -- K, Dist: -938.20 pc } \\
\multicolumn{11}{l}{Comments A: -- } \\
\multicolumn{11}{l}{Comments B: Single eclipse in Sector 3 that looked like a circumbinary planet;}\\
\multicolumn{11}{l}{~~~~~~~~~~~~~~~~~~~Two more eclipses in Sector 33 (one blended with A)} \\
\hline
439511833 & 143.595968 & -56.106778 & A &  6.594526 & 1548.0668 & 0.5291 & 40 & 4 & 3.9 & 4.2 \\
& & & B & 11.048213 & 1552.4717 & 0.5091 & 85 & 30 & 6.4 & 6.2 \\
\multicolumn{11}{l}{Additional information: TGV-86, Gaia EDR3 5307096100472398080, Tmag: 10.58, Teff: -- K, Dist: 863.20 pc } \\
\multicolumn{11}{l}{Comments A: -- } \\
\multicolumn{11}{l}{Comments B: -- } \\
\hline
441794509 & 263.598849 & 74.472259 & A &  4.668622 & 1716.1273 & 0.5014 & 30 & 4 & 4.0 & 4.1 \\
& & & B & 14.785724 & 1739.0617 & -- & 15 & -- & -- & -- \\
\multicolumn{11}{l}{Additional information: TGV-87, Gaia EDR3 1655666294396195200, Tmag: 12.18, Teff: 6221 K, Dist: 892.90 pc } \\
\multicolumn{11}{l}{Comments A: secondary too low SNR for reliable measurements} \\
\multicolumn{11}{l}{Comments B: potential ETVs } \\
\hline
443862276 & 121.618495 & 7.254959 & A &  3.073838 & 1494.1607 & 0.4993 & 90 & 85 & 4.2 & 4.2 \\
& & & B & 7.040997 & 1494.5507 & 0.5086 & 60 & 55 & 4.6 & 5.0 \\
\multicolumn{11}{l}{Additional information: TGV-88, Gaia EDR3 3097738913065529472, Tmag: 13.72, Teff: 6398 K, Dist: 2253.94 pc } \\
\multicolumn{11}{l}{Comments A: Either the period is half of the listed value or there are ETVs; }\\
\multicolumn{11}{l}{~~~~~~~~~~~~~~~~~~~Slight depth differences between S7 and S34; } \\
\multicolumn{11}{l}{Comments B: potential slight differences between the primary and secondary periods } \\
\hline
454140642 & 64.773473 & 0.900042 & A &  10.3928 & 1445.5657 & 0.4828 & 100 & 90 & 5.3 & 5.3 \\
& & & B & 13.623900 & 1454.4599 & 0.4552 & 105 & 100 & 6.0 & 5.9 \\
\multicolumn{11}{l}{Additional information: TGV-89, Gaia EDR3 3255659981455492608, Tmag: 9.85, Teff: 6592 K, Dist: 358.89 pc } \\
\multicolumn{11}{l}{Comments A: Confirmed quadruple \citep[]{2021ApJ...917...93K} } \\
\multicolumn{11}{l}{Comments B: -- } \\
\hline
458740670 & 160.808111 & -57.895051 & A &  6.2628 & 1573.606 & 0.4969 & 140 & 60 & 5.8 & 5.8 \\
& & & B & 7.020700 & 2288.216 & 0.5050 & 100 & 25 & 6.2 & 6.2 \\
\multicolumn{11}{l}{Additional information: TGV-90, Gaia EDR3 5351127310598407680, Tmag: 12.03, Teff: 8480 K, Dist: 1829.94 pc } \\
\multicolumn{11}{l}{Comments A: Depth changes between sectors; ephemeris might be slightly off; Contaminator for 458740798; }\\
\multicolumn{11}{l}{~~~~~~~~~~~~~~~~~~~Member of Cl VDBH 99 -- Open (galactic) Cluster; many nearby stars but too faint to}\\
\multicolumn{11}{l}{~~~~~~~~~~~~~~~~~~~produce detected eclipses as contamination} \\
\multicolumn{11}{l}{Comments B: Depth changes between sectors; ephemeris might be slightly off; Contaminator for 458740798; }\\
\multicolumn{11}{l}{~~~~~~~~~~~~~~~~~~~Member of Cl VDBH 99 -- Open (galactic) Cluster; many nearby stars but too faint to}\\
\multicolumn{11}{l}{~~~~~~~~~~~~~~~~~~~produce detected eclipses as contamination} \\
\hline
459959916 & 71.489665 & 4.829619 & A &  1.054483 & 1439.1191 & 0.5030 & 15 & 14 & 1.5 & 1.5 \\
& & & B & 8.768239 & 1438.7434 & 0.5022 & 50 & 30 & 4.4 & 4.7 \\
\multicolumn{11}{l}{Additional information: TGV-91, Gaia EDR3 3281768503532006016, Tmag: 12.82, Teff: 6197 K, Dist: 1078.48 pc } \\
\multicolumn{11}{l}{Comments A: -- } \\
\multicolumn{11}{l}{Comments B: -- } \\
\hline
461614217 & 129.293558 & -43.823310 & A &  2.288076 & 1518.9443 & 0.5414 & 20 & 15 & 3.7 & 4.0 \\
& & & B & 9.365512 & 1519.3945 & 0.5071 & 75 & 50 & 7.8 & 7.9 \\
\multicolumn{11}{l}{Additional information: TGV-92, Gaia EDR3 5523221328699009280, Tmag: 10.32, Teff: -- K, Dist: 1801.82 pc } \\
\multicolumn{11}{l}{Comments A: heavily-blended eclipses } \\
\multicolumn{11}{l}{Comments B: blended eclipses } \\
\hline
462322817 & 151.061119 & -58.968755 & A &  4.438139 & 1545.0788 & -- & 25 & -- & 2.9 & -- \\
& & & B & 7.278400 & 1564.2055 & 0.2229 & 92 & 11 & 3.8 & 4.7 \\
\multicolumn{11}{l}{Additional information: TGV-93, Gaia EDR3 5258268953743230976, Tmag: 12.97, Teff: -- K, Dist: 1184.02 pc } \\
\multicolumn{11}{l}{Comments A: depth differences between sectors } \\
\multicolumn{11}{l}{Comments B: depth differences between sectors } \\
\hline
470710327 & 357.329052 & 61.962787 & A &  1.104686 & 1765.1668 & 0.5025 & 60 & 50 & 5.8 & 5.5 \\
& & & B & 19.950922 & 1805.4542 & -- & 70 & -- & -- & -- \\
\multicolumn{11}{l}{Additional information: TGV-94, Gaia EDR3 2012876727243918336, Tmag: 9.23, Teff: 8986 K, Dist: 947.34 pc } \\
\multicolumn{11}{l}{Comments A: heavily-blended eclipses; Ellipsoidal variations } \\
\multicolumn{11}{l}{Comments B: heavily-blended eclipses; measurements might be slightly off } \\
\hline
141733685 & 129.855041 & -47.360673 & A &  5.290886 & 1528.2575 & 0.4986 & 12 & 8 & 3.2 & 3.2 \\
& & & B & 7.372395 & 1525.0685 & 0.5587 & 220 & 130 & 5.5 & 6.9 \\
& & & C & 43.621521 & 1553.7523 & 0.4820 & 75 & 35 & 9.5 & 9.5 \\
\multicolumn{11}{l}{Additional information: TGV-95, Gaia EDR3 5329397357374933760, Tmag: 11.82, Teff: 12937 K, Dist: 1815.40 pc } \\
\multicolumn{11}{l}{Comments A: Potential on-target or co-moving sextuple (or even a septuple); depth differences between}\\
\multicolumn{11}{l}{~~~~~~~~~~~~~~~~~~~Cycle 1 and Cycle 3; potential non-linear ETVs likely due to dynamical interactions with binary B;}\\
\multicolumn{11}{l}{~~~~~~~~~~~~~~~~~~~Measurements based on FITSCH pipeline } \\
\multicolumn{11}{l}{Comments B: Prominent non-linear ETVs likely due to dynamical interactions with binary A;}\\
\multicolumn{11}{l}{~~~~~~~~~~~~~~~~~~~Measurements based on FITSCH pipeline } \\
\multicolumn{11}{l}{Comments C: Measurements based on FITSCH pipeline  } \\
\hline
168789840 & 63.520209 & -31.922876 & A &  1.305883 & 2151.193 & -- & 10 & -- & -- & -- \\
& & & B & 1.570013 & 2151.868 & 0.5002 & 40 & 20 & 3.2 & 3.0 \\
& & & C & 8.217111 & 2151.446 & 0.4991 & 55 & 10 & 3.4 & 3.3 \\
\multicolumn{11}{l}{Additional information: TGV-96, Gaia EDR3 4882954370433742336, Tmag: 10.80, Teff: 5730 K, Dist: -- pc } \\
\multicolumn{11}{l}{Comments A: Confirmed sextuple \citep[]{2021AJ....161..162P} } \\
\multicolumn{11}{l}{Comments B: -- } \\
\multicolumn{11}{l}{Comments C: -- } \\
\hline
1337279468 & 252.514389 & -44.034898 & A &  4.446303 & 1631.8797 & 0.4992 & 40 & 25 & 5.2 & 4.9 \\
& & & B & 5.939225 & 1635.2211 & 0.4999 & 25 & 15 & 5.3 & 4.9 \\
& & & C &  10.572127 & 1650.3 & 0.3624 & 60 & 20 & 4.1 & 14.4 \\
\multicolumn{11}{l}{Additional information: TGV-97, Gaia EDR3 5964415796876077952, Tmag: 12.30, Teff: -- K, Dist: 944.23 pc } \\
\multicolumn{11}{l}{Comments A: Potential sextuple/septuple; Target nearly blended with 1337279457 and 1337279471 }\\
\multicolumn{11}{l}{~~~~~~~~~~~~~~~~~~~(${\rm \Delta T = 6.53}$ mag, ${\rm \Delta T = 7.25}$ mag); potential co-moving with 1337279471}\\
\multicolumn{11}{l}{~~~~~~~~~~~~~~~~~~~as a co-moving sextuple or even septuple } \\
\multicolumn{11}{l}{Comments B: Potential ETVs } \\
\multicolumn{11}{l}{Comments C: -- } \\
\hline

\label{tbl:fitpar}
\end{longtable}
}


\begin{table}
\begin{tabular}[t]{cc}   
\begin{tabular}[t]{c|c|c|c} 
\hline
\hline
TIC & Tmag & ${\rm Depth_A (\%)}$ & ${\rm Depth_B (\%)}$ \\
\hline
97356407 & 6.46 & 2 & 14 \\
357810643 & 7.03 & 2 & 8 \\
344541836 & 7.77 & 4 & 3 \\
367448265 & 7.83 & 2 & 20 \\
200094011 & 8.66 & 18 & 23 \\
146810480 & 8.67 & 2 & 13 \\
470710327 & 9.23 & 6 & 7 \\
257776944 & 9.42 & 12 & 3 \\
286470992 & 9.60 & 2 & 20 \\
307119043 & 9.77 & 7 & 9 \\
311838200 & 9.82 & 15 & 5 \\
260056937 & 9.85 & 23 & 2 \\
454140642 & 9.85 & 10 & 10 \\
278352276 & 10.00 & 12 & 8 \\
317863971 & 10.24 & 14 & 6 \\
250119205 & 10.25 & 5 & 8 \\
461614217 & 10.32 & 2 & 7 \\
392229331 & 10.34 & 15 & 14 \\
255532033 & 10.34 & 14 & 8 \\
389836747 & 10.36 & 18 & 2 \\
139944266 & 10.36 & 1 & 4 \\
399492452 & 10.39 & 0 & 21 \\
434452777 & 10.56 & 7 & 7 \\
73296637 & 10.57 & 1 & 2 \\
328181241 & 10.57 & 15 & 17 \\
439511833 & 10.58 & 4 & 8 \\
414026507 & 10.61 & 7 & 13 \\
52856877 & 10.62 & 22 & 20 \\
322727163 & 10.65 & 4 & 11 \\
139650665 & 10.74 & 11 & 3 \\
123098844 & 10.79 & 4 & 8 \\
168789840 & 10.80 & 1 & 4, 6 \\
204698586 & 10.84 & 4 & 12 \\
\hline
\end{tabular} &  
\begin{tabular}[t]{c|c|c|c} 
\hline
\hline
TIC & Tmag & ${\rm Depth_A (\%)}$ & ${\rm Depth_B (\%)}$ \\
\hline
309025182 & 10.87 & 8 & 11 \\
89278612 & 10.88 & 5 & 11 \\
79140936 & 10.90 & 2 & 40 \\
63459761 & 10.93 & 2 & 7 \\
274791367 & 10.97 & 11 & 11 \\
78568780 & 11.05 & 6 & 3 \\
408147984 & 11.10 & 6 & 2 \\
239872462 & 11.14 & 6 & 6 \\
25818450 & 11.14 & 1 & 8 \\
314802266 & 11.18 & 7 & 26 \\
370440624 & 11.34 & 5 & 5 \\
125952257 & 11.36 & 14 & 3 \\
336882813 & 11.40 & 1 & 6 \\
348651800 & 11.40 & 1 & 6 \\
283940788 & 11.47 & 3 & 16 \\
251757935 & 11.51 & 11 & 11 \\
266771301 & 11.51 & 1 & 0 \\
284814380 & 11.54 & 18 & 7 \\
178953404 & 11.62 & 4 & 11 \\
190895730 & 11.63 & 8 & 4 \\
264402353 & 11.65 & 2 & 1 \\
271186951 & 11.78 & 8 & 28 \\
414475823 & 11.79 & 14 & 21 \\
285681367 & 11.80 & 14 & 13 \\
141733685 & 11.82 & 1 & 8, 22 \\
321471064 & 11.91 & 8 & 6 \\
161043618 & 11.91 & 8 & 5 \\
130276377 & 11.93 & 7 & 4 \\
391620600 & 11.95 & 9 & 2 \\
285655079 & 11.96 & 22 & 11 \\
9493888 & 11.98 & 15 & 9 \\
427092089 & 11.98 & 1 & 1 \\
\hline
\end{tabular} \\
\end{tabular}

\caption{Quadruple candidates brighter than Tmag = 12 and eclipse depths greater than 1\% (sorted by magnitude).}
\label{tab:good_for_obs}
\end{table}

\end{document}